\documentclass[sigconf, nonacm]{acmart}
\usepackage[cjk]{kotex}

\usepackage{graphicx}
\usepackage{amsmath}
\usepackage{color}
\usepackage{epsfig}
\usepackage{url}	
\usepackage{array,ragged2e}
\usepackage{mathtools}  
\usepackage{subfigure}	
\usepackage{ifpdf}
\usepackage{endnotes}
\usepackage{xspace}	
\usepackage{multirow}	
\usepackage{tabularx}	
\usepackage{balance}	
\usepackage{marginnote}
\usepackage{enumitem}
\usepackage{breakurl}	
\usepackage{flushend}
\usepackage{xcolor}

\usepackage{algorithm} 
\usepackage{algpseudocode}
\usepackage{booktabs}
\usepackage{hhline}

\newcommand{\ie}{{\em i.e.}}
\newcommand{\eg}{{\em e.g.}}

\definecolor{blue}{rgb}{0.0,0.0,1.0}
\long\def\comment#1{}
\newcommand{\reminder}[1]{[\mbox{$\Longrightarrow$}{\textcolor{blue}{#1}}{\mbox{$\Longleftarrow$}}]}
\newcommand{\module}[1]{\vspace{1ex}{\bf {#1}}\:\:}

\def\l2p{{\em L2P}\xspace}

\newcommand\vldbdoi{10.14778/3461535.3461538}

\usepackage{comment}
\newcommand{\falloc}{{\em FlashAlloc}\xspace}

\begin{document}

\title{Enlightening Flash Storage to Stream Writes by Objects}
\comment{ 
GROUP Interface for Enabling Flash Storage To Stream Writes by Objects \\
FlashGroup: Enabling Flash Storage to Stream Writes by Objects \\
(FlashAlloc:) Enlightening/Hinting Flash Storage ... \\
Hinting Flash Storage to Stream Writes by Objects \\
FlashAlloc: Making Flash Storage Object-Aware \\
FlashAlloc: Preempting Physical Blocks in Flash Storage \\ 
FlashAlloc: Notification of Same Logical Data in Flash Storages \\
FlashAlloc: Clustering (or Grouping) Writes By Logical Objects into Physical Flash Blocks \\ 
FlashAlloc: Clustering (or Grouping) Writes By Logical Objects in Flash Storages\\ 
FlashAlloc: De-multiplexing logical write streams into physical blocks in flash storages
For the given logical ranges, contiguous physical (erase unit) flash block에 write하도록 한다는 의미 영단어\\
FlashAlloc(Trim/Discard-like):Dedicating Physical Blocks to Logical 
Eager (or Pre-) L2P mapping in flash storage ; Write-Ahead (Pre-)Mapping in SSD to Group Data with same Death Time \\
FALLOC: Eager Allocation of Clustered Physical Space for Logical Objects \\  
FlashAlloc: Efficiently de-multiplexing(or streaming) writes (logical write streams) (into different blocks) in flash storages}

\author{Jong-Hyeok Park Gihwan Oh Sang-Won Lee}
\affiliation{Sungkyunkwan University Suwon, Korea}
\email{{akindo19, wurikiji, swlee}@skku.edu}




\begin{abstract}
For a write request, today's flash storage cannot distinguish the logical object it comes from.\comment{belongs to has no knowledge about the object it belongs to.} In such object-oblivious flash devices, concurrent writes from different objects are simply packed in their arrival order to flash memory blocks; hence objects with different lifetimes are multiplexed onto the same flash blocks. This multiplexing incurs write amplification, worsening the performance. 



Tackling the multiplexing problem, we propose a novel interface for flash storage, \falloc. It is used to pass the logical address ranges of logical objects to the flash storage and thus enlighten the storage to stream writes by objects. The object-aware flash storage can de-multiplex writes from different objects with distinct deathtimes into per-object dedicated flash blocks. Given that popular data stores separate writes using objects (\eg, SSTables in RocksDB), we can achieve, unlike the existing solutions, transparent write streaming just by calling \falloc upon object creation. 
Our experimental results using an open-source SSD prototype demonstrate that \falloc can reduce write amplification factor (WAF) in RocksDB, F2FS, and MySQL by 1.5, 2.5, and 0.3, respectively and thus improve throughput by 2x, 1.8x, and 1.2x, respectively. In particular, \falloc will mitigate the interference among multi-tenants. When RocksDB and MySQL were run together on the same SSD, \falloc decreased \comment{device-level} WAF from 4.2 to 2.5 and doubled their throughputs. 
 \comment{In addition, given that its cumulative WAFs have sustained less than 1.3 for both workloads, \trf performs close to an ideal FTL with WAF of 1.}
\comment{To address the multiplexing problem, we propose a novel interface for flash storage, \falloc. Its purpose is to pass the host-side semantic about object's logical address range to the flash storage and thus make the storage object-aware in handling writes. Enlightened by \falloc, flash storage can stream writes by objects into per-object dedicated flash blocks, thus preventing the multiplexing. \comment{\falloc makes flash storage object-aware in writing and thus stream writes by objects into distinct blocks.} Considering that popular data stores separate writes explicitly using objects (\eg, SSTable files in RocksDB), \falloc, compared to the existing solutions, enables to achieve transparent write streaming at no or minimal program modification. Compared to MS-SSD, \falloc is simpler and more intuitive abstraction which does not resort to the concept of stream-id redundantly. ;;;; allows applications to transparently realize the full potential of write streams at no program modification or further optimization overhead.  }
\comment{The semantic is valuable to flash storage while not such useful to harddisk where overwrites are allowed. At this moment, we know that the range belongs to the object. Unfortunately, this valuable semantic is not, however,  passed to the storage but wasted.  (HDD where logical vs. physical address align --> 불필요; SSD서는 valuable) } 

\end{abstract}

\maketitle




\section{Introduction}


Most software systems, including LSM (Log-Structured-Merge) tree-based KV (Key-Value) stores, relational DBMSs, and file systems manage data using logical objects: to name a few, SSTables in RocksDB, DWB (double write buffer) in MySQL, and segments in F2FS. And, upon each object creation, its logical space is secured in advance before writes are made to the object. For instance, RocksDB calls {\tt fallocate()} right after creating an SSTable file so as to secure the logical address space for the file. The logical address range allocated from the call belongs to the file object. For this reason, from an LBA address, the host-side data stores can identify its belonging object. In addition, when an object is deleted, all its data will be invalidated together at once; thus the data set belonging to an object is said to have the same deathtime. Meanwhile, different objects are, though created and populated simultaneously, usually destructed at different points in time; they have different deathtimes. 
In summary, host software stacks manage data by objects; each object is the unit of logical space allocation and, in many cases, all its pages will have the same deathtime.


\begin{figure}[h]
\centering
\includegraphics[clip,width=0.43\textwidth]{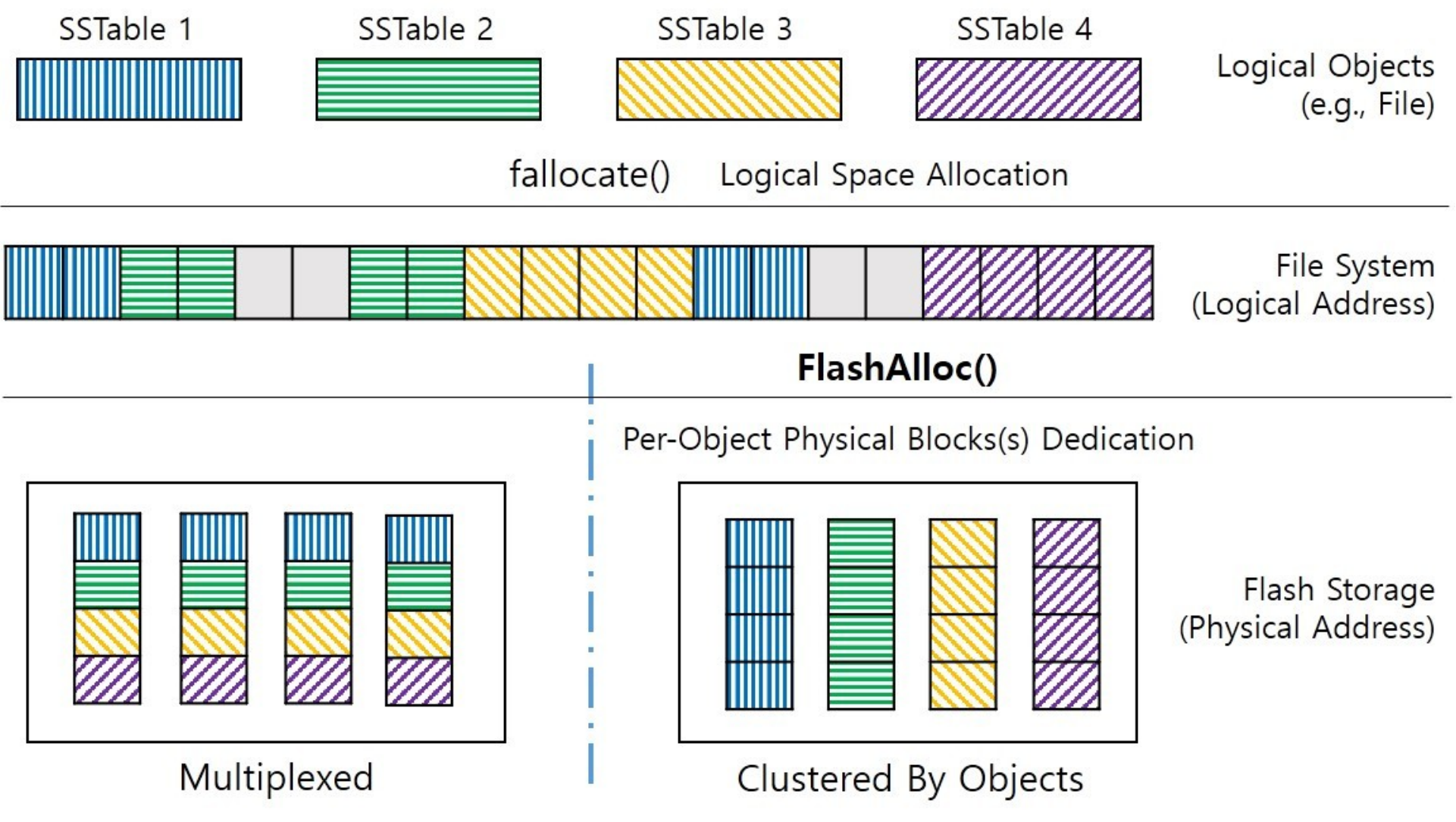}
\caption{Object Placement: Multiplexing vs. Clustering}
\label{fig:sallocate-archi}
\end{figure}

Though host software stacks can distinguish objects by their logical address ranges, the host-side semantic about objects' logical address ranges cannot cross the storage interface wall simply because no interface exists to pass it to the storage. As a result, today's flash storage has no knowledge about the belongs-to relationship between LBA address and object. Therefore,  when concurrent writes from different objects interleave, the conventional {\em object-oblivious} flash storage cannot distinguish each write's object so that it has no choice but to simply append new data in their arrival order into flash blocks. As a consequence, writes from different objects colocate in the same flash blocks. That is, each flash block will be multiplexed by data from multiple objects with different deathtimes, as illustrated in Figure~\ref{fig:sallocate-archi}. We call this situation as {\em multiplexing}. 

The multiplexing is the main culprit of physical write amplification in flash storage, worsening the performance and lifespan of flash storage. Since logical objects have different deathtimes each other in most cases (\eg, four SSTables in Figure 1 will be deleted at different points of time), pages in a multiplexed flash block will be incrementally invalidated at different points of time. When the block is chosen as a victim for garbage collection, the remaining valid pages have to be relocated to another block, amplifying physical writes inside flash storage. In particular, contrary to the common belief, the multiplexing will make even RocksDB and F2FS taking sequential write patterns suffer from severe device-level write amplification~\cite{DONOTSTACK,AMF,VLDB21FlashSSDs}. 
According to our experiments using commercial SSDs, write amplification factor (WAF in short) becomes even greater than 
eight (Figure~\ref{fig:f2fs-on-micron}). 
In this sense, the so called {\em flash-friendly} sequential writes are {\em not less harmful} than random ones in the conventional flash storage~\cite{SFS}.
\comment{ In particular, since {\bf random writes are considered harmful} in Flash~\cite{SFS}, numerous database/file systems such as RocksDB and F2FS taking flash-friendly sequential writes have been proposed and become popular. Contrary to the belief/expec. that seq WR is good for flash, however, even such softwares also suffer from high WAF~\cite{AMF, VLDB21FlashSSDs, ++}. 많은경우 sequential write 선호 systems서도, OLTP random write comparable WAF 관찰됨: >= 2, in some cases, up to 6!.  the unexpected multiplexing, despite of logically sequential writes against each object, is attributable to several reasons such as concurrent writes, split, append-only writes, striped write, ever-increasing block and channel size; 각 objects with different lifetimes and sequential write patterns are also multiplexed, thus experiencing excessive write amplification.
} 
In addition, the multiplexing can exacerbate the performance interference in multi-tenant environment~\cite{SeattleReport}. When multiple applications, each with its set of objects, are run together on a flash device, more objects with more deviating deathtimes will be multiplexed onto the same flash blocks, worsening write amplification.  
To sum up, since no interface exists to offload the host semantic about objects' logical address range to the storage, the valuable semantic is discarded and the object-oblivious flash device can not stream writes by objects, incurring multiplexing and write amplification. 



Addressing the multiplexing problem, we propose a novel interface, {\em FlashAllocate} (\falloc in short), which is used to pass the host semantic about object's logical address range to the flash storage and thus to enlighten the storage to be {\em object-aware} in handling writes. 
To be concrete, after creating an object, a data store calls \falloc with the object's local address range as a parameter to inform the flash device that the address range belongs to the same object. Then, upon receiving \falloc, the flash device creates a corresponding FlashAlloc (FA in short) instance, which keeps the given range information and thus is now aware of the address range of the object. At the same time, the flash device will secure physical flash block(s) where to place writes from the object and dedicate those blocks to the FA instance. Hereafter, the logical address range of an \falloc call has its corresponding physical space. For each write request, \comment{(from an \falloc-ed range)} flash storage probes its matching FA instance, if it exists, using the starting LBA address of the request and then stream the write to the dedicated blocks for the FA instance. 


Once logical objects are \falloc-ed, their writes will be, as illustrated in the right bottom of Figure~\ref{fig:sallocate-archi}, de-multiplexed into per-object dedicated flash blocks.\comment{Also, different objects will not share the same flash block but each object's data is isolated into distinct block(s)} In this regard, we say that flash storage, enlightened by \falloc, can stream writes by objects. The write streaming by objects is a natural way to achieve {\em grouping data by deathtime}~\cite{UnwrittenContract} because logical objects have different deathtimes while all pages of an object become dead together at once upon the object destruction. 
Thus, the ultimate benefit of streaming writes by objects is to avoid GC-induced write amplification. For instance, as each SSTable file is deleted in Figure~\ref{fig:sallocate-archi}, all its pages are invalidated, and thus, its dedicated block can be erased in its entirety, not causing any page relocation.
In summary, \falloc provides an alternative in {\em controlling the physical placement of writes} inside flash devices: it enables to de-multiplex logical objects into the different flash block(s) just resorting to the already-existing host-side object semantic and thus without introducing other intermediate concepts such as stream-id or zone-id~\cite{MultiStreamSSD,ZNS}. 


Whilst effectively de-multiplexing objects with sequential write-once pattern, \comment{ into different flash blocks,} \falloc is not a panacea for the write amplification problem. In particular, \falloc is not intended to be used for objects with random overwrites (\eg, relational database's tablespace under OLTP workload). For non-\falloc-ed objects, their data will be stored in non-FA instance and managed by the conventional FTL. In this way, flash storage supporting \falloc can support two types of objects, objects with sequential writes and objects with random overwrite, concurrently. In this respect, \falloc is in stark contrast with the Zoned Name Space (ZNS) interface, which forces applications to follow the strict sequential write rule~\cite{ZNS}. 


\comment{
\begin{itemize}
\item {\bf bf Concurrent Write Threads, Write Split, fallocate후 append } 또한, 동시 write, write split, append 되기 때문에, 위 ON-DEMAND allocation과 결합되어서, 서로 다른 file page들이 한 flash block에 섞이게 되고, thus pages belonging to a file or a logical object are multiplexed onto different flash blocks (Figure~\ref{fig:sallocate-archi} (a)). 
\item {\bf On-demand physical space allocation in SSD} 또한, flash storage는 out-of-place write 특성으로 인해, 실제 write 요청이 왔을 때, ON-DEMAND로 physical address space가 allocation된다.
\item {\bf SSD RAID striping} split 2MB --> into 8 * 256KB  
\item FS Fragmentation:  1) LBA fragmentation; with buffered writes, when a file consists of N fragmented LBAs, then N writes will be made scattered? 2) Unless a file is LBA-fragmented, a large sequential write made at page cache layer can be split into multiple write calls, along the IO stack, thus causing N write commands to the storage.
\item unit mismatch between fallocate and write, and write 시점이 다름.  
\end{itemize}
}

The main contributions of this paper are summarized below.

\begin{itemize}
\item We make an observation that data stores with flash-friendly write patterns, contrary to the common belief, experience severe write amplification on conventional SSDs and investigate the write multiplexing as the main culprit for the problem. We also show that the write multiplexing makes multi-tenants interfere with each other. 

\item We motivate that existing flash devices are object-oblivious \comment{in handling writes} simply because the host semantic about the object's logical address range cannot cross the storage interface wall. 

\item Based on this motivation, we propose a new interface, \falloc, which allows to offload the host semantic about object's logical address range to the storage and thus to enlighten flash device to be able to stream writes by objects, reducing device-level write amplification. We also suggest an extended design of page-mapping FTL to support \falloc. 

\item The abstraction used in \falloc is so simple and intuitive that existing software stacks can realize the write streaming by objects \comment{nearly in a transparent way} with minimal change in its codebase.
\comment{or even no change in its codebase}.



\comment{in modern data intensive applications (even known as flash-friendly database storage engine/file system). }


\item  We have prototyped \falloc using the Cosmos open-source SSD. Our experimental results show that \falloc can reduce WAFs in RocksDB, F2FS, and MySQL by 1.5, 2.5, and 0.3, respectively, and accordingly improve throughput by 2x, 1.8x, and 1.2x, respectively. In particular, when RocksDB and MySQL are run together, \falloc can reduce WAF from 4.2 to 2.5, and doubled their throughputs. \comment{improve RocksDB's OPS by 2x and MySQL's TPS by 2x, respectively.}
\comment{Our experimental results across a comprehensive set of representative workloads confirm that \falloc can, compared to the existing on-demand allocation, significantly reduce write amplification, thus considerably improve throughput and latency. In particular,  when applied to multi-tenant workloads of Rocksdb and MySQL, \falloc can reduce total WAF from 5 to 1.5, and improve OPSs by 3x and 2x for RocksDB and MySQL, respectively. ...} 
\comment{For this, we modified F2FS and MySQL/InnoDB minimally to call the \falloc while RocksDB at no change.} 
\comment{Using various representative use-case applications as well as multi-tenant scenarios. Our result are very promising., including 1) explicit fallocate call, 2) segment write at LFS file system, 3) DWB@InnoDB/RBJ/WAL@SQLite etc., 4) synthetic FileBench/varmail ( large sequential write(s) (even w/o explicit fallocate call) @ page cache; 5) multi-tenant (RocksDB / MySQL). 성능개선정도 언급!
fallocate()통해 \falloc을 호출하면, 해당 응용 자체 성능개선 외에, 동일 SSD를 공유하나 fallocate을 호출하지 않는 다른 응용들의 데이터들과 다른 플래시블록에 분리 저장되어서 SSD 쓰기증폭이 줄어들어 궁극적으로 모든 응용의 성능 개선에 도움이 될 수 있다. 응용 with \falloc would be altruistic to other applications. 
}
\end{itemize}

\falloc is similar to the {\tt trim} command~\cite{trim}: both aim at reducing write amplification by hinting flash device about host semantics (\ie, logical address range allocated upon object creation and staled upon object deletion, respectively). Interestingly, when used together, they are synergetic to reduce write amplification further. 
Meanwhile, the \falloc interface might not be such novel considering there already exists a write streaming interface, MS-SSD~\cite{MultiStreamSSD}. Unlike MS-SSD resorting to the stream-id concept, \comment{intermediate concept of stream-id,} we can stream writes \comment{at the grain of objects} simply by calling \falloc after object creation. Thus, developers are free from the burdens of determining the proper number of streams, grouping objects into limited streams, and assigning stream-id to every write call. Also, \falloc will not incur the stream-id conflict among multi-tenants. In this sense, \falloc is a simpler and more advanced abstraction than MS-SSD, enabling transparent write streaming.

\comment{
\reminder{
(R3:D2) "FlashAlloc will not incur the stream-id conflict among multi-tenants" does not seem an important advantage, as it could be easily avoided by either having a unique prefix for each application or using psudo-random stream IDs.";
(R1:D2) "Can't one use NVMe namespaces to separate address spaces for different tenants already"; 
namespace로 분리하는 것은 logical isolation만 달성할 뿐이며, physical isolation (e.g., IOD device)은 달성할 수 없음. 
FlashAlloc이 MS-SSD에 비해가지는 장점 상세히 명시 필요; host application이 stream id assign을 잘 했더라도 
GC 또는 wear-leveling 과정에서 stream id가 섞일 가능성이 있음. stream ID 갯수는 제한된다는 점 강조 필요!!! 
}
}

\comment{
\module{vs. Related Work - better abstraction and more practical} unlike MS-SSD, \falloc is 1) no limits in stream \#, 2) no mapping overhead b/w object-stream \#, 3) no (or minimal) s/w modification. \comment{In addition, though pages from one file are streamed into the same stream, they can be still spread over multiple flash blocks, thus causing waf!;} stream abstraction instead of object.(though object can be used as stream) - not-intuitive (stream - object mapping by application developer) + \# of streams << object; Unlike the existing stream separation scheme such as MS-SSD, \falloc will store different streams into different flash blocks in a very natural way, thus requiring no additional user intervention in assigning stream ids to each write stream and ++. (in theory unlimited number of streams, thus no limitation x) The \falloc interface can be easily and seamlessly integrated into the existing software eco systems. That is, as long as the existing applications call for {\tt fallocate} file system call with logical address, (and the system call internally invokes the corresponding \falloc command, the benefits of \falloc can be exploited. In addition, compared to {\bf host-managed SSD approaches} such as ZNS, OC-SSD, and AMF~\cite{AMF,OCSSD,ZNS}, no host/application overhead, no sw echo change (sequential write)  

{\bf Related work:} Compared to MS-SSD, \falloc provides {\bf better abstraction} for streaming and is {\bf more practical} solution particularly for multi-tenant environment with dynamic and unpredictable streaming requirement. In addition, compared to host-managed SSD solutions such as OC-SSD and ZNS which require radical and holistic changes across the stacks, \falloc is evolutionary and compliant to the existing stack. 
- unlike MS-SSD: 1) no limits in stream \#, 2) no mapping overhead b/w object-stream \#, 3) no (or minimal) s/w modification -- in addition, though pages from one file are streamed into the same stream, they can be still spread over multiple flash blocks, thus causing waf!; - unlike application/host managed solutions such as ZNS, OC-SSD, and AMF; no host/application overhead, no sw echo change (sequential write)  
}
\comment{아래 내용 들
We make an important observation about the multiplexing of sequential write streams in many applications. and show that the effect of multiplexing on write amplification, performance degradation/performance variability and shortened lifetime is quite serious. Even in a single application such as RocksDB with of multiple concurrent write threads, Sequential writes at the application layers are not sequential at the block trace or FTL layers but instead splited and thus mixed with other stream chunk. In DWB; in log structured file systems, In the case of multi-tenants, the multiplexing is more outstanding. (HotStorage20 submit참고@Fallocate폴더@논문서버) ++++++ 그리고, investigate why logical objects are multiplexed into different flash blocks. 

\begin{itemize}
\item {\bf Concurrent Sequential Writes are Multiplexed:} We made an {\bf important observation} that logical objects with sequential write patterns are multiplexed to different flash blocks and thus they also suffer from high WAF in many softwares; Also, we investigate/explain why sequential write streams are multiplexed. 
{\bf Multiplexing~\cite{DONOTSTACK}:} ``Mixed workload from logs and other traffic destroys sequentiality – Each log writes sequentially, but the device gets mixed workload, most likely to be random – Underlying flash-based SSD also writes its own metadata.Multiple logs on a shared device results in random traffic seen by underlying device.'' (see also the slides at https://www.usenix.org/sites/default/files/conference/protected-files/inflow14\_slides\_yang.pdf)
\item {\bf Multiplexing effect:} Later, they are invalidated upon deletion or overwrite. As a result, contrary to the expectation, many applications suffer from high write amplification (NOTE: This observation is not widely known. Simply they assume that sequential writes are good for flash storage.) +  {\bf Log-on-log~\cite{DONOTSTACK} 언급} {\bf Why Multiplexex:} 보기에 sequential write (즉, 같은 object에 대해 긴 write가 주어졌을때, 이들이 같은 flash blocks에 clustering 되지 않고 다른 block들에 흩어짐, for various reasons such as on-demand allocation, file system fragmentation, SSD architecture, and write split~\reminder{2장 관련 정리 후, 여기서는 simply enumerate할 것}. For this reason, 실제 seq- wr 요청하는 다양한 application들에서 서로 다른 deathtime/lifetime을 가지는 file or objects들의 data pages들이, not clustered but multiplexed into flash blocks;
\end{itemize}
}

\comment{\module{Misc.}  
1.Terminology: page and its belonging object; object and its pages
2. Object(File, Table) vs. segment(or chunk/extent, unit of logical space allocation at FS or database)  vs. pages, blocks VS. host에서 logically fallocate 단위. 같은 file도 two or more fallocate 가능. finer granularity!! eventually, pages from different objects will not mix with each other in the same flash block.;
3. An object may consist of one or more segments, and a segment has a set of logical pages (mostly contiguous, but not always) .. {\bf For the simplicity of discussion, we regard every object calls one {\tt fallocate} at its creation time, which is true in many cases such as RocksDB SSTable, F2FS segment, and DWB. But, \falloc does not exclude the case where an object call multiple \falloc. Recall that each segment is streamed into diff blocks, we still guarantee that {\bf group by objects}}  
4. falloc vs. fallocate(): 다른 layers/roles, 단, fallocate -> \falloc 호출
5. that logical segment whose pages will be dead together at once to SSD, segregating those pages into dedicated physical block NAND. 6. (regardless whether pages from a file are written scatteredly in time (by application) or one large sequential write is split into multiple segments by kernel (fragmentation or IO stack), a write segment from a file object is eventually merged into the same flash block.} 
\section{Background and Motivation}~\label{sec:bg} \comment{and Problem Definition}
This section reviews several key concepts about flash storage, presents a few motivating examples about write amplification in flash-friendly data stores, and makes a problem statement on why flash storage is object-oblivious. 

\subsection{Flash Memory SSD}

Here we review how the existing flash storage works and explain two key concepts of page deathtime and stream write by time. 

\comment{ \module{The role of FTL} 
\item FTL = intermediate software module, mapping 정보 관리
\item No overwrite / out-of-place update, block 단위 erase -> mapping 필요 
\item Mapping은  logical page address(LSN) -> physical page address(PPN)
\item FTL 종류: Page-/block-/Hybrid-mapping (FAST/BAST) -- 동작 방식 간단 설명
\item Flash chip: blocks, pages, invalid / valid page 설명 필요
\item WAF (write amplification factor) 정의/설명: internal write / logical write at host  (어디서? FaCE journal paper 참고) 
}

\comment{\module{Page-Mapping FTL}  An FTL is responsible for several key functionalities such as address mapping, GC and wear level management~\cite{FTLsurvey}). Because overwrites are not allowed in flash memory, a new data page write should be handled in an out of place manner (\ie, log-structured) - the old version of the page will be marked as invalid and new version will be stored in a new clean page. Thus, FTL has to manage the ever-changing address mapping between each page's logical address at the file system layer and its physical address in flash memory chips. Since the address mapping scheme is critical to the performance and lifespan of flash storages, numerous address mapping schemes such as block-mapping, page-mapping ,and hybrid-mapping~\cite{FTLsurvey} have been proposed since the inception of FTL techniques.} 

\module{FTL}  An FTL (Flash Translation Layer) is responsible for several key functionalities such as address mapping, GC and wear level management~\cite{FTLsurvey}. Because overwrites are not allowed in flash memory, a new page writes should be handled in an out of place manner (\ie, log-structured) - the old version of the page will be marked as invalid and new version will be stored in a new clean flash page. Thus, FTL has to manage the ever-changing address mapping between each page's logical address at the file system layer and its physical address in flash memory chips. Since the address mapping scheme is critical to the performance and lifespan of flash storages,
most flash storage prefers the page-mapping FTL scheme among numerous address mapping schemes, mainly for performance reason at the cost of memory resource for managing the logical-to-physical mapping at the page granule~\cite{2R,DFTL,FTLsurvey}. 

\module{Garbage Collection} When clean space for new writes runs out, FTL has to reclaim new clean space by the garbage collection (GC in short) procedure. Upon GC, a victim block $V$ is chosen, then its valid pages are relocated to a clean block $B$ (\ie, valid pages are read out from $V$ and written back to $B$), and then $V$ is erased and returned to the free block pool. After GC, new writes from the host will be appended to the remaining space in $B$. 
Relocating valid pages during GC amplifies physical writes inside flash storage. Informally, {\em write amplification factor (WAF)} represents the ratio of physical writes to flash memory over logical writes from the host. 
\comment{Let us denote the total number of pages in a flash block, the number of valid and invalid pages in a victim block as $N$, $V$, $I$, respectively. During GC, $V$ pages have to be copied to another clean block $B$. After GC, the block $B$ now has $I$ (that is, $N$ - $V$) free pages. To write $I$ logical pages from the host, $I+V$ (\ie, $N$) physical pages have to be written. Hence, WAF can be defined as follows: 

\[WAF = \frac{I+V}{I} = 1 + \frac{V}{I}\]

Note that the last term in the above equation, $V/I$, represents the fraction of additional physical writes over the logical writes from the host. Ideal WAF will be one when the value of $V/I$ becomes zero. The above equation indicates that minimizing $V$ is crucial to reducing write amplification.} 

\module{Page Deathtime} When a page copy in flash block is overwritten by new write or discarded by the trim command, the old copy is said to be dead. And the point of time it becomes invalidated is called as its deathtime~\cite{UnwrittenContract}. The distribution of deathtimes of pages in flash blocks is critical to determining the write amplification. For instance, let us assume that a flash block {\tt fb1} stores only data pages from the same SSTable in RocksDB. All the pages in {\tt fb1} will be dead when the SSTable is deleted after compaction. Then, the GC procedure can secure a new clean block without relocating any page but simply by erasing {\tt fb1}, incurring no write amplification. In contrast, when a flash block stores pages with quite distant deathtimes and later is chosen as a victim for GC, many valid pages should be relocated to another block. Therefore, grouping pages by deathtimes is paramount to reducing write amplification. 

\comment{
\item By Space: Since adjacent logical pages are more likely to have distant death times as file systems are fragmented, grouping by space could not be a practical solution, either. vs. \falloc: 어떤 logical address range가 동일 file/object에 속한다는 정보 기반으로 clustering 가능케 함.; : file 단위, Extent 단위(when multiple tables reside in one OS file), 좀 더 fine grain한 object 단위(?? if any), 응용단위, VM 단위  ==> 논리적 종류 상관없이 물리적으로는 결국 fallocate aligned with EU size로 대응가능함. 
\item By Order/time: in servers (not PC/mobile mostly with single user) with concurrent users, multi writers -> split-> GB order 지원불가.(although FTl append data to a log,);  서로 다른 논리적 객체에 속하는 page들이 semantic 이유로 같은(비슷한 시점에 같이 update되어 같이 persistent하게 될때 (예: SQlite처럼 force 정책을 사용하는 경우); 또는 checkpoint시 disk write할 때 같이 update되는 page들을 가급적 묶어서 write하는 전략 필요; 또는 결국 random write라써 2R 같은 FTL에서 다루어야 하나??
}

\comment{Hence, the larger the OP factor is, the faster an SSD performs.} 
\comment{\[OP Factor = \frac{P-U}{U}\]}


\module{Stream-Writes-by-Time}  Consider how today's flash storage handles writes when  
concurrent write requests from different objects interleave. For each write request with logical address (\ie, start\_lba), the existing flash storage cannot distinguish the object the data belongs to. Therefore, the conventional {\em object-oblivious} flash storage will simply append writes from different objects in their arrival order at the clean flash memory space~\cite{Chen-TOS16}; we call this write policy as {\em stream-writes-by-time}. 
\comment{
\begin{figure*}[!ht]
\centering
\subfigure[EXT4 (4 RocksDB Tenants)]{
\label{fig:rocksdb}\includegraphics[clip,width=0.22\textwidth,height=3cm]{figure/graph_ext4_rocksdb.JPG}
}
\subfigure[F2FS (4 RocksDB Tenants)]{
\label{fig:f2fs}\includegraphics[clip,width=0.22\textwidth,height=3cm]{figure/graph_f2fs_rocksdb.JPG}
}
\subfigure[DWB in MySQL]{
\label{fig:dwb}\includegraphics[clip,width=0.22\textwidth,height=3cm]{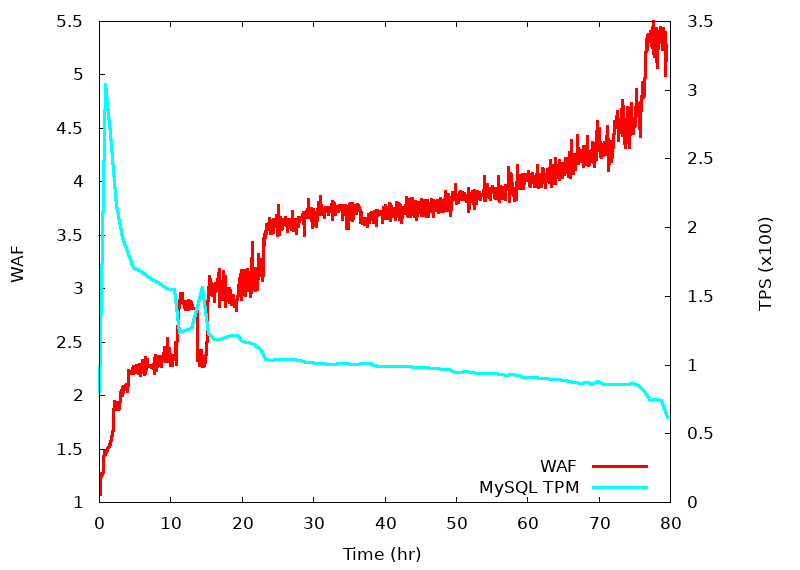}
}
\subfigure[Multi-tenant DB (TPC-C + RocksDB)]{
\label{fig:multi-tenant}\includegraphics[clip,width=0.22\textwidth,height=3cm]{figure/graph_multi.JPG}
}
\caption{Write Amplifications and Performance Trends on Flash-Friendly Workloads}
\label{fig:types-of-multiplexing}
\end{figure*}
}

\begin{figure*}[!h]
\centering
\subfigure[4 DB\_Bench Tenants (Ext4)]{
\label{fig:dbbench-on-micron}\includegraphics[clip,width=0.235\textwidth,height=3.2cm]{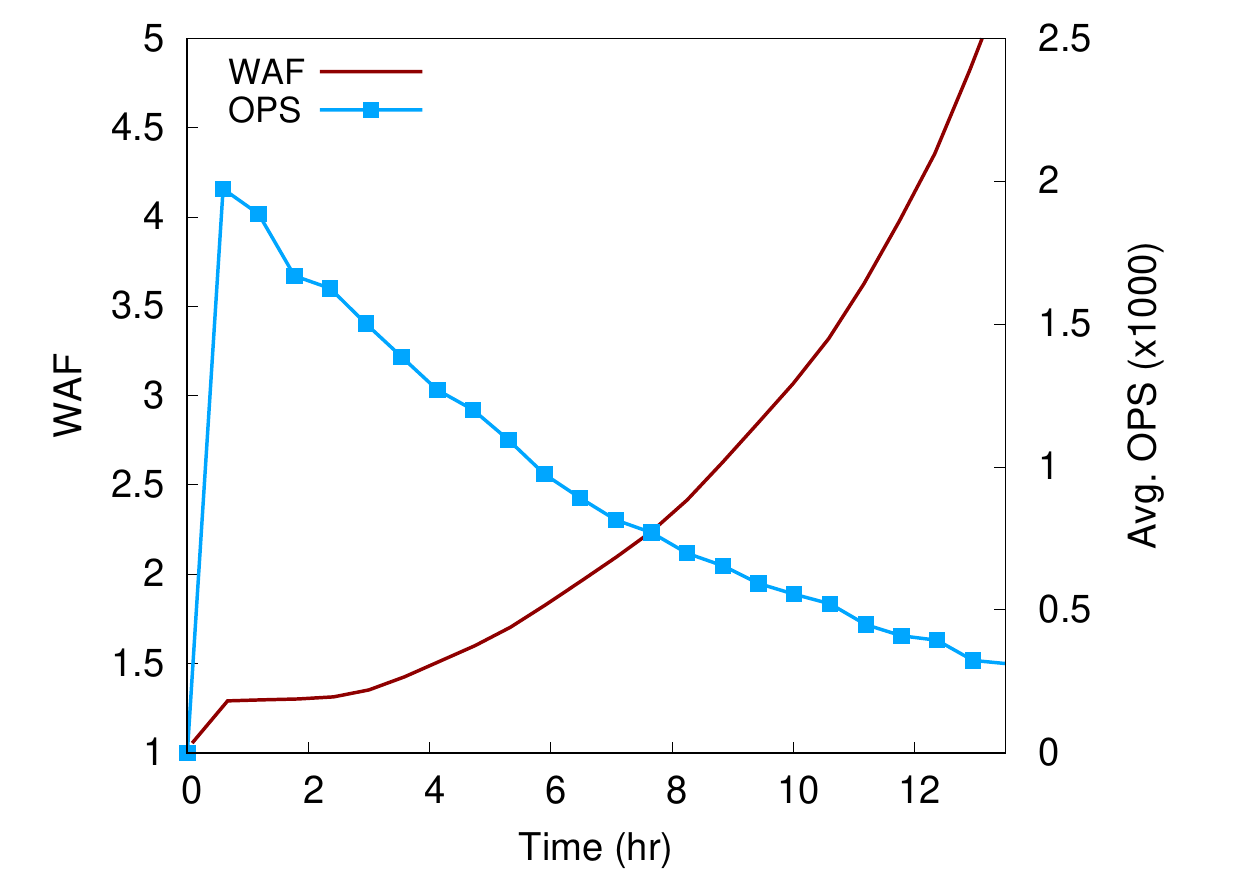}
}
\subfigure[4 DB\_Bench Tenants (F2FS)]{
\label{fig:f2fs-on-micron}\includegraphics[clip,width=0.235\textwidth,height=3.2cm]{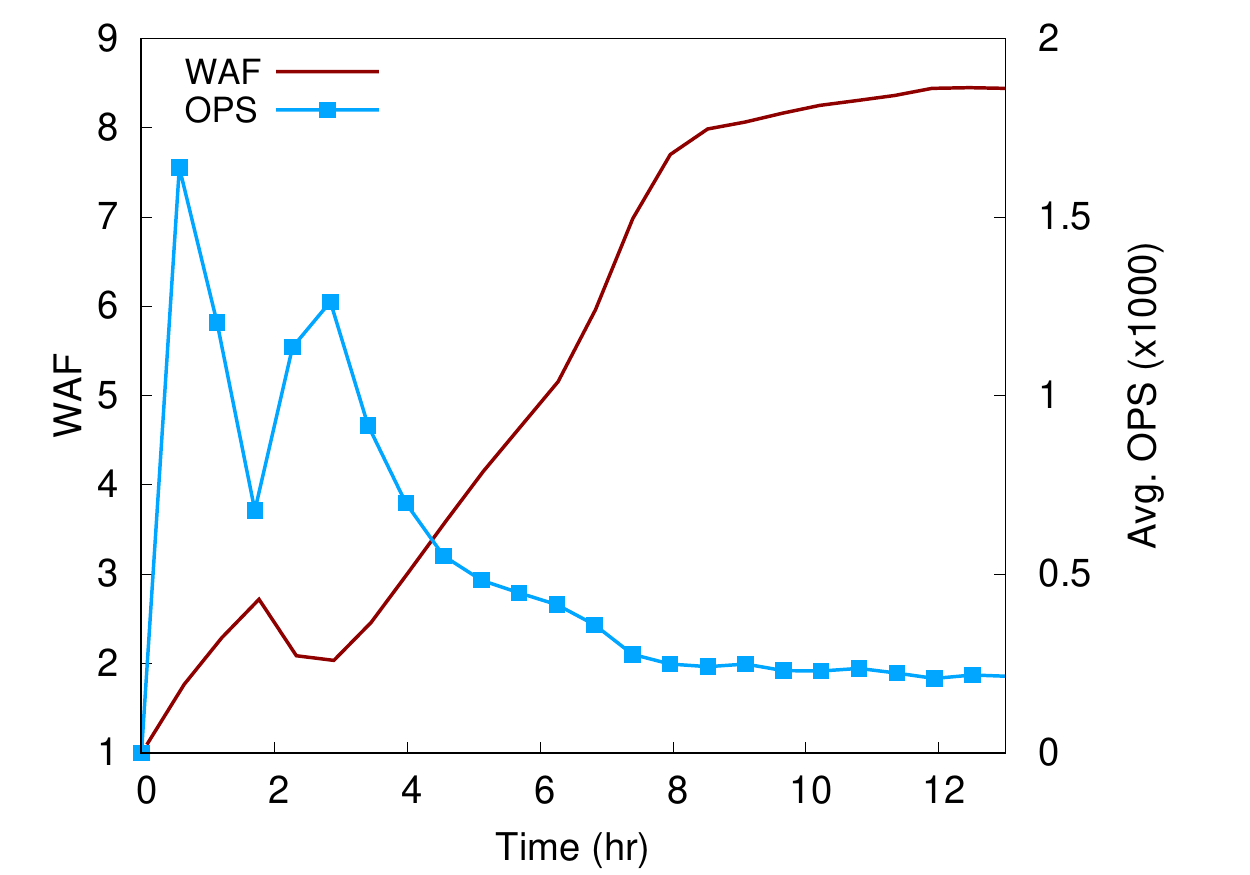}
}
\subfigure[TPC-C (MySQL, Ext4)]{
\label{fig:dwb-on-micron}\includegraphics[clip,width=0.235\textwidth,height=3.2cm]{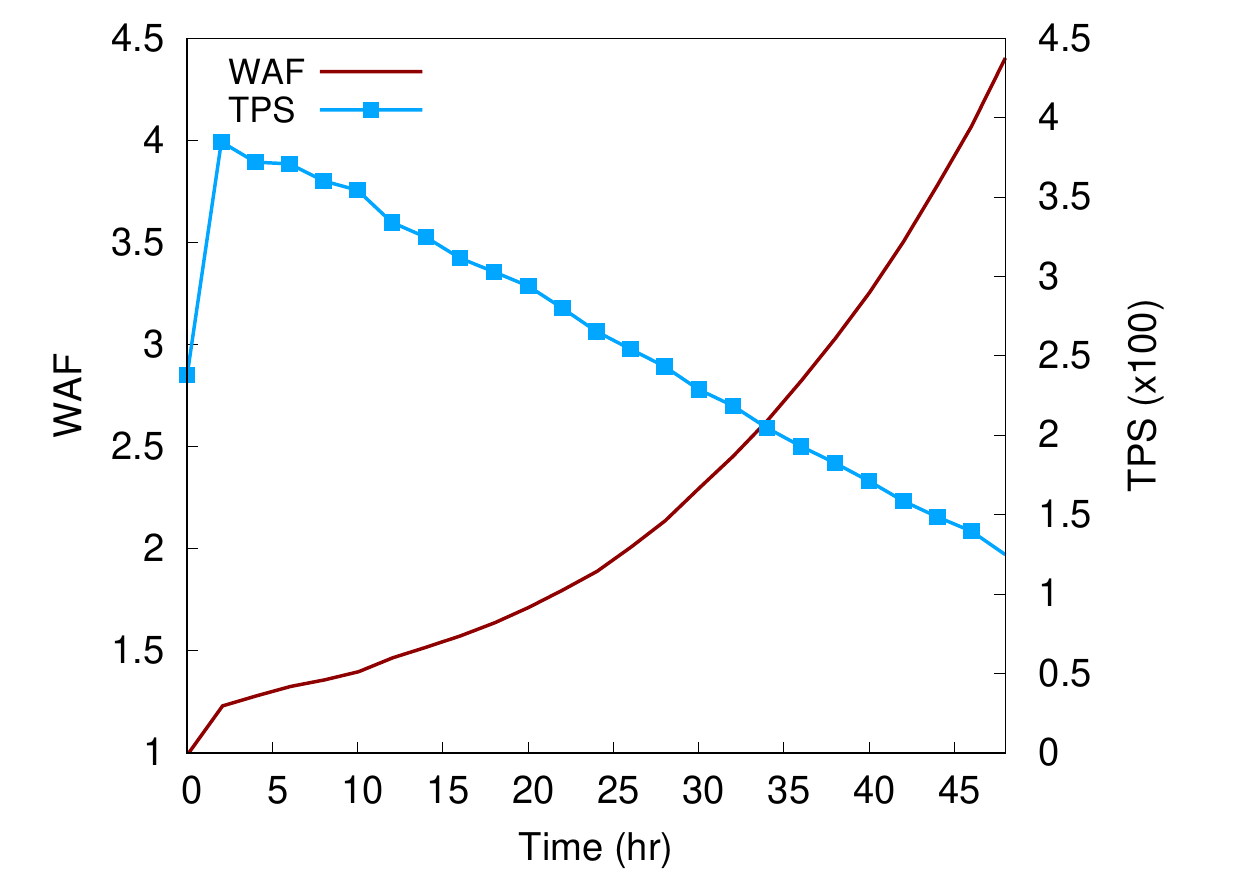}
}
\subfigure[DB\_Bench + TPC-C (Ext4)]{
\label{fig:multi-tenant-on-micron}\includegraphics[clip,width=0.235\textwidth,height=3.2cm]{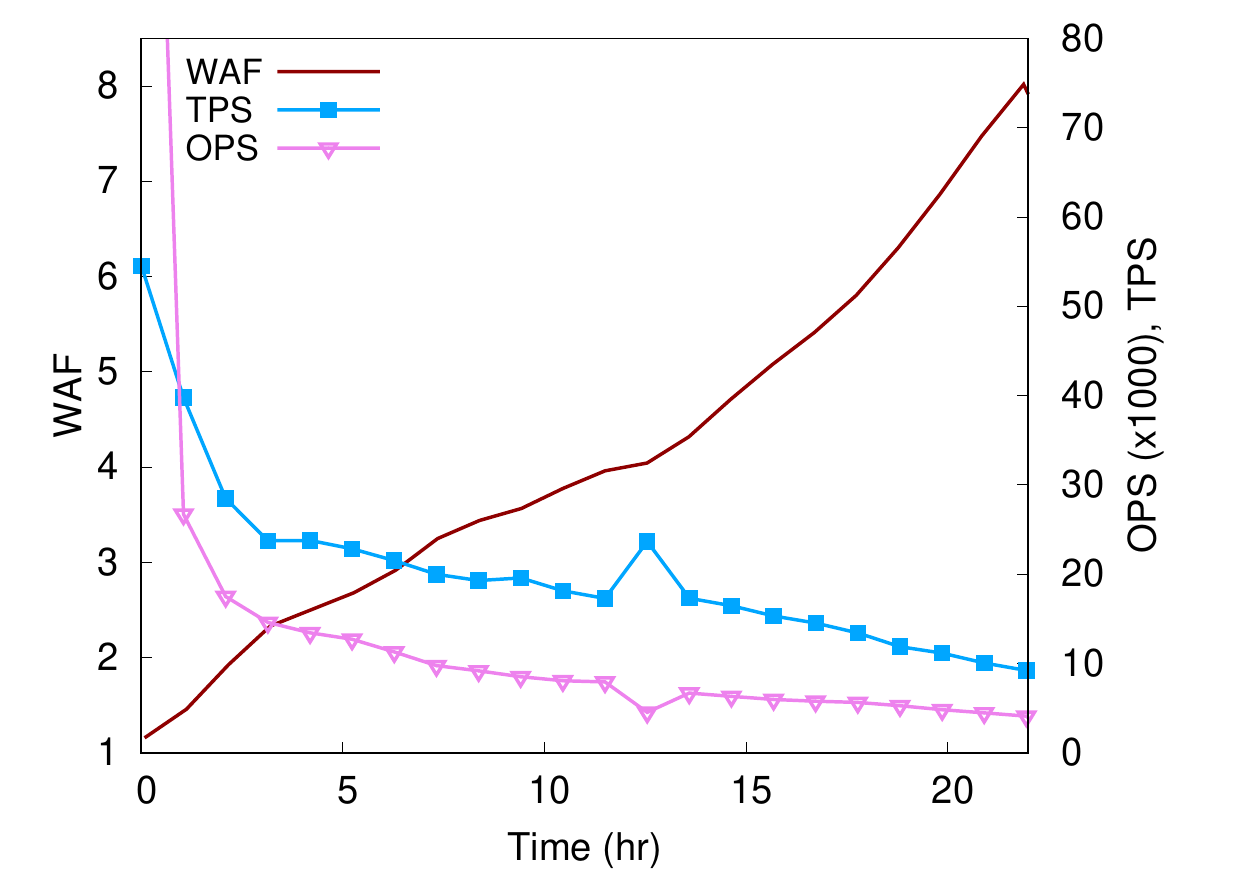}
}
\caption{Four Realistic Workloads on a Commercial SSD: WAF and Throughput}
\label{fig:micron-exp}
\end{figure*}


\subsection{Motivation}~\label{sec:motivation}


It has long been believed that sequential writes are {\em flash-friendly}: less harmful than random writes in terms of write amplification~\cite{SFS}. With this expectation in mind, sequential writes have been opted for by many data stores such as LSM (log-structured merge) tree-based KV (key-value) stores~\cite{MultiStreamSSD, LevelDB, LSM, RocksDB} and F2FS (Flash-Friendly File System)~\cite{F2FS} which is a variant of log-structured file system~\cite{LFS}. 

Contrary to the belief, however, log-structured sequential writes at such flash-friendly software stacks are not effective in reducing write amplification. In reality, our experimental results below will clearly reveal that such sequential writes are as harmful as random ones (\eg, OLTP workloads on RDBMS) in terms of WAF in flash storage. Similar observations have been made consistently across different software stacks~\cite{UnwrittenContract,AMF,VLDB21FlashSSDs,DONOTSTACK}, albeit they did not explain the reasons. 

\comment{Thus, blocks chosen as victim upon garbage collection will have more valid pages, thus {\bf causing more page relocation upon.}  the mixture by GCed page and normal writes. \reminder{In particular, SSDs write log-structured to the flash media that requires GC.-> Multiplex data streams onto the same GC units -> WAF Increase, OP, and thereby cost -> Decrease throughput and latency predictability.}}


In this section, we demonstrate that two data stores, RocksDB and F2FS with the so-called {\em flash-friendly} write patterns, suffer from severe write amplification. We also motivate a write multiplexing problem in MySQL. In addition, we point out that two tenants of RocksDB and MySQL interfere each other and thus exacerbate the write amplification further. 
While running each workload on top of a commercial SSD of 256GB, 
we measured its throughput and also the running WAF at the device-level using {\it smartmontools}~\cite{smartctl}
\footnote{We refer to technical note from vendor\cite{micron} which 
describes how to measure the write amplification factor for client SSDs using the Self-Monitoring, Analysis, and Reporting Technology (SMART) attributes.}
and present the results in Figure~\ref{fig:micron-exp}.
The figure clearly shows that WAFs increase over time across all cases, and application throughputs inverse-proportionally degrade. In particular, it is noteworthy that WAF values in RocksDB and F2FS with sequential writes are no less than that in MySQL with random writes. Below, for each workload, in turn, we describe the experimental setting and explain its IO architecture and dominant write patterns. In particular, we elaborate on why they experience severe device-level WAF despite their {\em flash-friendly} write patterns. 


\module{\bf RocksDB(Figure~\ref{fig:dbbench-on-micron})} 
RocksDB is a popular KV store used in many large-scale data services as well as databases~\cite{RocksDB,RocksDB-Exp}. Since it uses Log-Structured-Merge (LSM) tree~\cite{LSM} as the primary data structure, the dominant write pattern from RocksDB is sequential in the unit of SSTable (Single Sorted Tables). 
Upon memtable flush or compaction, RocksDB creates new SSTable file(s), allocates a logical space of (by default) 64MB to each file via the {\tt fallocate()} call, writes data, and then flushes the file.\comment{using the {\tt fsync} call.} SSTables will be later deleted after compaction; all pages of an SSTable will be invalidated together upon the file deletion. Note that SSTable files which are created and populated simultaneously will be compacted and deleted at different points of time. Using the sequential batch write for each SSTable, RocksDB expects pages from the same SSTable and thus with the same deathtime to colocate in the same flash blocks and thus to barely cause write amplification. 

To verify the WAF problem in RocksDB, we ran four RocksDB instances with  db\_bench's {\tt fillrandom} workload~\cite{dbbench}
on Ext4 file system until the commercial SSD was full. 
\comment{
\reminder{(R1:D1) "more details about your experimental setup and results."
db\_bench fillrandom 실험에서 data loading을 하고 수행하지 않았습니다. 
해당 부분은 삭제해도 될 것 같습니다. 
대신 각 인스턴스당 4개의 user thread가 있음을 명시하면 될 것 같습니다.
Each instance was populated with initial database of 40GB. }
}
Each RocksDB instance has four concurrent user threads and 
uses four concurrent compaction threads to utilize storage better
and thus mitigate the compaction overhead~\cite{RocksDB-Exp}.
In addition, the discard option was enabled while mounting Ext4.\comment{ to minimize write amplification.} 
While running the workload, we measured 
the average OPS (Operations per Second) of four benchmarks and also running WAFs of the SSD in every five minutes, and plot the result in Figure~\ref{fig:dbbench-on-micron}. To our surprise, the WAF has continued to soar over time, ending around five. Note that the WAF value of five is even greater than that observed in OLTP workloads with random writes~\cite{2R}. This unexpected result about flash-unfriendly RocksDB 
has been reported consistently by other researchers~\cite{AMF,DONOTSTACK,VLDB21FlashSSDs}. All other LSM tree-based KV stores will we believe\comment{including LevelDB and Cassandra} experience the same WAF problem~\cite{LevelDB,MultiStreamSSD}.
\comment{(?? Each RocksDB instance has levels.\reminder{What was the maximum number of levels while running RocksDB?})}
\comment{DO NOT DELETE!!! For EXT4, the default journaling mode was used and the discard option was enabled to use TRIM commands.\footnote{The discard/trim option is very effective in reducing write amplification in RocksDB. For the same workload used for Figure~\ref{fig:rocksdb}, the cumulative write amplification was lowered from 5.?? to 2.?? when the discard option was enabled. Similar observations were consistently reported by others~\cite{2R,PTSonFlash}.}} 

\comment{
\begin{figure}[!t]
\centering
    \includegraphics[clip,width=0.4\textwidth]{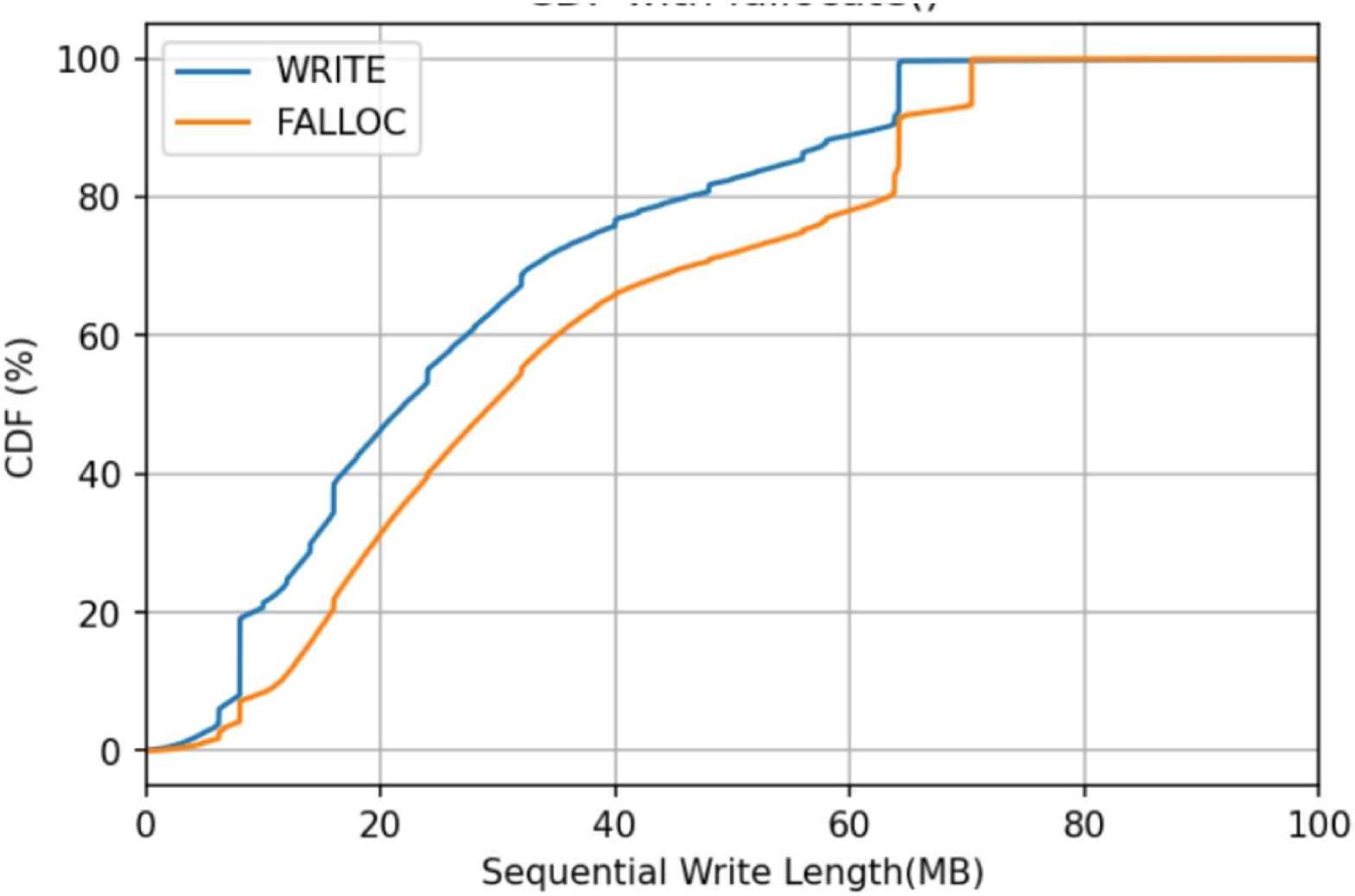}
    \label{fig:rocksdb-cdf}
    \caption{Write Size Distribution in RocksDB}
\end{figure}}

Consider why RocksDB suffers from high WAF. Though each SSTable file is sequentially written, four compaction threads will flush their SSTables concurrently. In addition, each flush of 64MB SSTable file tends to split into smaller write requests due to file system fragmentation and kernel IO scheduling~\cite{fragmentation, SoyeeFastPoster20}. Thus, write requests from multiple SSTables will interleave at the flash storage, and they are handled according to the stream-writes-by-time policy. Further, the striped architecture will divide each write request into smaller write chunks (\eg, 4KB) and distribute them over multiple channels~\cite{Cosmos2020,PageAllocationsInSSD,RAID}. As a result, pages from multiple SSTables being flushed concurrently tend to be stored together in the same flash blocks. \comment{Namely, data from multiple SSTables are multiplexed onto the same by multiple SSTables.} Recalling that SSTables have distinct deathtimes, the multiplexing will incur write amplification.

\comment{
First of all, recall that each RocksDB instance is enabled with 16 compaction threads. Thus, 64 write threads in total from four RocksDB instances are likely to flush SSTables which belong to different levels at different instances. This implies that though written concurrently they are very likely to be deleted at distant points of time and thus their death times are quite different. 
Next, each flush request of 64MB SSTable file is splited along the IO stack~\cite{SoyeeFastPoster20}  
Next, when {\tt fallocate} call with 64MB length is made, the file system will allocate a set of smaller fragments, instead of a large contiguous region, to each SSTable file because of the notorious file system aging~\cite{fragmentation}. Once fragmented, file system has to issue separate per-fragment write request to the storage. As clearly illustrated in Figure~\ref{fig:rocksdb-cdf}, we confirm that SEVERELY FRAGMENTED. 
Lastly, when small splited or fragmented write chunks are given, SSD FTL will further split the chunk into multiple 4KB-sized subchunks and, like RAID~\cite{RAID}, stripe them across massive channel/way/die-level parallelism~\cite{Cosmos2020,PageAllocationsInSSD}. 
RAID-like striping in the unit of 4KB over 
SPLIT!!!! Concurrent sequential writes are {\bf SPLIT} and each SSTables file is multiplexed into different flash blocks.; 
}



\module{F2FS (Figure~\ref{fig:f2fs-on-micron})} 
F2FS aims at reducing write amplification by taking log-structured writes~\cite{F2FS}. In F2FS, a segment is a basic unit of management
and, the whole storage volume is divided into fixed-size segments (by default 2MB). To turn random writes into sequential ones, F2FS basically builds on {\em append-only logging} to segments. That is, it appends new page write to one of six active segments depending on the page's hotness. 
Hence, the write pattern in F2FS can be characterized as {\em multi-head logging}~\cite{F2FS}: writes are concurrently made against six segments while write to each segment is append-only logging. When an active segment is filled up, the segment becomes inactive and a new segment becomes active. The old inactive segment will be later chosen as a victim for garbage collection and be cleansed~\cite{LFS}.  Upon a segment cleansing, all logically valid pages will be relocated to a new segment and the old segment is returned as free segment. Note that all pages in a segment have the same deathtime because they are discarded altogether when the segment is cleansed.

To validate how WAF in F2FS varies, we measured OPS and device-level WAF while running the same four RocksDB instances \comment{each with the {\tt fillrandom} benchmark} used for Figure~\ref{fig:dbbench-on-micron} on F2FS, and present the result in Figure~\ref{fig:f2fs-on-micron}. 
The device-level WAF became quite high (\ie, greater than 8) around the mid-phase of the experiment. Considering that the log-structured file system itself incurs logical write amplification~\cite{DONOTSTACK,LFS}, it is not surprising to see that the average OPS of four RocksDB instances was lower than that on Ext4 (\ie, Figure~\ref{fig:dbbench-on-micron}). 
\comment{
Note that the write amplification at the right Y-axis in Figure~\ref{fig:f2fs} indicates that  the physical write amplification at the SSD layer and it does not include the logical amplification due to the compactions conducted made by the higher layers but about} 
Other recent studies have also reported the device-level WAF issue when running F2FS on top of flash storage~\cite{AMF,DONOTSTACK}. 
\comment{ 
In particular, Yang et al. attributed the write amplification to the {\bf log-on-log } problem~\cite{DONOTSTACK}: when a log-structured F2FS is run on yet-another log-structured flash storage, the physical write amplification at the lower storage could be high.
Refer to {\bf Slide 17} in~\cite{DONOTSTACK}.
Log-on-log(\eg, F2FS + SSD; RocksDB + Ext4 + SSD)Death time and file systems mix them! (Observation 18)~\cite{UnwrittenContract}}
Interestingly, the authors of F2FS acknowledged that its multi-head logging could mix pages from multiple active segments into the same flash block~\cite{F2FS}. Thus, considering that segments are cleansed at different points of time, pages with different deathtimes are multiplexed in the same flash blocks. This is the reason F2FS experiences such high WAF. 

\module{\bf MySQL (Figure~\ref{fig:dwb-on-micron})}
In order to guarantee the write atomicity in the presence of crashes, MySQL takes the redundant journaling approach using the special object, called {\em double write buffer} (DWB in short): before flushing dirty pages to their original locations, InnoDB engine first appends them sequentially to DWB. On system booting, contiguous logical address space of 2 MB is allocated to DWB. Though tiny in capacity, DWB will account for half of the writes in InnoDB to the storage. When full, the DWB space will be reused from the beginning. Therefore, the write pattern to DWB can be characterized as {\em sequentially appended and cyclically reused}. 
In addition, the pages written to DWB in the previous cycle will be quickly overwritten in the next cycle. Namely, the DWB object has short lifetime and its pages written during the same cycle will have almost the same deathtime. 

Another characteristics in InnoDB engine's write pattern is that sequential writes to DWB and random writes to original database locations will interleave at the storage. Thus, in the case of flash storage, according to the stream-writes-by-time policy, both types of data with different deathtimes will colocate in the same flash blocks, thus adversely affecting WAF. 
To confirm this, we measured TPS and running WAF while running the TPC-C benchmark~\cite{tpcc-paper} with 32 client threads and initial database of 150GB (\ie, 1,500 warehouses) on Ext4 file system until the SSD is full, and present the result in Figure~\ref{fig:dwb-on-micron}. As shown in the figure, the WAF has steadily worsened over time and accordingly the transaction throughput (TPS) has dropped continuously. The multiplexing of DWB pages and normal data pages on the same flash blocks is believed to contribute to the WAF in part, though not all.  \comment{used to provide data consistency} According to our experience~\cite{soyee:edbt18,hwpark:damon19}, we can lower write amplification moderately by separating normal and DWB writes into different flash blocks. 
  

\module{Multi-tenant Databases (Figure~\ref{fig:multi-tenant-on-micron})}
\comment{ As the capacity of an SSDs have been ever growing and the cloud computing becomes the main stream,}  With the ever-growing capacity of SSDs, it is not uncommon for multiple databases to share a single large SSD, which is particularly true for the cloud environment~\cite{AliCloud}. 
In multi-tenant workloads, as different tenants concurrently issue writes for their own objects, pages from {\em more} objects with {\em further distant} deathtimes are likely to be multiplexed onto the same flash blocks. Thus, the write amplification would be worse than that when each tenant was run alone, which is undesirable in guaranteeing performance isolation among tenants~\cite{SeattleReport}. 
\comment{Meanwhile, it is of paramount importance to guarantee performance isolation among tenants~\cite{SeattleReport}. The write multiplexing under multi-tenants is undesirable in terms of performance isolation, particularly in that noisy tenant(s) causing excessive write amplification can severely interfere the performance of neighbor tenants.
}

To verify how the device-level WAF behaves under multi-tenant databases, we ran db\_bench and TPC-C together on the commercial SSD. \comment{used for Figure~\ref{fig:rocksdb} and Figure~\ref{fig:dwb}} While running both benchmarks with initial database of 80GB until the SSD became full, we measured their throughputs and device-level WAF, and presented the result in Figure~\ref{fig:multi-tenant-on-micron}. 
By comparing WAF and throughput in the figure with those in Figure~\ref{fig:dbbench-on-micron} and ~\ref{fig:dwb-on-micron}, 
we confirm that the WAF over time in multi-tenant case are much higher (\ie, almost double) than that when either tenant was run alone. Also, note that as the WAF spikes at the initial phase during the multi-tenant experiment, both OPS and TPS drop more rapidly than the throughput in either single tenant. To sum up, the write multiplexing among multi-tenants will exacerbate the write amplification and thus worsen their performance interference. 

\comment{\reminder{Need to check - we can make two observations.} First, the WAF trend over time in multi-tenant case resembles to that in TPC-C, not in db\_bench. From this, we suggest a conjecture that the WAF trend in two tenants will be determined mainly by the tenant which shows higher WAF when ran alone. Second, as WAF continues to increase over time, db\_bench's OPS keeps decreasing. In contrast, as shown in Figure~\ref{fig:dbbench-on-micron}, note that db\_bench's OPS is quite stabilized when run alone. 
}
\comment{ 
첫번째 주장- micron 및 vanilla in comsmo+ board WAF 결과만 보면, 
multi-tenant 성능은 WAF 안좋은 tenant (RocksDB)를 따라간다고 할 수 있습니다.
하지만 falloc multi-tenant실험 결과를 보면, 
상대적으로 WAF가 안좋은 DWB (Figure 4(c) - up to 2.2)보다 
훨씬 더 높은 WAF를 기록한 것을 알 수 있습니다 (Figure 4(d)).
salloc관점에서는 neighbor instance 간섭으로 인해 
손해를 볼 수도 있다는 것에 초점을 맞춰야 할 것 같습니다.
두번째 주장 - falloc OPS의 경우 vanilla OPS 처럼 점점 줄어드는 것을 확인할 수 있습니다.
mysql database도 growing 하면서 write간섭이 발생해서 
성능 stability 언급하는 것은 리뷰어입장에서 헷갈릴 것 같습니다.
IOD 디바이스처럼 두 테넌트가 사용하는 하드웨어 물리적 영역을 구분하는 경우에는 
stabiltiy를 언급해도 될 것 같습니다만, falloc인터페이스가 
하드웨어 물리적 구분 효과 (stability 관점에서..)도 보여줄 수 있는지는 불분명 합니다.
(misc YCSB 를 RocksDB로 수정했습니다.)}


\comment{
\begin{figure}[!ht]
\centering
    \includegraphics[clip,width=0.46\textwidth]{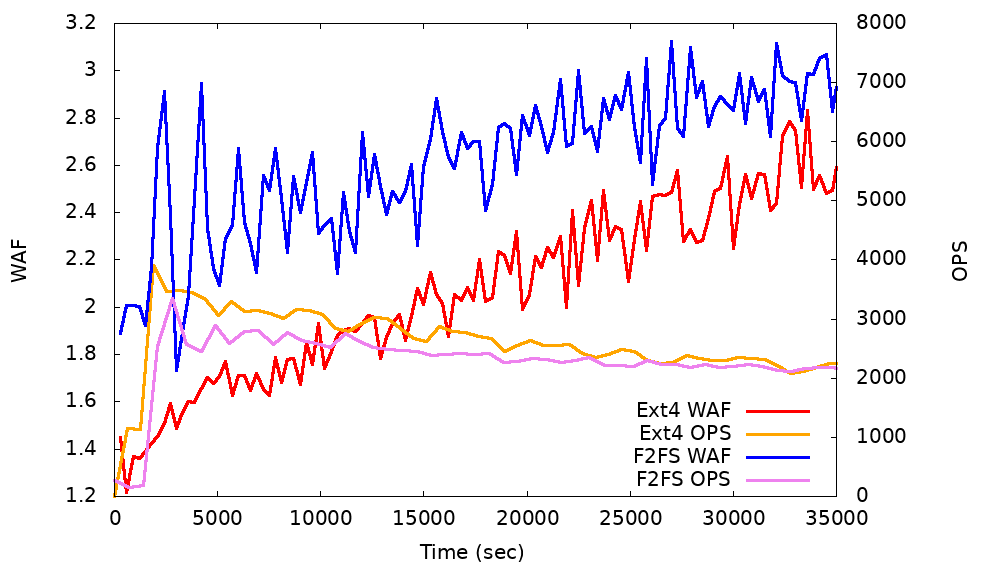}
    \label{fig:rocksdb-original-waf}
    \caption{RocksDB@Ext4/F2FS Multi-tenant}
\end{figure} }

In this subsection, we have demonstrated that, despite flash-friendly write patterns,  database engines and file systems still suffer from high write amplification which is comparable to or even greater than that in OLTP workloads. Though only three data stores are used to illustrate the multiplexing problem, the problem is we believe ubiquitous: to name a few, SQLite's RBJ or WAL journal, Ext4's journal area, and temporary files generated in Spark/Map-reduce applications or during join processing in relational databases. 


\comment{
\subsubsection{Multiplexing and its Causes}~\label{sec:multiplexing}

Every individual application intend to separate its logical objects and aims at making its write pattern flash-friendly either by writing files in their entirety (\eg, RocksDB) or appending write in sequence to logical objects (\eg, DWB or F2FS segments). However, these endeavors fail to group writes from different objects into different flash blocks, but they will be multiplexed, incurring higher WAF. 

Below we will explain why seemingly large sequential writes (\eg, 64MB SSTable write in RocksDB) are splited/scattered/multiplexed into many different flash blocks. In fact, there are numerous causes which can mix data pages from different logical objects with different death times in the same EU. 
concurrent wr, splitted by IO stacks, append-only writes 등; intermixed/multiplexed!
++ interplay of multilayers, complicated/복합적 <- log-on-log, fallocate. 
In addition, there are numerous other factors hindering clustering by objects: FS/block layer, different write time at app, and  file system fragmentation. \\
++ multi-tenancy, SSD architecture (RAID-like striping/multi-channel)

\begin{itemize}
\item  {\bf Multiple/Concurrent Writers} 
But, unfortunately, multiple write threads are active for better system utilization. For instance, multiple compaction threads in RocksDb, multi-log head in F2FS, and multiple buffer pool instances and multiple write threads in InnoDB. To be worse, it will be more common for multiple applications to share a single SSD (\ie, multi-tenants). Therefore, the assumption that pages from the same logical objects are clustered into the same flash block does not hold in reality, and the concurrent write threads will interfere each other. As a consequence, the pages from different logical objects are mixed in flash blocks and pages from one logical object are not less grouped instead multiplexed onto more flash blocks: we call this problem as multiplexing. 

\item {\bf Kernel-induced write splits along OS/IO stacks~\cite{SoyeeFastPoster20}: Concurrent Writers} rocksDB처럼 64MB write/flush해도, concurrent write threads 경우, 동시 write, write split, append 되기 때문에, in combination with SSD's ON-DEMAND allocation, pages from different files are mixed in the same flash block and also pages from a file/logical object are multiplexed onto different flash blocks (Figure~\ref{fig:sallocate-archi} (a)). In addition, due to the interplay of {\bf FS and block layer:} (, including IO scheduler, caching/replacement policy, and fairness~\cite{UnwrittenContract}), writes are split into smaller chunks, which are then shuffled and re-ordered along the path down to the SSD. A similar observation was made in {\bf Spark workloads}~\cite{Selecta}. 
\comment{DO NOT DELETE: ``Each Spark task reads/writes input/output data sequentially, streams from multiple tasks running on different cores interleave at the block device layer, thus the access pattern seen by storage devices consists of mostly random IO operations.''} 

\item {\bf Appending small writes:} In the case of DWB and F2FS segments with sequential write stream(s), unlike RocksDB, they fill the logical object in append-only manner (after allocated, but fill the lba ranges with data in scatter time; thus, mix data from diff apps into the same EU.  Each of multiple concurrent applications appends small-grained write to its write stream in an interleaved way. In particular, in combination with on-demand allocation and write-ordered mapping, application itself! - contiguous logical space에 대해 fin-grained chunk를 다른 시점에 쓰기 요청;  unit mismatch between fallocate and write, and write 시점이 다름. +++ concurrent writers \& single writers에서 write 시점 분리
\end{itemize}
}


\comment{ ##### FS Fragmentation 이슈? DO NOT REMOVE!!}

\comment{ On the other hand, copy-on-write로 인해 LBA-PBA의 mapping은 ever-changing해서, LBA-PBA가 align되지 않는 flash에서는 object에 대해 미리 contiguous allocation of LBAs at no fragmentation or 주기적 defragmentation is not effective at all in flash storage because XXX.  \reminder{SSD에서 LBA fragmentation 자체는 큰 문제 않됨. 대신 frozen page가 physically scattered되어 WAF 높이는 것이 더 문제임.} 따라서, 이런 문제점 개선을 위한 solution으로, \GATHER clusters and thus isolate objects of frozen pages into physical flash blocks so that the objects to be gathered are necessarily neither to be contiguous in terms of LBA nor to belong to the same object. The source of physical scattering of 'log structured write+FTL'}
\comment{Defragmentation - cluster data pages belonging to an object in terms of LBA so that their LBAs are to be contiguous as much as possible. vs. Gather - to cluster and thus isolate objects of frozen pages into physical flash blocks; the objects to be gathered are necessarily neither to be contiguous in therms of LBA nor to belong to the same object. }
\comment{The goal of defragmentation tools and file system defragmentation is to cluster data pages belonging to an object(e.g., file) in terms of LBA so that place object contiguously and make free space for one or a few contiguous regions. Note that 이게 동작하는 이유는,  in hdd, LBA/PBA가 align되기 때문.} 
\comment{what is the source of LBA fragmentation? LBA - PBA 1:로 mapping되기 때문에 fragmented LBAs means more movements of mechanical disk head and thus hurts read/write performance in HDD.  To rectify the performance degradation, ... }

\subsection{Object Characteristics in Data Stores}~\label{sec:pd-old}

Understanding workloads is essential to storage system design. However, little work has been conducted on characterizing objects in popular data stores from the perspective of the write multiplexing in flash storage.
In this section, we make three observations about object characteristics in flash-friendly data stores: logical space allocation, write pattern, and deathtime. The design of \falloc capitalizes on these characteristics. 

\module{Eager Logical Space Allocation} The host data stores manage data using logical objects such as RocksDB's SSTables, F2FS's segments and InnoDB's DWB. Each store will invoke the write system calls against such logical objects and those objects account for a dominant portion of total I/O in the store. To each logical object, data stores will allocate its logical address space at the file system layer in advance before writing data to the object. For instance, RocksDB calls the {\tt fallocate()} call prior to writing any data to a newly created SSTable. Hence, the logical address range per each SSTable is statically determined before writes are made to the object.  In this regard, we say that host-side stores know the relationship between objects and their logical address ranges and also {\em stream writes by objects}. 

\module{Write Pattern} The logical write pattern to each object can be characterized as sequential (in either batch or append). During flushing memtable in RocksDB, a write system call is made against L0 SSTable with memtable data as parameter, which we call batch sequential write. And, SSTables generated during the compaction in RocksDB will also be populated by batch sequential write. Meanwhile, the write pattern to individual segments in F2FS and DWB in InnoDB will be appended sequential write. It should be noted, however, that every data store usually armed with multiple write threads will issue writes from multiple objects concurrently to the storage. For example, RocksDB has by default four background compaction threads and F2FS employs multi-head logging for six active segments. Therefore,
even though the write pattern to individual logical objects is sequential, writes from different objects will interleave each other to the underlying storage.

\module{Deathtime} All pages of each individual object become dead together upon the object's deletion. For instance, when an SSTable is deleted upon compaction in RocksDB, all its pages will be invalidated. Likewise, upon a segment cleaning in F2FS, all its log pages will be discarded.  
Meanwhile, different objects will, though created at the same time, have different deathtimes. For example, two SSTables concurrently generated at different levels by two compaction threads are likely to compacted at quite distant points of time. The similar argument can be made with segments in F2FS, which were active together but cleansed at different times. 

\subsection{Problem Statement}~\label{sec:pd}
\comment{
repetitive issue!
\reminder{Introduction을 1)Multiplexing, 
2)Missing interface subsection, 3) Contribution 3가지 subsection으로 나눠서 
2.4 내용을 introduction과 합치는 방식은 어떨까요?}
}
\module{Multiplexing} 
The characteristics of object deathtime discussed above can bring to flash storage the chance of realizing stream-write-by-deathtime and avoiding write amplification.  
That is, given that pages from the same object tend to have the same deathtime and objects differ in their deathtimes, once flash storage can stream pages from different objects into per-object dedicated flash blocks, its effect is to stream writes by deathtimes. However, despite the eager logical space allocation by objects and log-structured sequential write to individual object in flash-friendly data stores, concurrent writes from different objects in single or multiple tenants will interleave to the flash storage. Thus, according to the stream-writes-by-time policy, the existing flash storage will pack those writes in their arrival order into flash blocks. An undesirable consequence is that pages from different objects are packed onto the same flash blocks. We call this phenomenon as {\em multiplexing}. As flash blocks are multiplexed with pages with different deathtimes, write amplification is inevitable. For example, consider the case where four SSTable files are multiplexed (left-bottom in Figure~\ref{fig:sallocate-archi}). When the SSTable1 file is deleted, each of the four flash blocks still keeps three valid pages. Thus, if a block becomes victim, three pages have to be relocated. In contrast, when four files are de-multiplexed into different files (right-bottom in Figure~\ref{fig:sallocate-archi}), a flash block with all pages invalidated is available. Thus, a clean block can be obtained without relocating any page.
\comment{ 
Further, the multiplexing problem will worsen under the striped write architecture.  For higher I/O bandwidth, most flash storage has taken the RAID-like multi-channel architecture and stripes data over different channels~\cite{SSDDesignTradeoff,UnwrittenContract,Chen-TOS16}. Thus, a write request will be striped in the unit of page (\eg, 4KB-long) in a round robin manner across many channels to exploit the internal parallelism in flash storage. 
As flash blocks are multiplexed by pages from larger number of objects with more distant death times each other~\cite{UnwrittenContract}, flash storage will experience higher WAF. 
}
\module{Missing interface} 
Consider why existing flash devices should be object-oblivious in handling writes and thus incur the multiplexing problem. The main reason is that, from the logical address in a write request, flash storage can not distinguish the logical object the address belongs to.
This is in turn because the conventional block interfaces do not provide any mechanism to pass the relationship between objects and their logical address ranges to the storage. That is, while the host software stacks are aware of the relationship, the valuable semantic cannot cross the storage interface wall but is simply discarded. In this sense, we say that a crucial semantic interface is missing which can make flash storage object-aware. Note that the absence of such useful interface is due to the legacy that the existing storage interfaces have been developed assuming harddisk. In the case of harddisk where overwrites are allowed, logical space allocation via the {\tt fallocate()} call at the file system layer implicitly indicates physical space allocation at the storage. 
Thus, writes from the host will go to their corresponding physical location simply according to their {\tt start\_lba}s: harddisk is object-aware in handling writes without any hint from the host. For this reason, no storage interface was in need, which can pass the host semantic about object's address range to the storage. This unwritten contract fails on SSDs where overwrites are not allowed.


\comment{ 
In summary, because there exists no interface to offload the valuable host-side semantic about object's logical address range, many data stores with flash-friendly write patterns suffer from device-level write amplification on flash device. We believe that it is compensating to turn the object-oblivious flash storage to be object-aware. 
In the next section, we suggest a new interface which can enlighten flash storage to be object-aware in handling writes, thus preventing the multiplexing problem. 
}

\comment{
\begin{verbatim} 
fallocate(2) - Linux man page Synopsis
<fcntl.h>int fallocate(int fd, int mode, off_t  offset/"", off_t " len ");

Description
 .. specified method of ensuring that space is  allocated for a file, see posix_fallocate(3). fallocate() allows the caller to directly manipulate  the allocated disk space for the file referred to  by fd for the byte range starting at offset and 
continuing for len bytes.

The mode argument determines the operation to be  performed on the given range. .. 

Allocating disk space
The default operation (i.e., mode is zero) of  fallocate() allocates and initializes to zero  the disk space within the range specified by  offset and len... This default behavior closely  resembles  the behavior of the posix_fallocate(3) 
library function, 
.. 
After a successful call, subsequent writes into the  range specified by offset and len are guaranteed  not to fail because of lack of disk space.

... Preallocating zeroed blocks beyond the end of  the file in this manner is useful for optimizing  append workloads.

Because allocation is done in block size chunks, fallocate() may allocate a larger range of disk  space than was specified.

Deallocating file space
Specifying the FALLOC_FL_PUNCH_HOLE flag (available  since Linux 2.6.38) in mode deallocates space (i.e., creates a hole) in the byte range starting at offset  and continuing for len bytes. Within the specified  range, partial file system blocks are zeroed, and  whole file  system blocks are removed from the file.  After a successful call, subsequent reads from this  range will return zeroes.
\end{verbatim}
}

\section{Design of \falloc}~\label{sec:design}

In this section we propose new interface for flash storage, called \falloc, and presents its design principles, semantics, and architecture. In addition, we explain its use cases and benefits. 



\subsection{Key Idea and Design Principles}

As discussed in Section~\ref{sec:bg}, grouping data by deathtime is effective in reducing write amplification~\cite{UnwrittenContract}. Considering that pages of individual object have the same deathtime while different objects have different deathtimes, 
grouping data by objects will have the effect of grouping data by deathtime. 
However, as pointed out in Section~\ref{sec:pd}, the existing flash devices can not group writes by objects simply because they are unaware of the relationship between objects and their logical address ranges. This is, in turn, because no interface exists to allow host to convey the semantic to the flash storage. 


Recognizing the missing interface, we introduce a new interface,  \falloc, to hint flash devices about the host-side semantic that all pages in a logical address range belong to the same logical object. With the help of the simple hint, flash devices should be able to place writes into distinct flash blocks by objects. The design objectives of \falloc are threefold. First, it takes advantage of the existing concept at the host layer, per-object logical address range. This is in stark contrast with other approaches (\eg, multi-stream SSD and ZNS~\cite{MultiStreamSSD,ZNS}) which introduce new concepts (\eg, stream-id and zone-id) and thus force applications to adapt to their interfaces. Second, host-side data stores should be able to leverage \falloc with minimal change. In particular, required changes, if any, must be limited to the use of abstraction provided by \falloc. 
Third, \falloc aims at passing the host semantic to the storage without being limited to any specific application domain. So the abstractions of \falloc must introduce minimal changes to the standards such as NVMe and the changes must not disrupt existing applications. 
This approach is novel in that it turns the common knowledge at host (\ie, a logical address range constitutes an object) into a strong point for flash storage (\ie, to be able to stream writes by objects). 

\comment{ 
\module{Bimodal block utilization(여기?)} Once \falloc-ed, different objects are isolated into dedicated flash blocks. In addition, those object will be dead at different point of time. Therefore, at a point of time, while all pages of dead objects' flash blocks are invalid their entirety, all pages of live objects' flash blocks remain valid. Thus, the extreme bimodal distributions of block utilization are expected for \falloc-ed objects.}

\subsection{Interface} 

As a way to pass the information that a logical address range constitutes an object from the host to the flash storage,  we propose new {\tt FlashAlloc(logical\_addr\_range)} interface, as detailed below.

\begin{description}
\item[FlashAlloc (\{LBA, LENGTH\}*)] \falloc informs flash storage that the logical address range denoted by the parameter, {\tt \{LBA, LENGTH\}*}, belongs to one object. As indicated by *, an address range can consist of one or multiple logical chunks. Recall that an object may consist of multiple chunks because of the file system fragmentation~\cite{fragmentation}. Each chunk is presented by a pair of {\tt LBA} and {\tt LENGTH} which represent its starting address and length, respectively, and chunks should not be either continuous or overlapping. 

\end{description}

\comment{CYCLIC flag없이 명시적 alloc/dealloc반복 {\tt \falloc ( \{LBA, LENGTH\}*, CYCLIC)})\\ The second parameter, {\tt CYCLIC} is an {\bf optional flag} indicating whether this address space will be continuously recycled. 
} 

Since the storage command is not always available to applications (\eg, database engines) that access objects through a file system,  we exploited the {\tt ioctl} infrastructure so that the \falloc command can pass through the file system to the storage device, instead of invoking the new command directly from applications.


More importantly, calling a \falloc command, the host expresses its intention that it will perform operations on the given logical address range, {\em LS}, as an {\em integral unit for writes}: once a portion of the dataset is written, all the dataset is going to be written once, and later they will be invalidated together nearly at the same time.  
This is a useful hint for write optimization in flash storage. Upon receiving a \falloc command, flash storage will dedicate the corresponding physical flash block(s) ({\em PS}). 
\comment{FA's key role (by mw.oh)} Then, flash storage will store all writes from {\em LS} in the arrival order into {\em PS}; In this way, flash storage will guarantee the physical clustering of all pages from the same logical object. In particular, note that even when the writes from an \falloc-ed object are {\em spatially fragmented} (\eg, due to file system aging~\cite{fragmentation}) or {\em temporally split} (\eg, due to log-appending in F2FS), they are guaranteed to be eventually clustered into the same flash block(s). \comment{ though arriving as splitted or at different time;; or split along the kernel I/O stacks.} Thus, when properly \falloc-ed, concurrent write streams from different logical objects will be de-multiplexed into each own dedicated flash block. Recalling the object deathtime characteristics that objects tend to be dead at different times while pages in each object are invalidated at the same time, the physical clustering by logical objects will have the effect of stream-write-by-deathtime, thus minimizing write amplification. The beauty of \falloc lies in that it can achieve the transparent write streaming without the hassle of assigning stream-id or zone-id to each write request~\cite{MultiStreamSSD,ZNS}.    
\comment{ 
{\bf MS-SSD, ZNS 대비 \falloc 장점 (mw.oh):} 옵젝트별 same deathtime 보장-> 리소스 할당 후 호스트가 굳이 batch/seq write 않아도 같이 invalidate;  seq steams의 demux 라기 보다, {\bf logically clustered streams의 demux}가 more general 표현; stream id 부여같은 번거로움 없이!!}
\comment{FA guarantees the physical clustering of pages belonging to the same logical clustered obj; group by death time 달성;; Once FA-ed upon a logical obj creation, all its new writes will be eventually physically clustered even when its logical space is spatially fragmented due to file system aging or when its writes are temporally split (due to either log-append at applications (F2FS) or splitted along the kernel I/O stacks),}

\module{Use Cases} As illustrated in Figure~\ref{fig:falloc-archi}, popular database engines and file systems have write-intensive objects (\eg, SSTable, DWB, segment) whose IO patterns fit well with the purpose of \falloc: each object is written sequentially just once and later becomes dead in its entirety at the same or similar time. In addition, objects with such ``write-once and dead-at-once'' pattern are ubiquitous in most data stores: numerous LSM-based KV stores, WAL log files in relational databases, two journaling modes (RBJ and WAL) in SQLite, and file system journaling.  
\comment{
TODO(jhpark): 
appending for unknown size 내용 comment로 남겨서 다음 제출때 내용 보강
종규형 sosp 논문 reference + fallocate 사용하지 않으면 fragmentation 심한 문제 언급
fallocate사용해야 하는 논리 보강
application level에서 fallocate를 사용하는 경우를 타겟
random writes ( + unkonw size) write내용 언급
}
\comment{one or multiple append points (each with contiguous address space), append/write once, and later all pages are invalidated/overwritten at the same or very similar times.} In addition, \falloc will naturally stream writes from different tenants so that it is, as demonstrated in Section~\ref{sec:eval}, quite beneficial in reducing write amplification and performance interference in multi-tenant database environments~\cite{SeattleReport}. Meanwhile, \falloc is not a panacea for write amplification problem in flash storage. 
In particular, \falloc would be inappropriate for objects with random overwrites (\eg, OLTP tablespace). Thus, data from such objects can be better handled, as discussed below, using conventional FTL techniques.

\comment{In particular, \falloc would be inappropriate for objects with random overwrites or unknown block size (\eg, OLTP tablespace). Thus, data from such objects can be better handled, as discussed below, using conventional FTL techniques.
(내용 흐름 어색해지지 않는 선에서 다음 문장 추가 고려?)
However, most applications are already attempting to cluster object writes 
by directly manipulating the disk space allocated via posix\_fallocate().
We would like to note that \falloc, in a sense, enables transparent 
write streaming upon posix\_fallocate() call.
}
\comment{ 
Though we have discussed only a few use cases of \falloc, we believe it can be leveraged by many {\bf other numerous applications} and can also effectively address many inefficiencies in flash storage ~\cite{DONOTSTACK,UnwrittenContract,defragger}. \comment{, which have been raised in system communities as both the block size in flash chip and the parallelism degree in flash storage are ever increasing, flash storage's segment size가 계속 증가. 이는 the Log-on-log problem in multi-log environment 심화주범~\cite{DONOTSTACK}.}  \reminder{이 이슈 F2FS at 2.2 or eval. section서 언급?} For instance, we can mitigate the the {\bf log-on-log problem} (what?) by (fine-grain) controlling each of active log objects at the upper layer (\eg, segment in F2FS) to align its corresponding erase unit at the underlying flash storage (thus, providing multiple aligned append points) and thus making different log objects isolated into different flash block.
across mobile and desktop applications and even for cloud/VM/multi-tenancy environment. It is apparent, as discussed so far, that \falloc can reduce write amplification in applications with large and concurrent sequential writes (\eg, RocksDB), multiple append write streams(\eg, F2FS), and cyclic logs (\eg, DWB).  In addition, we expect that \falloc can fully or partially (more or less) address many of observations on the SSD-related inefficiencies recently made~\cite{UnwrittenContract}.\comment{(including 7, 10-20, and 22-24).} It would be interesting to characterize the workloads to which \falloc is helpful and to develop a comprehensive taxonomy for \falloc use cases, but which is beyond the scope of this paper. 
}

\subsection{Architecture}~\label{sec:archi}

\comment{ \includegraphics[clip,width=0.4\textwidth]{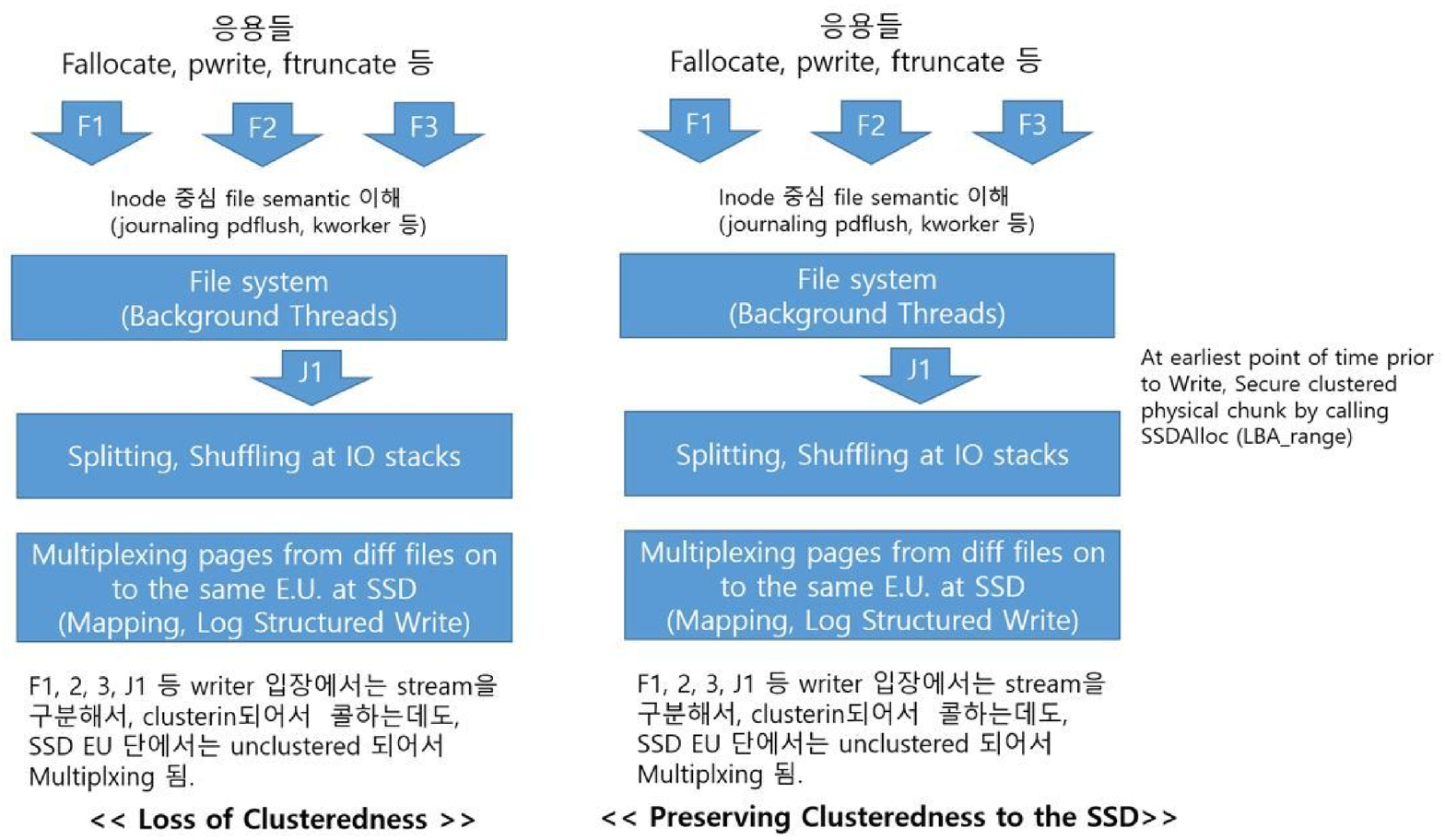} \label{fig:idea} }
\begin{figure*}[!ht]
\centering
    \includegraphics[clip,width=0.7\textwidth]{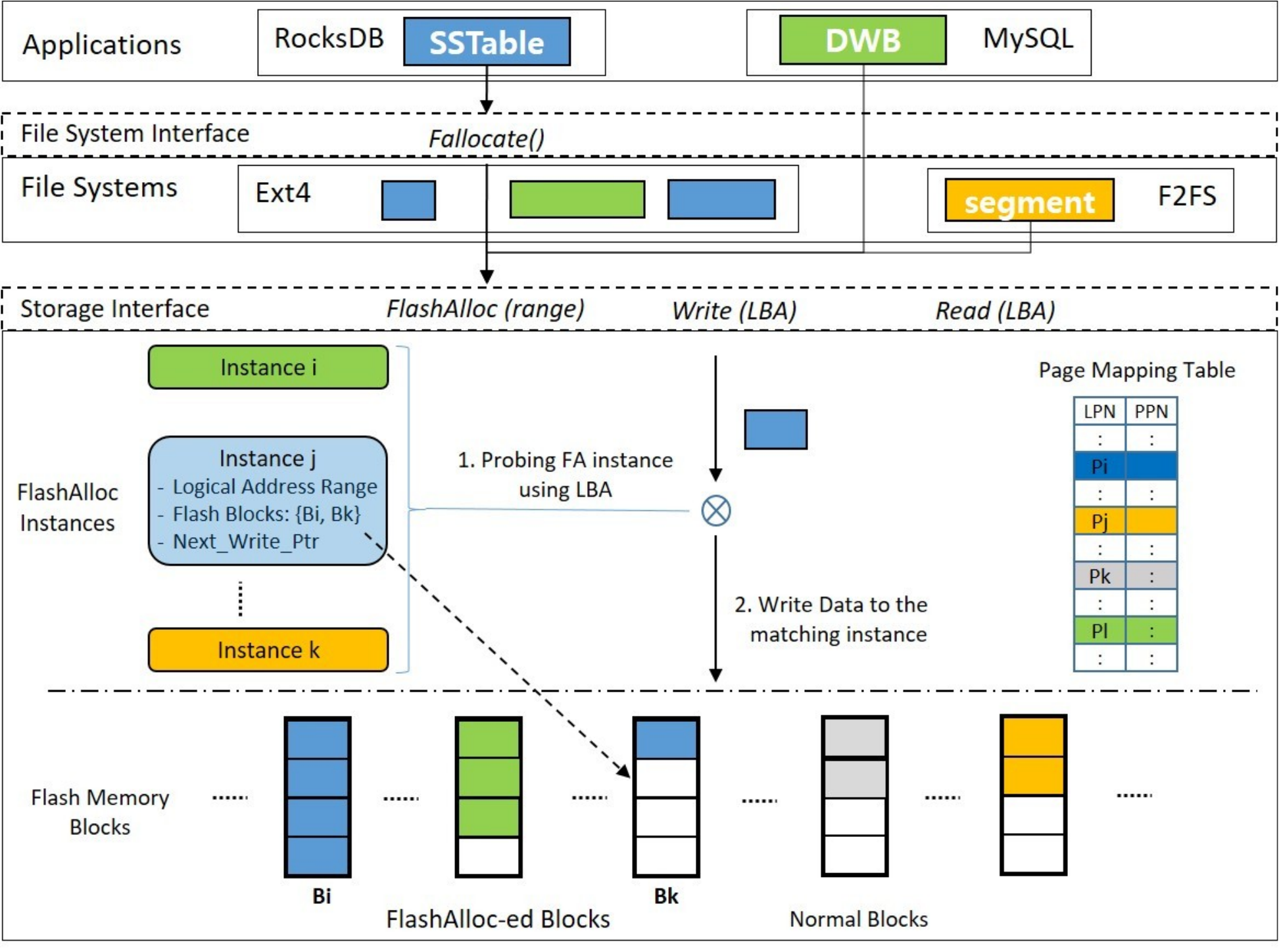}
    \caption{\falloc Architecture}
    \label{fig:falloc-archi}
\end{figure*}

Figure~\ref{fig:falloc-archi} illustrates the architecture of \falloc. Using the figure, we explain the concept of FlashAlloc instance and its physical space management. Also, we explain how write and read operations work and describe how the logic of trim and garbage collection is extended to support \falloc. Though the page-mapping FTL is assumed throughout this paper, \falloc can be seamlessly supported by other FTL schemes such as block and hybrid mappings~\cite{FTLsurvey}. 

\module{\falloc Instance} Upon receiving a \falloc command with a logical address range, flash storage will create its corresponding \falloc instance (in short, FA instance). In addition, flash storage will secure the corresponding physical space (\ie, one or more clean flash memory blocks whose total size amounts to the given logical address range's size), and dedicate the space to the instance\footnote{For the simplicity of discussion, we assume synchronous physical space allocation throughout this paper. Note, however, that the \falloc interface does not mandate synchronous allocation. Instead, as long as the physical space is secured prior to the first write to the instance, either asynchronous or even on-demand physical space allocation (or securing) is acceptable, which may bring some optimization opportunities. We leave such an interesting optimization topic as future work.}. 
The physical space to an FA instance is allocated in the unit of flash memory block (\eg, usually 2MB in flash memory chips) so as to isolate writes from the instance into the dedicated blocks
Though we assume flash memory block as the unit of physical space allocation for the simplicity of discussion, we do not exclude the case where two or more FA instances can share one physical flash block, when their logical address sizes are less than the size of a flash memory block. We would like to stress that, even when a flash memory block is shared by two or more FA instances, it is always beneficial to cluster pages from one logical object into the same physical flash block and thus to prevent them from scattering over different flash blocks
\comment{
\footnote{Though we assume flash memory block as the unit of physical space allocation for the simplicity of discussion, we do not exclude the case where two or more FA instances can share one physical flash block, when their logical address sizes are less than the size of a flash memory block. We would like to stress that, even when a flash memory block is shared by two or more FA instances, it is always beneficial to cluster pages from one logical object into the same physical flash block and thus to prevent them from scattering over different flash blocks}. 
}
The dedicated blocks for FA instances are called as \falloc-ed. Once multiple blocks are allocated to an instance, data for the instance will be striped across channels for better bandwidth~\cite{Cosmos2020}. As illustrated in Figure~\ref{fig:falloc-archi}, an FA instance consists of three metadata: logical address range, physical flash blocks, and physical location for next write. Next write for an FA instance will be appended to the page pointed by {\tt next\_write\_ptr}. 

\comment{ 
THUS, given an \falloc command, SSD/FTL will probe its logical address range against those of existing active instances, and return errors if any overlapping instance exists. In fact, user/application/FS should guarantee that the logical address ranges does not overlap that of any active instance.} 

\module{Active FA Instances}  Once created, every FA instance and its metadata will remain active and also be managed persistently until its destruction. An FA instance and its metadata will be destructed once its physical space is filled up. As depicted in Figure~\ref{fig:falloc-archi}, multiple FA instances could be active at a point of time. Note that the logical address ranges of all instances should be disjoint from each other. Meanwhile, the number of active FA instances will remain rather small in practice, though unlimited in principle. For instance, considering that an SSTable file in RocksDB is, once created, quickly filled by memtable flushing or compaction, its FA instance will be destructed shortly after created. Thus, when running a RocksDB, the number of active FA instances will not be greater than that of concurrent compaction threads(\ie, four by default). Similar argument can be made for F2FS segments.  

{\bf Non-\falloc-ed Objects} While good for objects with {\em sequential or log-appending write patterns}, \falloc is not intended for objects with random write pattern. For the latter type of objects, \falloc is not recommended to be called. Thus, flash devices supporting \falloc are desirable to be able to support such non-\falloc-ed objects as well; writes from non-\falloc-ed objects can be handled same as in the conventional SSD and stored in {\em normal} blocks. For this reason,  
in our \falloc architecture, there are two types of blocks, \falloc-ed and normal, as illustrated in the bottom of Figure~\ref{fig:falloc-archi}. In this regard, our approach is in contrast with the ZNS interface which strictly assumes all writes to be sequential and thus cannot support workloads with random writes. 

\module{Write} For a write request with two parameters, {\tt LBA\_start} and {\tt length}, indicating the starting LBA and length of the data respectively, flash storage first probes its matching FA instance using the {\tt LBA\_start}, as depicted in Figure~\ref{fig:falloc-archi}. If a matching instance is found, the write data will be appended to a {\em \falloc-ed block} pointed by the instance's {\em next\_write\_ptr}, and the corresponding entry in \l2p table and {\tt next\_write\_ptr} value will be accordingly adjusted. Otherwise, the request comes from non-\falloc-ed objects. In this case, as in the conventional SSDs, the writes will be stored in {\em normal} block. 
In this way, our approach supports two write policies: stream-write-by-object for \falloc-ed objects and stream-write-by-time for normal writes.

\comment{ 
\reminder{위치 여기?} All such writes with no matching active FA instance will be directed to the special instance, {\bf non-\falloc instance}. two types of normal and FA blocks: FA instance data는 다른 object data와 섞이면 안되니, flash block(s) allocated to a specific FA instance is dedicated to store data for the instance. We distinguish such FA block from normal blocks which non-FA write를 저장. Thus, flash blocks can be classified into two types: normal and \falloc-ed blocks.} 

\comment{When {\bf active FA instances} are too many, the space overhead of maintaining their metadata and the runtime overhead of probing LBA to FA instance will be proportional to the number of instances. Such overheads will be acceptable since the number of active instances, as explained above, is quite limited in practice.} 

\module{Read} While \falloc aims at reducing WAF by controlling the placement of writes into different blocks by objects, its read operation will proceed exactly same as in the conventional page-mapping FTLs: after looking up the physical page number (\ie, block-id + page-offset) from the mapping table with the given LBA address, FTL reads out the page from the corresponding flash block.

\module{Trim} The trim command was introduced to notify the flash storage that a set of logical pages is no longer valid. Upon receiving a trim command, flash storage will invalidate the relevant pages, preventing them from being unnecessarily relocated. In our \falloc architecture, the trim command will be handled same as in the conventional flash storage with one exception. Considering that pages in an \falloc-ed object have the same deathtime and 
\comment{: they are invalidated together at once when the object become dead  (\eg, RocksDB SSTable deletion and F2FS segment cleansing).}  
are also clustered in the same physical flash block(s), the trim command against the object can complete by erasing the block(s) instead of invalidating individual pages. \comment{, and returning them to the free block pool.} In this sense, \falloc enables to achieve nearly {\em zero-overhead trim}. In fact, the trim command is, though effective in lowering write amplification, known to induce non-trivial run-time overhead (\eg, trim spikes~\cite{RocksDB-Exp}); \comment{, especially when pages being trimmed are scattered over different flash blocks.} thus data stores such as RocksDB and F2FS take the {\em delayed discard} policy to mitigate the effect of trim spikes on the write latency~\cite{MyRocks, F2FS}. The trim optimization enabled by \falloc will remove the {\em trim-induced stalls} so that the developers are free from the burden of devising deliberate {\em rate limiting} to file deletion~\cite{MyRocks} or {\em delayed discard}~\cite{F2FS}. 
\comment{In fact 이하 trim optimization: 공간 부족시 삭제} 
\comment{ 
{\bf Cyclic Flag, DeAlloc, and Instance Lifecycle:}
 When \falloc is called with the {\bf cyclic} flag enabled, then flash storage will allocated a large chunk (\eg, 64MB = 2MB flash block x DoP). In addition, when the current chunk becomes full, another chunk will be automatically allocated without further \falloc call from the host. A flash block dedicated for cyclic FA instances will be erased when all its pages are {\bf invalidated by cyclic overwrites.} In addition, an FA instance with cyclic flag has to be explicitly deallocated. For instance, MySQL engine has to deallocate the DWB area upon a database shotdown or destruction.
 
\module{DEALLOC and \falloc Instance's Life cycle} Every \falloc instances has to be {\bf destructed when} its object is deleted(?). A normal FA instance without cyclic option will be {\bf destructed once its physical space runs out}. In contrast, cyclic FA instances (\eg, DWB/ext4 journal area) will not be automatically destructed even when its physical space is ran out, but instead new physical space will be automatically allocated. \reminder{status: creation, active, closed, destructed??} 
{\bf Explicit or Implicit de\falloc?? When in case of Implicit?}  Note that all the physical pages of a destructed \FA instance remains valid until invalidated/overwritten/discarded/deleted; 
\comment{recall that 해당 데이터들은, in most cases, file deletion 또는 명시적 discard(예: DWB)에 의해 한꺼번에 전체가 invalidated되고, 따라서, upon de-\falloc, all \falloc-ed blocks will be erased and returned as free block. NOTE: zero TRIM overhead}} 

\comment{ 
\falloc design can be compatible with all mapping policies?? or, \falloc enforce a mapping policy or vice- versa?  for flexible page write, dynamic (vs. static) page allocation is assumed, which is fortunately taken by every SSD with page-mapping FTLs ~\cite{DynamicPageAllocationInSSD}.
The effectiveness of increasing data parallelism can be highly dependent on the physical data layout. And, the mapping between LBA and physical space can determine the physical data layout. There are a few alternative mapping policies. Dynamic write-order-based mapping is commonly used, and thus assumed throughout this paper. 
\begin{itemize}
    \item Static = LBA mod C\# 
    \item Write-order-based dynamic = i-th write mod C\#
    \item cf. log-structured mapping = appends write in a flash block 
\end{itemize}   
Cases: \falloc can cause {\bf unbalanced- or ill-mapped data layout}, which can in turn lead to suboptimal write/read performance.
\begin{itemize}
\item alloc size < 2MB;  2MB < size < 64MB; >64MB;;  
\item 2MB == dedicate a physical flash block in a channel/die --> this can limit the parallelism in read as well as write.; but with highly concurrent IOs, this is not problematic at all. 
\item 64MB == full striping == works at no performance loss. 
\item between 2MB and 64MB === limited form of striping and thus limited performance gain for write; but for read, same to  
\end{itemize}} 

\comment{\module{Life Cycles of Two Block Types} Before explaining GC mechanisms for \falloc in detail, let us explain the {\bf Life cycle of two block types.} First, {\bf FA blocks} will experience three stages: {\bf clean, partial full(active), and full}. That is, initially completely clean when allocated, partially filled as writes are appended, and finally ) full with all valid pages. After remaining full while the corresponding object is alive(?), an FA block will be erased and returned as free upon FlashDealloc. In this way, all the pages of the \falloc-ed objects will not be trimmed individually so that the trim overhead can be compltely avoided. In reality, an FA block could be in {\bf full-then-partially invalidated status} when pages will be overwritten individually at different times and thus incrementally invalidated. But, this scenario is very rare in real applications and, though happen, at very short duration. Thus, for the simplicity of discussion, we assume that all pages in FA block will be invalidated when object deletion. In the case of  {\bf normal blocks:}, ... \reminder{언급 필요?} 
}

\module{Garbage Collection (GC)} When no space is available for new writes, FTL has to conduct the GC operation: copying valid pages from a victim block to a clean block, {\em fb} and then returning {\em fb}. Unfortunately, the conventional GC algorithm returning a partially clean block {\em fb} is inappropriate for FA instances. Recalling that \falloc intends to disallow pages from different objects to mix in the same flash block(s), each FA instance needs total-clean block(s). \comment{and also to secure them in advance.} Thus, FTL has to handle two cases of GC differently: The first case is when \falloc is called, and the second one is when free space is unavailable for normal write. 
Recall that when a \falloc-ed object is trimmed, all its blocks will be returned as free and thus the free block pool is likely to have some free blocks. 
In the case when GC is triggered for normal write, FTL will first check the free block pool for a free block and, if found, use it. Otherwise, it will carry out the conventional GC: it chooses a victim block with the least valid pages and merges the victim. In the second case, when a \falloc call is made, FTL checks whether clean blocks are available in the free block pool and, when unavailable, has to merge multiple blocks to secure total-clean block(s)~\cite{2R}. 

In addition, while selecting victim blocks for each of two GC cases, block types have to be taken into account so that pages from normal blocks are not mixed with those from \falloc-ed blocks. 
In the case of GC for normal writes, only normal blocks can be victims. Otherwise, once an \falloc-ed block is chosen as victim, non-\falloc-ed normal writes should co-locate with the relocated pages from the old \falloc-ed victim block.  In the case of GC for \falloc, as discussed above, multiple blocks have to be merged to secure total-clean block(s) for \falloc. In this case, those multiple blocks should have the same type. In this sense, we call our algorithm as GC-By-Block-Types. 
\comment{{\bf First of all}, \falloc needs {\em eager GC} while the existing FTL will trigger GC {\bf on demand} when there is no space for new writes.}
Interestingly, the GC-By-Block-Type algorithm will {\em adaptively allocate} the space of \falloc-ed and normal blocks: depending on victim block type, \comment{
which is in turn determined by the write pattern from the host,} one region grows while the other shrinks. For instance, if a new \falloc block is in need and no free block is available, multiple normal blocks have to be merged so as to secure a total-clean block, enlarging \falloc-ed region. In this way, the space allocation to both regions will adapt to the changing workloads, requiring no static allocation or tuning. 



\comment{\falloc aims at improving {\bf sustainable performance}, not immediate, short-term one; e.g. channel-wise allocation. Therefore, since it is clustering pages with same death times into the same flash blocks, 
its performance benefit is not immediate instead comes later. In some cases, \falloc may result in suboptimal performance for seq read/write. For instance, BW of seq write may be lower than non-falloc case because it can not exploit full parallelism when single block is allocated. Though, \falloc will eventually reduce interference among multi-tenants. so that it will outperform non-falloc version in terms of sustainable performance.} 
\comment{ Third, in terms grouping pages into flash blocks, the conventional FTLs can be categorized as {\em group-by-time} ~\cite{UnwrittenContract}  in that new page writes are merely appended to flash memory blocks in the time order requested from the host. In contrast, with \falloc, new page writes can be grouped into dedicated flash blocks by their segments. Considering that pages of the same segments tend to have the same deathtime, \falloc thus enables the {\em group-by-deathtime} policy, thus lowering write amplification~\cite{UnwrittenContract}. 
\comment{2) Group by logical address space: {\bf Segmented FTL}~\cite{J.U. Kang Super Block, C.I.Park REconfigurable FTL 등},  3) Data by Death Time: this type is missing. \falloc can support this concept efficiently - fully or partially(?). {\bf vs. Segmented FTL / Fragmentation:}  (J.U. Kang Super Block, block mapping도 비슷한 논리?!), where N+M physical blocks are allocated to contiguous N logical blocks.}
}


\subsection{Advantages} 

To the best of our knowledge, \falloc is the first work which allows host applications to hint the storage that a logical address range belongs to the same object. The advantages of \falloc are threefold. First, enlightened by \falloc, flash storage can now cluster data from the same object and thus with the same deathtime into the same physical flash blocks, thus minimizing write amplification.\comment{Recall that data from the same logical object tend to be physically fragmented into different blocks in the conventional object-oblivious flash storage.} In particular, data from objects which are even {\em logically fragmented} will be {\em physically de-fragmented} into the same flash block(s). In this regard, \falloc could be a clean solution to the logical and physical fragmentation problem in flash storage~\cite{AMF,defragger}.
Second, since the abstraction used in \falloc complies to the popular abstractions (\eg, {\tt fallocate()} call and segment in F2FS), the existing software stacks can achieve write streaming with minimal code changes, as detailed in Section~\ref{sec:impl}. \comment{less than 10 lines were added to each software's codebase for \falloc.}
\comment{and further those changes are limited to modules interacting flash storage via the \falloc command.}
\comment{
아래 내용 삭제; 대신 이미 fallocate를 사용하는 RocksDB 같은 응용들은 
소스코드 수정 최소로 할 수 있음. 
In particular, some applications such as RocksDB will do so at no code change. Such {\em application-transparent write streaming} is possible because ext4 file system will automatically invoke \falloc upon {\tt fallocate()} call.
}

\comment{
\reminder{
(R2:D2) "It’s unclear whether the proposed scheme incurs any penalties if turned on by default on file-system level."
fallocate()를 사용하는 근본적인 이유는 write clustering이 주 목적이여서 
default로 켜져 있어도 큰 패널티는 없을 것 같음. (file system에서 모든 write를 \falloc-ed write로 변환하는 것이 아님.)
하지만, flash block size보다 작은 경우 자주 호출되는 경우 문제가 될 수 있음을 명시?
해당 write 들의 lifecycle 및 flash block utilization을 고려하여 선택적으로 적용?
}
}

\comment{As a design principle, \falloc is designed to be transparent as much as possible to applications/users so that the changes required to be made at existing software stacks should be none or minimized .}
Third, since the \falloc semantic is simple and does not require involving intermediate layers such as the kernel block layer, it is easy to incorporate the command into the existing storage interface standards such as SATA, SAS and NVMe (\eg, using the vendor specific command).  

\comment{ 
More importantly, unlike the existing multi-stream abstractions~\cite{MultiStreamSSD, PCStream} which requires to find every write calls and to modify call to add the proper streams. In addition, they poses several technical issues on developers such as {\bf \# of necessary streams and object-stream mapping},\comment{through deathtime analysis} each of which is challenging considering that \# of streams is quite limited in practice. In contrast, \falloc will completely free developers from such burdens because of the fine-grained FA instance-per-object. Please refer to Section~\ref{sec:MS-SSD} for details.}

\comment{ \begin{figure}[!ht] \centering
    \includegraphics[clip,width=0.3\textwidth]{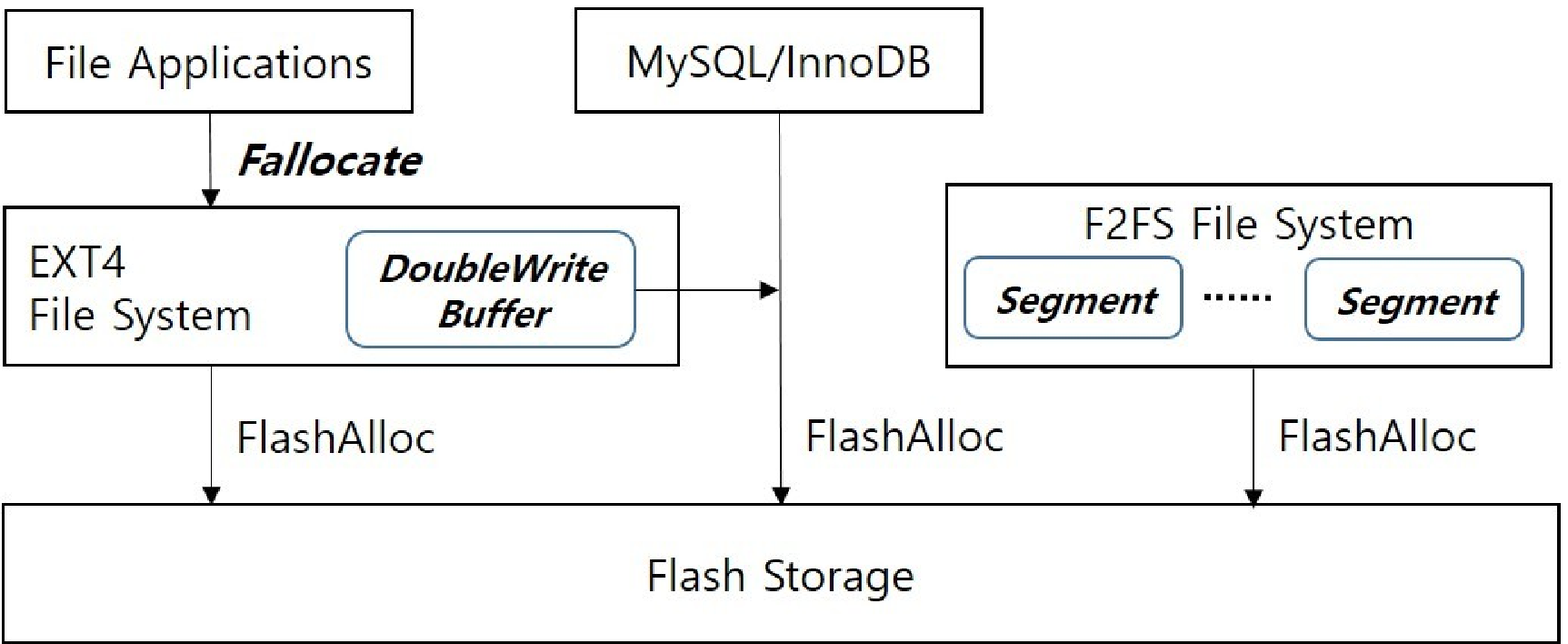}
    \label{fig:usage-archi} \caption{\falloc: Use Cases} \end{figure} }

\comment{
Now let us discuss three additional benefits of physically clustering each logical object's data using \falloc such as physical de-fragmentation. 

\module{Physical de-fragmentation of logical objects:} 
In the object-oblivious flash storage, as explained in Section~\ref{sec:multiplexing}, {\bf writes against logically contiguous} spaces of an object will be {\bf physically fragmented and scattered} across different flash blocks. Although sequential/appending write for the same logical object can, split/diverge/scatter TEMPORALLY and SPATIALLY for various reasons such as multiple concurrent write threads, write splits along the IO stacks, and small-appended writes at different points of time, the \falloc mechanism can address all the root causes of multiplexing by eventually physically de-fragmenting/
de-multiplexing them into the same physical flash blocks.
In addition to contiguous LBAs, \falloc can achieve {\bf physical defragmentation even for logically fragmented objects}. Due to the inevitable file system aging over time~\cite{fragmentation}, the logical address space of an object (\eg, RocksDB SSTable) could be, though perceived as contiguous from the application perspective, 
in reality allocated by the file system as fragmented. Even for such logically fragmented object, \falloc can naturally defragment their pages into the same physical flash blocks. Hence, it could be a {\bf clean and fundamental solution} to the logical and physical fragmentation problem in flash storage~\cite{AMF,defragger}.
\comment{Note that logical fragmentation will cause physical fragmentation even under {\em segmented FTL}, where N+M physical blocks are allocated to contiguous N logical blocks~\cite{UnwrittenContract,SUPERBLOCK}.} Note that such physical defragmentation is impossible even with harddisks once an object is logically fragmented~\cite{fragmentation}. 
}

\comment{영어 참고 
-- 하지만, {\bf FS fragmentation} and F2FS cause pages of a file to be scattered across logical address so that even segmented FTL could not guarantee that those pages are not likely to be clustered in the same physical flash blocks. different files' pages can be stored in one logical block\cite{defragger}. By contrast, with \falloc, even when the address space of a logical segment is fragmented (\ie, a fallocate call is allocated to multiple, non-contiguous LBA ranges), pages from the same segment will be clustered into the same physical space.}

\comment{ 
Until now, we have explained the benefit of \falloc in terms of write performance (reducing GC and thus improving write performance). Interesting, once properly \falloc-ed, we can also expect {\bf better read performance}: the fact that all pages from the same file are clustered in the same flash blocks will make \eg, compactionSSTable에서 sequential read faster.(근거?);; In addition, {\bf Prefetching for Read:} \falloc 앞부분 순차 읽히면, can prefetch next reads in sequence! For instance, this idea is applicable to each input SStable during RocksDB compaction.}


\comment{ \item By space: 논리적 구분: file 단위, Extent 단위(when multiple tables reside in one OS file), 좀 더 fine grain한 object 단위(?? if any), 응용단위, VM 단위  ==> 논리적인 종류 상관없이 물리적으로는 결국 fallocate aligned with EU size로 대응가능함.  \item By time: 서로 다른 논리적 객체에 속하는 page들이 semantic 이유로 같은(비슷한 시점에 같이 update되어 같이 persistent하게 될때 (예: SQlite처럼 force 정책을 사용하는 경우); 또는 checkpoint시 disk write할 때 같이 update되는 page들을 묶어서 write하는 전략 필요; 또는 결국 random write라써 2R FTL에서 다루어야 하나?}


\comment{ 
- dealloc 없이 다 쓰고 나면 자동 dealloc? 이 경우, 다 사용해도 FA instance가 persist하게 하는 cyclic reuse option 필요. 이는 또 explict dealloc 필요?  As will be discussed later,  an \falloc instance remain active (\ie, kept persistent) at SSD/FTL until destructed explicitly or implicitly. An FA instance will be {\bf destructed} once its physical space is ran out. 
\reminder{trade-off: 매번 falloc call, org wr후 바로 dealloc VS. 자동 reuse - \falloc call 회수 줄임?} ? However, some logical objects such as DWB and ext4 journal area will be cyclically reused (that is, overwritten). In this case, it would be better for flash storage to automatically reallocate new physical flash blocks space whenever the allocated physical space is consumed even without explicit \falloc call from the host. }

\comment{ Attribute Integral Dataset for Write (IDW): DSM Word 11 @ NVMe Spec 1.4:
If set to ‘1’, then the dataset should be optimized for write access as an integral unit. The host expects to perform operations on all ranges provided as an {\bf INTEGRAL UNIT for writes}, indicating that if a portion of the dataset is written it is expected that all of the ranges in the dataset are going to be written.  This is an important information and can be used to segregate the ranges of data given by the DSM command into a particular stream so that they are written together in physical Nand.
#### ``https://medium.com/\@saswatidas13/importance-of-dataset-management-command-in-ssd-6dbb77146edd'' April 2019; 
The property of Nand Flash is such that data cannot be written on flash without erasing and erase size (called of block i.e. consists of hundreds of pages) in much larger than write size which is in terms of pages. So if some part of the data in a block becomes invalid or stale and the space occupied by the data needs to be reclaimed, the valid data in the block had to be read from the block and written to a new block. Thereafter the old block may be erased and space occupied by invalid/stale data in the old block can be freed up and used for new data. This process is called garbage collection. Garbage collection is a necessary feature of SSDs but it increases the write amplification in a drive. 
If dataset for which has IDW attribute set to 1, are {\bf written together} in the Nand, they also can be {\bf erased together} thereby reducing write amplification, increasing write throughput and enhancing the lifetime of the drive.
}

\comment{ 
    \item Can reduce the number of flash-level transactions inside SSD controller~\cite{Sprinkler} and thus increase efficiency ?
    \item 나중 data access시 parallelism 극대화 또는 interference 최소화 space allocation 가능?\cite{Sprinkler, PageAllocationsInSSD,DynamicPageAllocationInSSD}
}

\comment{ 
\comment{ \module{On-demand vs. Eager allocation:}  
Note that the {\bf Eager pre-allocation} of physical space in \falloc can improve write latency considerably. Since every incoming writes to \FA instances can be stored, unlike on-demand GC, at its corresponding physical space with no garbage collection, we can expect faster and stable write latency. In particular, the tail latency will be considerably cut~\cite{TailAtScale}.} 
\comment{ \module{The unit of physical space allocation to an FA instance is flash memory block (\eg, 2MB), which is also the unit of erase.} 
For instance, when an \falloc call is made while creating an SSTable in RocksDB, whose size is by default 64MB, SSD/FTL will secure 32 clean flash blocks of each 2MB-long and allocate them to its FA instance. In fact, in most practical use case of \falloc including RocksDB, DWB and F2FS segment where the size of logical objects is at least 2MB, each flash block is dedicated to only one instance and thus each instances has its distinct set of dedicated flash memory blocks. \reminder{what about ever-increasing  size of flash blocks (\eg, 8MB and 16MB)} ++ Assume that {\bf the size of range} should be equal to the minimum erase unit in the flash storage or its multiplicative. (\eg, physical flash block size usually 2MB) so that when the corresponding object deleted ...}
\comment{ {\bf striping for better bandwidth when multiple flash blocks.} 
{\bf Striped Block Allocation Strategies (vs. per-channel)} When an FA instance need two or more flash blocks (\eg, N), there are two strategies to allocate flash memory blocks across multi-channels in flash storages: per-channel and striping. ++ When the size of range >> 2MB; thus multiple blocks will be striped over multiple channel for higher write throughput by exploiting SSD parallelism. + Each flash block will be allocated round-robin manner over the RAID0-like parallel dies (.} 
Now let us discuss in detail about the {\bf unit of physical space allocation and allocation strategies.} and also explain its {\bf flexibility}. extent (when multiple tables reside in one OS file; table당 space alloc은 extent  단위로); Note that the unit of \falloc can be made at finer grain than file: the granularity of object can vary from finer-grain object(if any?), individual applications even to VM. \falloc및 FTL은, irrespective of object types, can flexibly adapt/respond by allocating physical space to FA instance in the unit of physical flash memory block. The unit of physical space allocation to an FA instance is flash memory block, which is also the unit of erase. For instance, when an \falloc call is made while creating an SSTable in RocksDB, whose size is by default 64MB, SSD/FTL will secure 32 clean flash blocks of each 2MB-long and allocate them to its FA instance. In fact, in most practical use case of \falloc including RocksDB, DWB and F2FS segment where the size of logical objects is at least 2MB, each flash block is dedicated to only one instance and thus each instances has its distinct set of dedicated flash memory blocks. \reminder{what about ever-increasing  size of flash blocks (\eg, 8MB and 16MB)} 
\module{<< 2MB (block size) 처리:} 
자연스럽게 static way(column)-wise allocation이라 나중 접근시 blocking 최소화;  
Meanwhile, we do not exclude the case where two or more FA instances can share one physical flash block, when their logical sizes are less than the size of a flash memory block. \reminder{뭔가 보충설명?}  We would like to stress that, even when a flash memory block is shared two or more FA instances, it is always beneficial to cluster pages from one logical object into the same physical flash block and thus to prevent them from scattering over different flash blocks. Length of \falloc (relative to the size of EU) has any relationship with its effectiveness? Although this topic is beyond the scope of this paper, {\bf 동일 block clustering 항상 유리}

\subsection{When and Where/How to \falloc} 

{\bf When: Eager vs. On-demand; Where: A channel-wise block or striped}
{\bf Striped Block Allocation Strategies (vs. per-channel)} When an FA instance need two or more flash blocks (\eg, N), there are two strategies to allocate flash memory blocks across multi-channels in flash storages: per-channel and striping. 
To exploit the abundant parallelisms at channel, way, die, and plane levels for better performance~\cite{Chen-TOS16,SSDDesignTradeoff}, flash storages have aggressively been taking RAID-like striping architecture in storing and retrieving data~\cite{RAID}. The {\bf striped allocation over channels} allocates blocks evenly from P channels. For instance,  given \falloc(LBA, 16MB) and assuming 8 channels and 2MB / block, there are two allocation strategies: 1) 8 blocks from 1 channel and 2) 1 blocks from each of 8 channels.\reminder{장단점 - Taking into account the multi-tenant scenario where random IOs are blocked by sequential reads/writes, channel-wise allocation can reduce their interference - trade-off. But, this topic is beyond the scope of this paper, thus left as future research topic.} This uniformly striped block allocation are 1) 현재 SSD seq RW 동작방식과 유사, 2) 따라서, provide high performance for sequential RW operation against an FA instance. \reminder{언급 필요? What if there is no free block in each channel?} conduct GC in the channel. It is critical to secure a completely clean block. Thus, for this purpose, FTL has to conduct the multi-block-merge scheme~\cite{2R}. Meanwhile, we can take the {\bf per-channel allocation}, where all N flash blocks for an FA instance are obtained from one channel. \reminder{pros: ??, cons.: RW bandwidth drop } 1 < N < P -> partial striping 필요 or block-wise RW 수행 (simple though inefficient); 

아래 existing SSD's page/block allocation scheme과 compatible? this might contradict with the static page allocation, but compatible at least with dynamic page allocation scheme~\cite{DynamicPageAllocationInSSD}. 
\begin{itemize}
\item 1 block case: per-channel, block-wise physical space allocation; read/write  <- compensate performance by thread level parallelism
\item 2 or more block case: striped over P: N blocks are striped to exploit the parallelism to maximize the write and read bandwidth. {\bf physical space contiguity 개념 애매하고 그 보장이 복잡해짐. (next-ptr로만 유지 안됨?)}
N = multiple of P: striped read/write, as existing does (most common?)
\item limitation(?): 위 문제들 때문에, N = P 배수일때 가장 단순; Otherwise (N != P), per-channel block 단위로만 유지? 이 경우, 해당 FA instance에 assigned blocks 정보 어떻게 관리? 
\end{itemize}
Note that the fine-grained page mapping FTL and dynamic page allocation scheme is flexible enough to seamlessly support even per-channel physical space allocation at the unit of flash block. 

Striped EU Unit: logical object size대비, ever-growing flash block size and DoP로 striped EU size(\eg, 180MB)가 효과 반감? 
Ideal case: object size = DoP x flash block size?; Nonetheless, 무조건 \falloc통해 object page들을 동일 EU에 colocate시키는 것이 하상 이익?

\module{Rationale for channel-Wise physical allocation of blocks:}  Due to the complexity For this reason, \falloc takes the channel-wise physical block allocation.eg, 2MB in Cosmos+). Therefore, when writing 2MB, the whole write will be directed to only one channel/die(?) so that its bandwidth will be limited by the chip performance, thus failing to exploiting the parallelism. 
{\bf \falloc-induced trade-off in terms of throughput and latency} What is EU-wise Physical Layout? cf. Write-ordered mapping policy for maximizing write bandwidth by exploiting parallelism. This is a {\bf trade-off between WAF minimization and parallelism maximization}. {\bf Effect of \falloc on the Interference between concurrent IOs} 1) \falloc된 1MB * 8개 동시수행 vs. 8MB single write request: 동일 bandwidth가능? 2) 1MB write 와 random write/read 7개 동시수행시, 1MB \falloc vs. non-\falloc 버전 random read/write 장단점?

{\bf need to mention the trade-off between RAID-like striping and write-order-based mapping policy.} 
\comment{ 
In the case of PC SSD where concurrent IOs from multiple applications are rare and thus the performance of single application does matter, this design may result in sub-optimal read and write performance. In contrast/However, in the case of DC or Cloud storage where multi-tenant applications are concurrently running, the impact of parallelism in multi-tenancy might not be less than that in single-tenant case. 
NOTE: even in the case of PC SSD, when the application calls the fallocate system call in a large unit (that is, degree of parallelism x EU size, e.g. 64MB), we can achieve best result in terms of both WAF and parallelism.} 


\comment{ SYNC vs. ASYNC physical space pre: allocation::
\falloc aims to improve performance by securing dedicated physical blocks prior to writing. Therefore,  Thus, even when flash storage did not or could not secure physical blocks for the \falloc-ed ranges, it is not such problematic except for the performance degradation. \comment{Note that \falloc은 무시해도 상관없는 성능상 이슈}  Thus, it is okay for flash storages to handle the \falloc call as follows: \falloc(offset, length) 통해 논리주소범위만 전달하고, the physical allocation can be made in either of three ways: {\bf sync, async, or on-demand (that is, upon the first write to the address range)}. While sync. alloc.은 \falloc 명령어 처리시 미리 공간 확보, the asynchronous allocation will simply return ack. upon \falloc 시점, and later 시간 될때 but 최초 write 전, 물리공간 확보. \remidner{any further optimization chance with async over sync? if so, what?}  이 두 할당 시점은 구현 난이도, 최적화 기회요인 등에서 장단점을 가진다. 또한, depending on the availability of free blocks per channel, either allocation scheme can be SELECTIVELy chosen. In this paper, for the simplicity of discussion, the sync. allocation is used in our implementation. 
In fact, the async allocation will in effect work same as the synchronous allocation for two reasons.  First, app에서 file allocate 시점 t1과 최초 write call 시점 t2사이 t2-t1사이에 alllocate을 conduct할 수 있음. 사실 일이 몰려서 t2-t1 사이에 allocate을 확보하지 못한 겨우에, 이는 기존 post-allocate방식의 GC overhead로 인해 write delay와 동일할 뿐이다. Second, \falloc에 의해 빠른 GC가 가능하기 때문에, 미리 clean block pool을 충분히 확보하고 있기 때문에 많은 경우 바로 바로 assign 가능할 것임.}
 }

\comment{ %
When a set of logical LBAs become invalid (\eg, file deletion), the {\em trim (or discard)} command is used to inform flash storage of XXXX so that their corresponding physical pages are invalidated and no longer relocated upon GC~\cite{trim}. Assuming that the whole logical address range of a file are \falloc-ed and its all pages are, as intended, isolated in dedicated flash blocks, all pages will be discarded together at once upon the file deletion, which is typified by SSTables in RocksDB~\cite{trim2,TrimInRocksDB}. When all valid blocks in a flash block have to be discarded, can we simply erase the block instead of carrying out trim operation for every pages?)?  Those blocks can be simply erased without causing any GC and returned as free block. In addition, note that the discard call need not to be even made. How much time can we save by avoiding discards per pages? 
(((( 위와 결합: when the object is deleted or overwritten, all flash blocks in their entirety can be invalidated and thus simply erased at no copyback, thus causing no write amplification.; One additional benefits of \falloc is that most flash block will be erased in its entirety when deallocated. {\bf \eg, trim overhead when each of 1024 pages constituting a file is scattered over 1024 diff block:} ((From ~\cite{RocksDB-Exp}: While the TRIM command can improve overall performance and flash endurance, updating the address mapping table persistently may also cause disruptive performance issue such as foreground IO latencies, thus RocksDB has to rate-limit file deletions to avoid the trim spike activity~\cite{RocksDB-Exp}.)) 
{\bf In contrast, when all pages are stored  together in the same block, the block can be erased even without calling the trim command for individual pages: perf. gap?} 

Rate-limited file deletions: RocksDB typically interacts with the underlying storage device via a file system. These file systems are flash-SSD-aware; e.g., XFS, with realtime discard, may issue a TRIM command [28] to the SSD whenever a file is deleted. TRIM commands are commonly believed to improve performance and flash endurance [21], as validated by our production experience. However, it may also cause performance issues. TRIM  is more disruptive than we originally thought: in addition to updating the address mapping (most often in the SSD’s internal memory), the SSD firmware also needs to write these changes to FTL’s journal in flash, which in turn may trigger SSD’s internal garbage collection, causing considerable data movement with an attendant negative impact on foreground IO latencies. To avoid TRIM activity spikes and associated increases in I/O latency, we introduced rate limiting for file deletion to prevent multiple files from being deleted simultaneously (which happens after compactions).

MyRocks relies on SSD’s Trim command to reduce SSD’s internal write amplification and improve performance [16]. However, we noticed that performance for some SSDs may temporarily drop after a spike of Trim commands. Compactions may create hundreds of megabytes to gigabytes of SST files.
} 
\section{Implementation}\label{sec:impl}

This section presents the implementation details of the \falloc architecture. 
The dominant portion of the \falloc implementation is made into the FTL firmware of the Cosmos development board by cross-compiling the source code written in C. The \falloc command was prototyped via vendor unique command. A user-level library that implements a protocol for the new commands via the {\em ioctl} system call supports applications and SSDs. This approach not only allows to quickly prototype the concept in a development environment but also to make the prototype portable to most file systems. 
In addition, this section describes the changes we made in file systems and database engines to enable them to run on the \falloc interface. Note that
the changes are, as summarized in Table~\ref{tab:lines-modified}, marginal and moreover local to a few modules. 

\subsection{Changes made in F2FS}\label{sec-fs}
\comment{
\module{Ext4} When an {\tt fallocate()} call is invoked from application, the ext4 file is modified to invoke \falloc to the flash storage right after allocating the logical space. For this, just 24 lines were added to the {\tt extent} module, which obtains the logical address range for the {\tt fallocate()} call and then to call \falloc. Since the trim command is triggered on file deletion in Ext4, the \falloc-ed blocks for the file will be erased in their entirety. 
}
While initializing a file system volume, F2FS divides the whole volume into fixed-size segments. The segment is a basic unit of management in F2FS: it allocates storage pages in the unit of segments and also performs “cleaning” in the unit of segment.
F2FS maintains multiple active segments (by default, six segments). 
To stream writes by active segments inside flash storage, 
F2FS is modified to obtain, whenever a segment becomes active, its logical address range  using the super block data structure (\ie, f2fs\_sb\_info) and segment number information 
and then to call \falloc with the range as parameter. For this, only 26 lines were added to the {\tt segment-allocation} module. \comment{segment.c file} Like in Ext4, the \falloc-ed block for a segment will be erased in their entireties when the trim command is automatically called for the segment upon its cleansing.


\subsection{Changes made in {\large Applications}}\label{sec-appl}


\comment{
\module{RocksDB} Even without any change in its code, RocksDB can benefit from \falloc. As explained in Section~\ref{sec:bg}, RocksDB manages key-value documents using SSTables. In particular, after creating each SSTable file, RocksDB secures its logical address space (whose size is by default 64MB) in advance by calling {\tt fallocate()} for the file. In addition, our modified Ext4 file system will automatically call \falloc after allocating the logical space for the given {\tt fallocate()} call. For this reason, every SSTable's data will be streamed to its dedicated flash block(s). Flash blocks dedicated to each SSTable remain full of valid pages until the SSTable is later compacted and deleted. When an SSTable is deleted, all its pages will be trimmed and thus invalidated altogether at once and accordingly all flash blocks dedicated for the SSTable can be simply erased, thus causing no write amplification. Hence, as shown in Section~\ref{sec:eval}, RocksDB can achieve near ideal WAF (\ie, 1) transparently without any change in its codebase. \comment{\eg, \falloc with 64MB as in RocksDB SSTables, then FTL will allocate 32 flash blocks of 2MB each.} \comment{2. ZERO padding?; or, need to introduce a writing rule?: once \falloc-ed, a block should be filled to avoid the partial full \falloc-ed blocks.}
}
\module{RocksDB} 
As explained in Section~\ref{sec:bg}, RocksDB manages key-value documents using SSTables. In particular, after creating each SSTable file, RocksDB secures its logical address space (whose size is by default 64MB) in advance by calling {\tt fallocate()} for the file.
Once allocating the logical space for the given {\tt fallocate()} call, RocksDB engine calls \falloc.
For this reason, every SSTable's data will be streamed to its dedicated flash block(s). Flash blocks dedicated to each SSTable remain full of valid pages until the SSTable is later compacted and deleted. When an SSTable is deleted, all its pages will be trimmed and thus invalidated altogether at once and accordingly all flash blocks dedicated for the SSTable can be simply erased, thus causing no write amplification. Hence, as shown in Section~\ref{sec:eval}, RocksDB can achieve near ideal WAF (\ie, 1) transparently with minimal changes in its codebase.
\comment{without any change in its codebase.}

\module{MySQL/InnoDB} In order to isolate DWB pages from normal ones into different flash blocks and thus to reduce write amplification~\cite{soyee:edbt18}, we modified the InnoDB engine to call \falloc with the logical address range of {\em DWB} as parameter before writing to the journal area for the first time. To obtain the address range was used the {\tt FS\_IOC\_FIEMAP} ioctl call. 
Also, whenever DWB is cyclically reused, the trim call is made for the journal area so as to invalidate all the old pages and and thus to make the old \falloc-ed block to be erased in its entirety. As shown in Table~\ref{tab:lines-modified}, the changes made in InnoDB engine were minimal - less than 100 lines of code change were made at two modules of {\tt double-write-buffer} and {\tt file}.\comment{ buf0dblwr.cc and fil0fil.cc files}
\comment{Note that the use case of \falloc for DWB slightly differ from that for RocksDB. While an SSTable is deleted after compaction, the DWB area will be {\bf circularly overwritten.} Therefore, the \falloc call need to be made with the {\tt CYCLIC} flag set so that physical flash blocks are continuously dedicated whenever the \falloc-ed flash blocks are consumed by cyclic DWB overwrites. We assume that MySQL will ask to destruct the FA instance for DWB upon database shotdown or destruction.} 

\begin{table}[t]
\centering
\caption{Lines modified across applications to use \falloc}
\label{tab:lines-modified}
{


\begin{tabular}{clcc}
\toprule
\multicolumn{2}{c}{Applications} & Lines Added  & Lines Removed \\
\midrule
RocksDB & (v6.10)        & 72    & -      \\
MySQL/InnoDB & (v5.7)    & 74    & 16      \\  
F2FS & (v5.4.20)         & 26    & -      \\  
FTL (cosmos+) & (v3.0.0) & 1683  & 193    \\  
\bottomrule


\end{tabular}\\
}
\end{table}

\subsection{Changes made in FTL}\label{sec-ftl}


We have prototyped \falloc on the OpenSSD Cosmos board~\cite{Cosmos2020} by extending its firmware. The Cosmos board is an SSD development platform that is made publicly available by the OpenSSD Project to promote research and education. \comment{on the recent flash-based solid state drive technology~\cite{OpenSSD}.}  The board employs the HYU Tiger 4 controller based on Dual-Core ARM Cortex-A9 on top of Xilinx Zynq-7000 board, and 16GB MLC Nand flash memory. Thus it has the same performance characteristics as a commercial SSD equipped with the Barefoot controller. The Cosmos board adopts a page mapping scheme for flash memory management, as in most contemporary SSD products. The board is connected to a host system through the NVMe interface.
Main technical issues encountered while embodying \falloc on the board are summarized below. Note that the existing FTL can support \falloc with moderate changes in its codebase, as shown in Table~\ref{tab:lines-modified}. 

\module{FA Instances} For each \falloc command, a corresponding FA instance is created in Cosmos board's DRAM, which contains its logical address range, the list of flash blocks dedicated to the instance, and the current write pointer. The memory requirement per instance is, while slightly varying depending on address range and the number of flash blocks, just several tens bytes. Thus, considering that the number of active FA instances is in practice limited (\eg, less than 100), small amount of DRAM (\ie, several tens KB) will suffice to maintain active FA instances.

\module{GC and Block Type} The existing GC firmware in the Cosmos board was extended to support the {\em GC-By-Block-Type} policy. 
Also, since FTL need distinguish two types of blocks, normal and \falloc-ed, one bit flag, {\tt FA-BLK}, was added to the {\tt block\_header} struct in the Cosmos firmware. \comment{{\tt VIRTUAL\_BLOCK\_ENTRY} memory alignment padding 영역 1bit.}  A block's {\tt FA-BLK} flag will be set on dedicating the block to an FA instance and later reset when the block is erased and returned as free.


\comment{The number of necessary blocks will depend on the total size of the given logical address ranges.}

\module{Probing the matching FA instance} For a write request, FTL should be able to quickly probe the matching FA instance using the given LBA address. If the probing fails (\ie, no matching instance exists), the request is not for active FA instance thus will be written to non-FA instance. To determine whether the given write is for active FA instance or not, a flag bit was added to every entry in page-mapping table. The flag bit of every relevant logical page is set when an \falloc command is invoked and later reset when the page is overwritten or discarded. The next issue is, when the flag is turned on, how to probe the matching instance. While there should be alternative implementations such as hardware-acceleration and pipelining, rather a simple approach was taken for fast prototyping. That is, while scanning each of all active FA instances, we check whether its logical address range contains the start\_LBA in the given write request. Once a matching instance is found, the write will be appended at the physical space pointed by the {\tt next\_write\_ptr} of the instance. 


\comment{
\item overwrite 언급 필요?: fallocate를 부여한 LBA 영역에 대해  overwrite가 발생한 경우 현재 구현 버전에서는 한번만 처리하도록 처리되었음. overwrite를 허용하는 방식은 다음과 같음.  1. fallocate 블록으로 지정되면, fallocate 호출시 시작 LBA를 기록해둠. (fallocate에 쓰여지는 LBA 범위는 연속적이고,  현재 구현상 fallocate 크기를 2MB로 고정했기 때문에 fallocate 시작 LBA만 알고 있으면 범위를 알 수 있음.)  2. (overwrite가 발생한 LBA가 포함된) fallocate  block에 있는 페이지가 모두 쓰여진 경우, 보수적으로 해당 블록에 지정된 LBA mapping table에 fallocate flag를 invalidate (unset)시킴.
\item \falloc 호출 크기가 고정되지 않는 경우, 주어진 write size를 확인해서  코스모스 보드 physical block 크기에 맞춰서 On-demand 방식으로 FA instance를 생성한다. (FA instance size가 physical block 크기와 같은 경우)  FA instance utilization을 최대한 활용할 수 있도록 한다. 또한, MySQL DWB처럼 동일 영역에 "반복적"으로 "고정된 (write) 크기"로 \falloc을 호출하는 경우, pre-allocated FA instance를 활용한다.
}

\comment{
\reminder{(jhpark): LSA-VSA mapping table에 sallocate marking attribute 추가하였음. sallocated write는 해당 mapping table attribute 조회함. SSD내 DRAM utilization 문제 언급 필요?
현재 구현은 vssa 변수를 lsa-vsa mapping 테이블에 추가하였음. (구현의 단순성+편리성+디버깅 편의성 때문)
DRAM usage 크게 손해 보고 있음. 하지만 실제 디자인은 이렇게 바뀔 수 있음.
sallocate flag 1bit한개 필요+기존 vsa 변수를 활용하는 방법 있음. 
vssa (sallocted write에 대해 어느 블록에 쓰여야할지 알려주는 주소)만 지정되면 
salloced write에 대해 지정된 블록으로 보낼 수 있음. 
일단 블록에 지정되서 쓰여지면 더 이상 sallocate 정보는 필요 없음.  

++ overwrite인경우? 
sallocate flag == 0 && vsa == VSA_NONE : initialized (초기 상태 또는 erased된 상태)
sallocate flag == 1 && vsa != VSA_NONE : sallocated write (어디 블록에 쓰여질지만 알고 있음 vssa값만 배정됨)
sallocate flag == 0 && vsa != VSA_NONE : uninitialized (normal write, salloc write)
}
}

\comment{구현관련 misc. issues 
VM/Container/Disagregation서 지원가능? 
\module{\falloc in VM/Cloud:} NVMeOF기반 VM (Trim is recently supported in VM) 

Zeroed block semantic?: fallocate() allocates and initializes to zero the disk space within the range specified by offset and len --> \falloc시 read 연산 semantic은? erased flash block의 읽기는 zero return? zero-initialization semantic auto support?
}
\comment{ Because \falloc is not about data consistency, system crashes will not induce any \falloc-related data inconsistency. \reminder{Can crash make some instances to be invalidated?} Nevertheless, upon crashes, some recovery actions should be taken to restore xxx and also to reclaim the physical space occupied by {\em those invalidated} instances. How to recover active \falloc instances upon recovery? Need to develop the recover semantics according to individual application's fallocate?: 1) DWB: 그냥 persistent하게 사용?. 2) F2FS?, 3) RocksDB: 성능손해감수하고 simply ignore?; Do we need to devise specific solution for each use case or does a general solution exist? +++ how the RECOVERY module will handle active FA instances at failure time?
}

\section{Performance Evaluation}~\label{sec:eval}
In this section, we present performance evaluation carried out 
to analyze the impact of \falloc on key-value store, log-structured file system, relational database, and multi-tenancy. 

\subsection{Experimental Setup}

All experiments were conducted on a Linux platform with 5.4.20 Kernel running on an Intel Core i7-6700 CPU 3.40GHz processor with two sockets of 
four cores and 50GB DRAM.
The host machine has two storage devices,
16GB Cosmos OpenSSD and 256GB Samsung 850 Pro SSD. 
The Cosmos OpenSSD employs a controller based on Dual Core ARM Cortex-A9 
on top of Xilinx Zynq-7000 board with 256KB SRAM, 1GB DDR3DRAM, and 16GB MLC Nand flash memory ~\cite{cosmosboard}.
The Cosmos OpenSSD was used as the main storage device for the experimental data and 
connected to host using PCIe interface. The over-provisioning area in the board was set to 10\% (\ie, 1.6GB) for all experiments. The Samsung 850 Pro SSD was used as the log device when MySQL/InnoDB was run. 

\subsection{Workloads}

To demonstrate the benefit of \falloc, we used a synthetic workload and two realistic workloads, db\_bench and TPC-C. The {\tt fio} tool was used to generate a synthetic workload. And, to evaluate the effect of \falloc on key-value stores, we ran the db\_bench benchmark using RocksDB on ext4 file system. The same db\_bench workload was run also using F2FS~\cite{F2FS} to test the impact of \falloc on log-structured file system. In addition, the TPC-C benchmark was used to measure the effect of separating DWB in MySQL/InnoDB into dedicated flash blocks. Finally, to highlight the benefit of \falloc in multi-tenancy, we ran db\_bench using RocksDB and TPC-C using MySQL concurrently on Ext4 file system. In all experiments, the direct I/O option ({\tt O\_DIRECT}) was enabled to minimize the interference from file system's page caching and the {\tt TRIM} option was turned on for both file system. To compare the performance impact of \falloc, we ran those workloads using the vanilla databases and file systems on the Cosmos board running the original FTL and also ran them using the modified versions with the board supporting \falloc.
The three workloads used in the experiments are summarized below.

\begin{description}
\item[FIO]  The Flexible I/O (FIO) benchmark is commonly used to test the performance of file and storage systems~\cite{FIO}. It spawns a number of threads or processes doing a particular type of I/O operations as specified by the user parameters. 

\item[db\_bench] RocksDB provides db\_bench as the default benchmark program, consisting of several configurable workloads~\cite{dbbench}. 
The {\tt fillrandom} workload was used to evaluate the impact of \falloc, which writes key-value pairs in random key order.\comment{in async mode} Each key-value entry was sized to have a key of 16 bytes and a value of 100 bytes. 
The workload was run against empty database till the Cosmos board became full.

\item[TPC-C] 
The {\tt tpcc-mysql} tool~\cite{tpccmysql} was used for TPC-C benchmarking~\cite{tpcc-paper}. 
The benchmark was run using 32 clients against initial database of 80 warehouses until no space was left in the storage device. 
\end{description}

\begin{figure*}[!ht]
\centering
\subfigure[EXT4 (4 RocksDB Tenants)]{
\label{fig:ext4}\includegraphics[clip,width=0.235\textwidth,height=2.6cm]{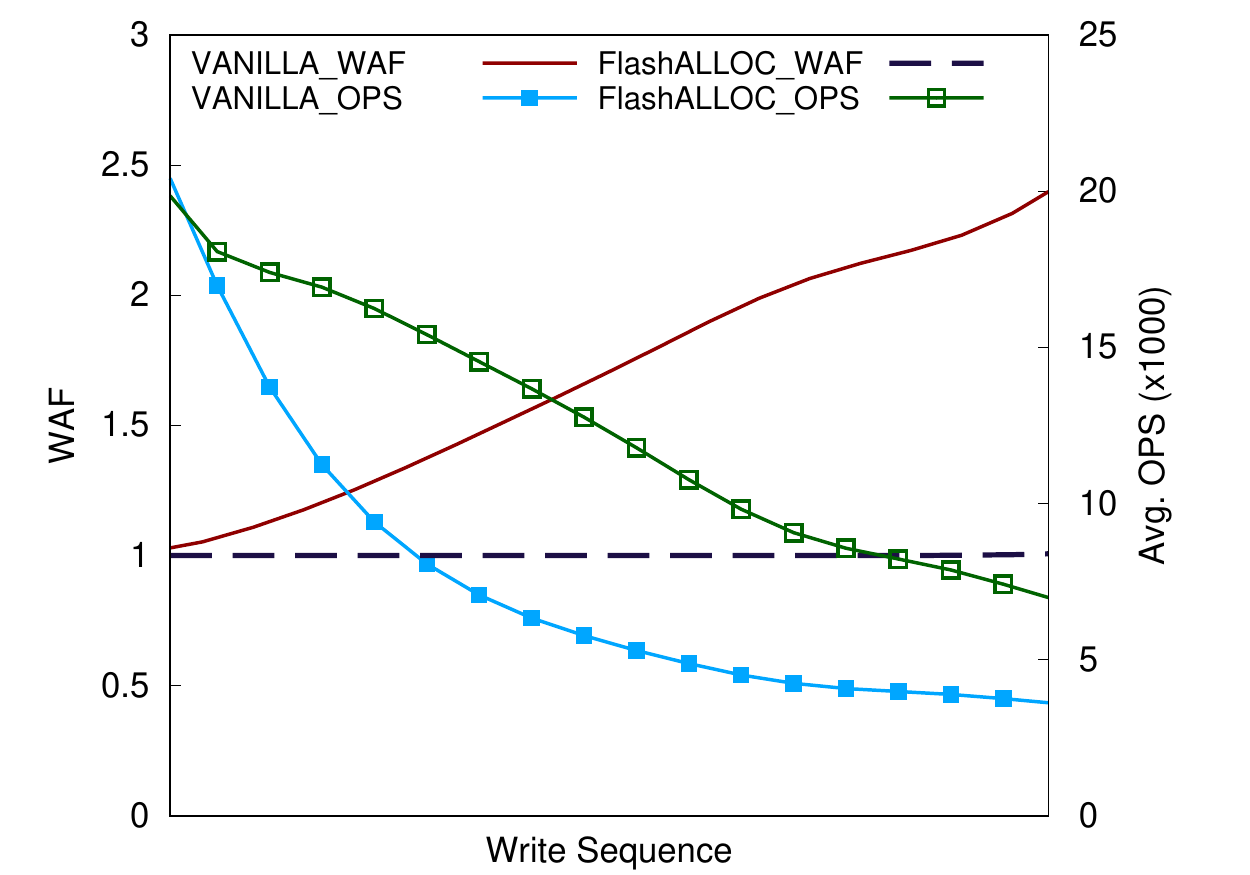}
}
\subfigure[F2FS (4 RocksDB Tenants)]{
\label{fig:f2fs}\includegraphics[clip,width=0.235\textwidth,height=2.6cm]{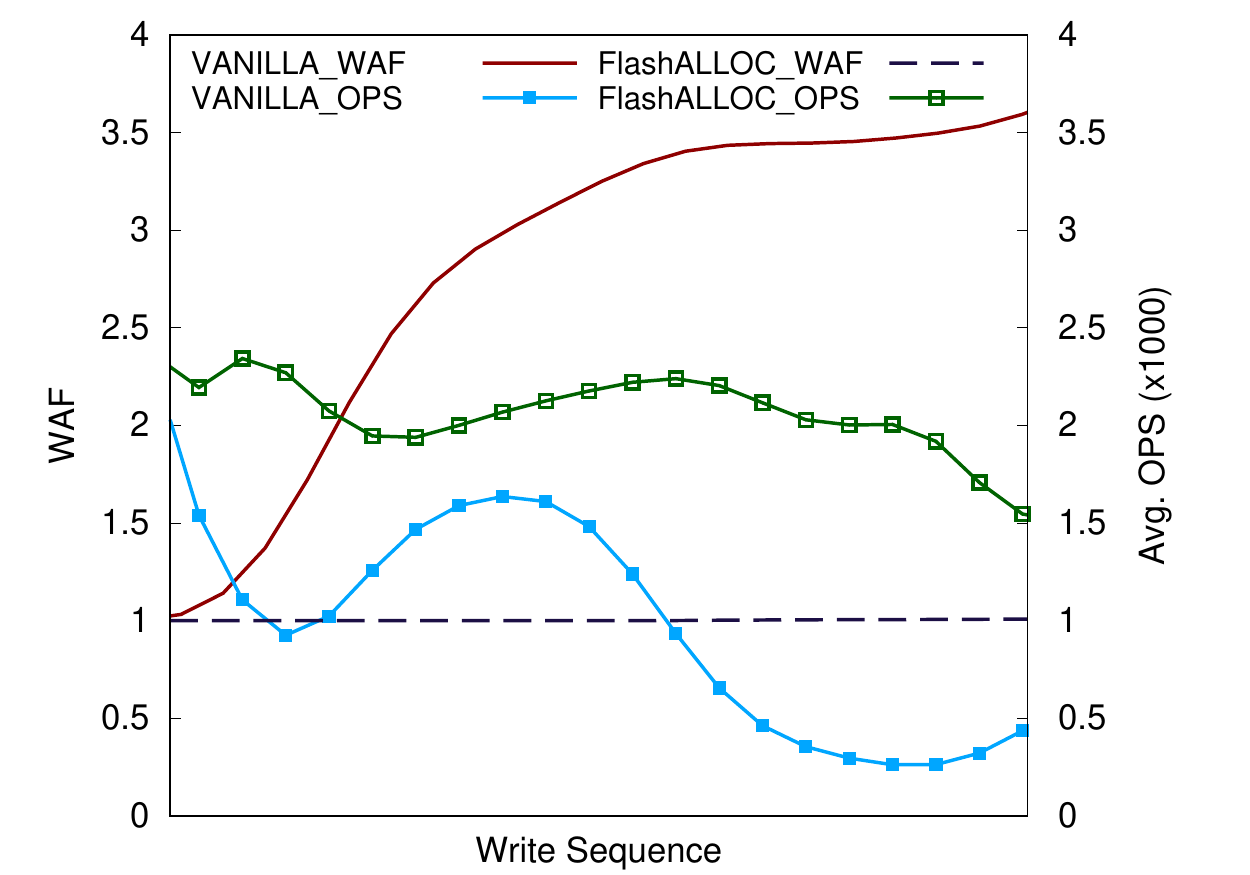}
}
\subfigure[MySQL (TPC-C) ]{
\label{fig:dwb}\includegraphics[clip,width=0.235\textwidth,height=2.6cm]{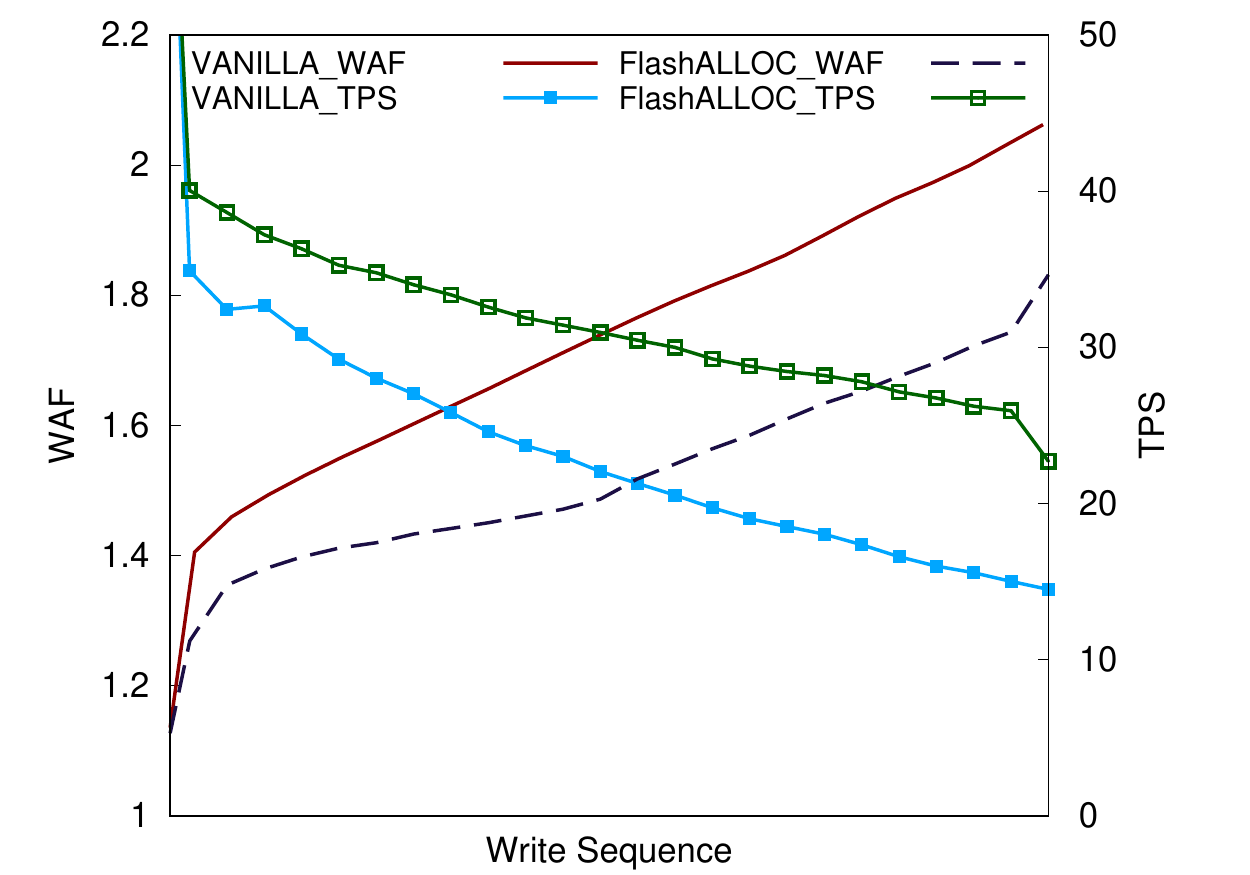}
}
\subfigure[Multi-tenant DB (TPC-C + RocksDB)]{
\label{fig:multi-tenant}\includegraphics[clip,width=0.235\textwidth,height=2.6cm]{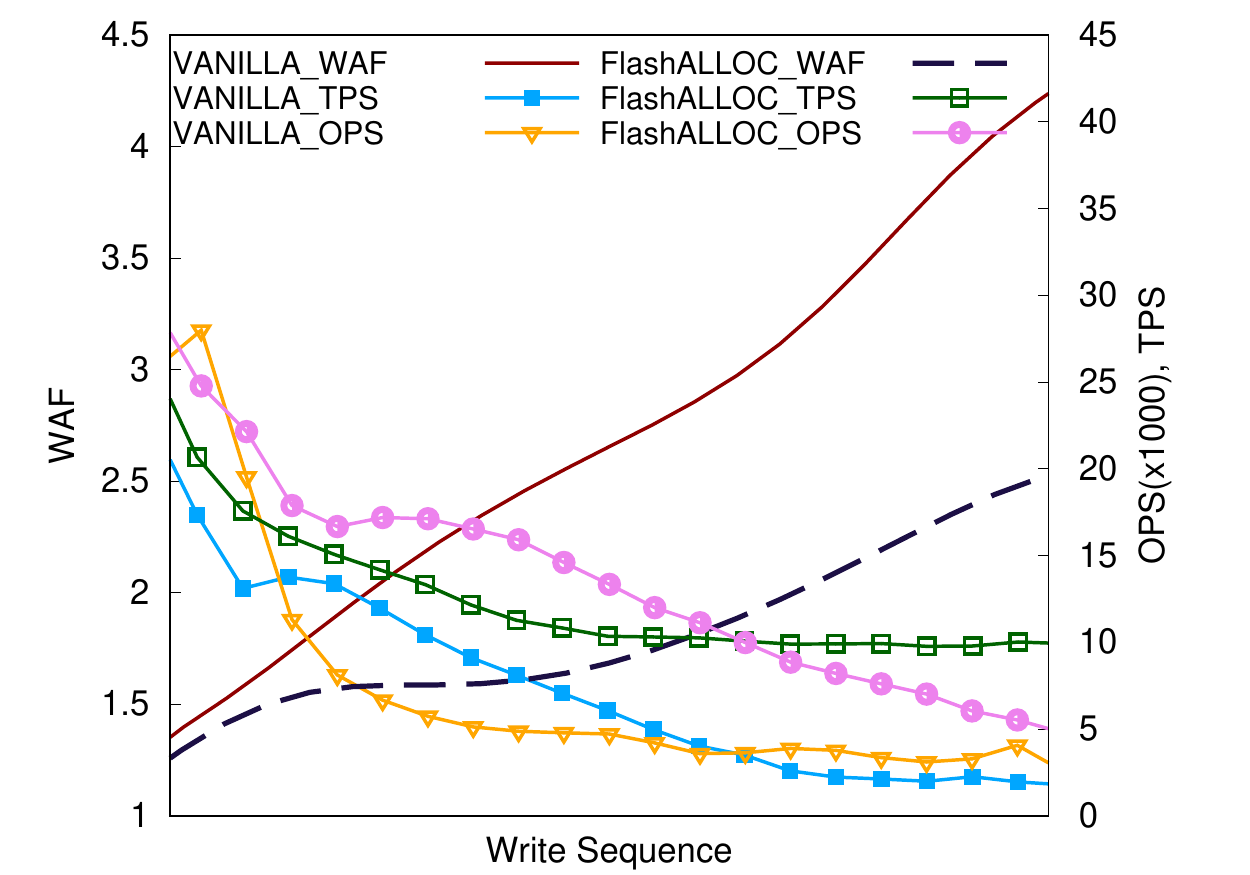}
}
\caption{Write Amplifications and Throughput in Realistic Workloads}
\label{fig:falloc-expr}
\end{figure*}

\subsection{Performance Analysis}

Let us briefly review the overall performance benefit of \falloc using Figure~\ref{fig:falloc-expr}. While running four experiments using vanilla and \falloc-ed configurations, we measured the throughput of each benchmark program and the running WAF at the Cosmos device every minute till no space is left in the Cosmos device, and present the results in Figure~\ref{fig:falloc-expr}. In the figure, the X-axis represents the time \comment{sequence of page writes in each benchmark} and the left and right Y-axis does the running WAF and the throughput of each benchmark, respectively.  
\comment{ We measured the KOPS(kilo operation per seconds) of the RocksDB  and TPS (transaction per seconds) of the MySQL instance on the left Y-axis.}
As shown in Figure ~\ref{fig:falloc-expr}, \falloc-ed version outperforms the vanilla one 
considerably in terms of throughput as well as WAF consistently across all four experiments. 
The running WAF gaps between two versions are ever-growing over time in all experiments. 
That is, as the Cosmos board is filled with data, the effect of de-multiplexing different objects into different blocks in \falloc becomes outstanding. In particular, the running WAF in \falloc-ed version remain close to 1 even at the ends of RocksDB and F2FS experiments. 


\begin{figure}[h]
\centering
\subfigure[8 Threads]{
\label{fig:fio-4thds}\includegraphics[clip,width=0.225\textwidth,height=2.6cm]{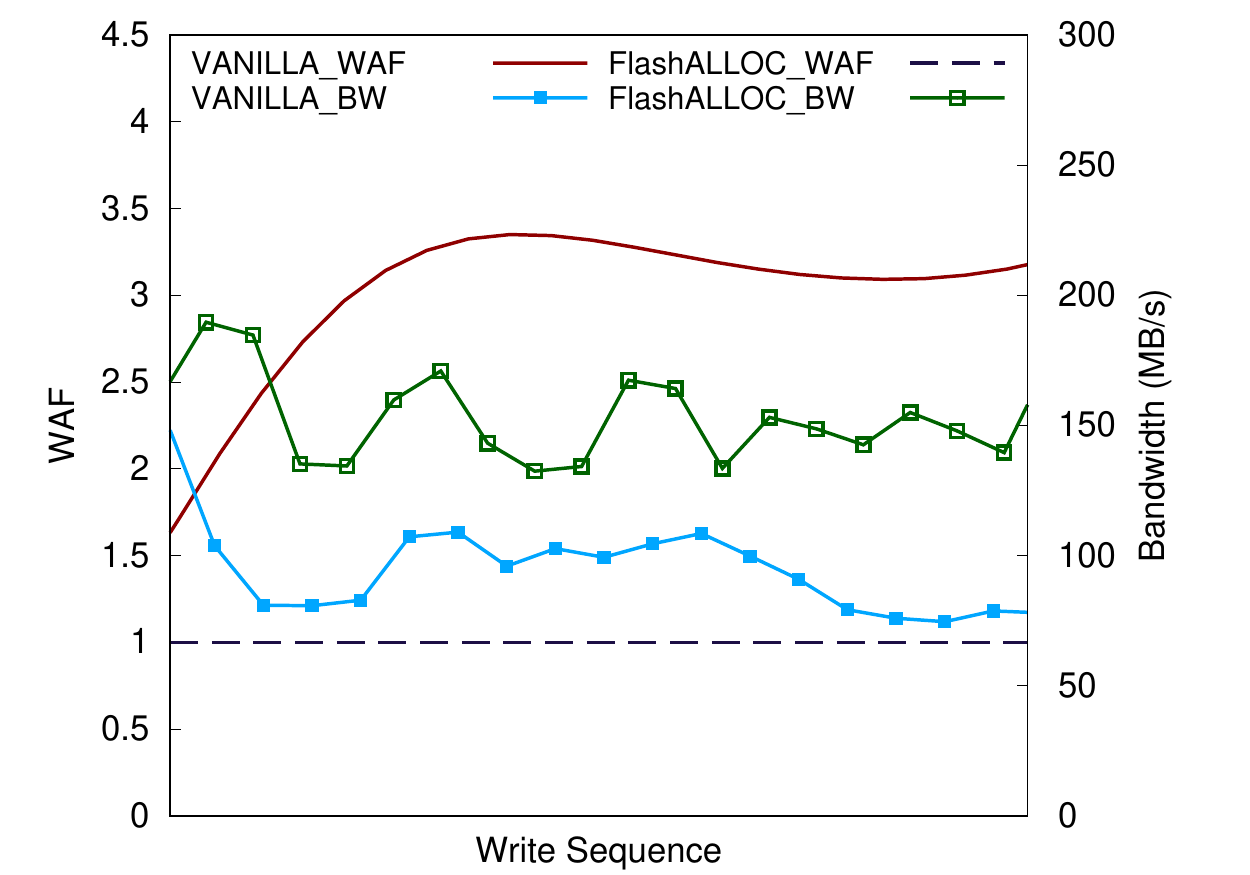}
}
\subfigure[32 Threads]{
\label{fig:fio-32thds}\includegraphics[clip,width=0.225\textwidth,height=2.6cm]{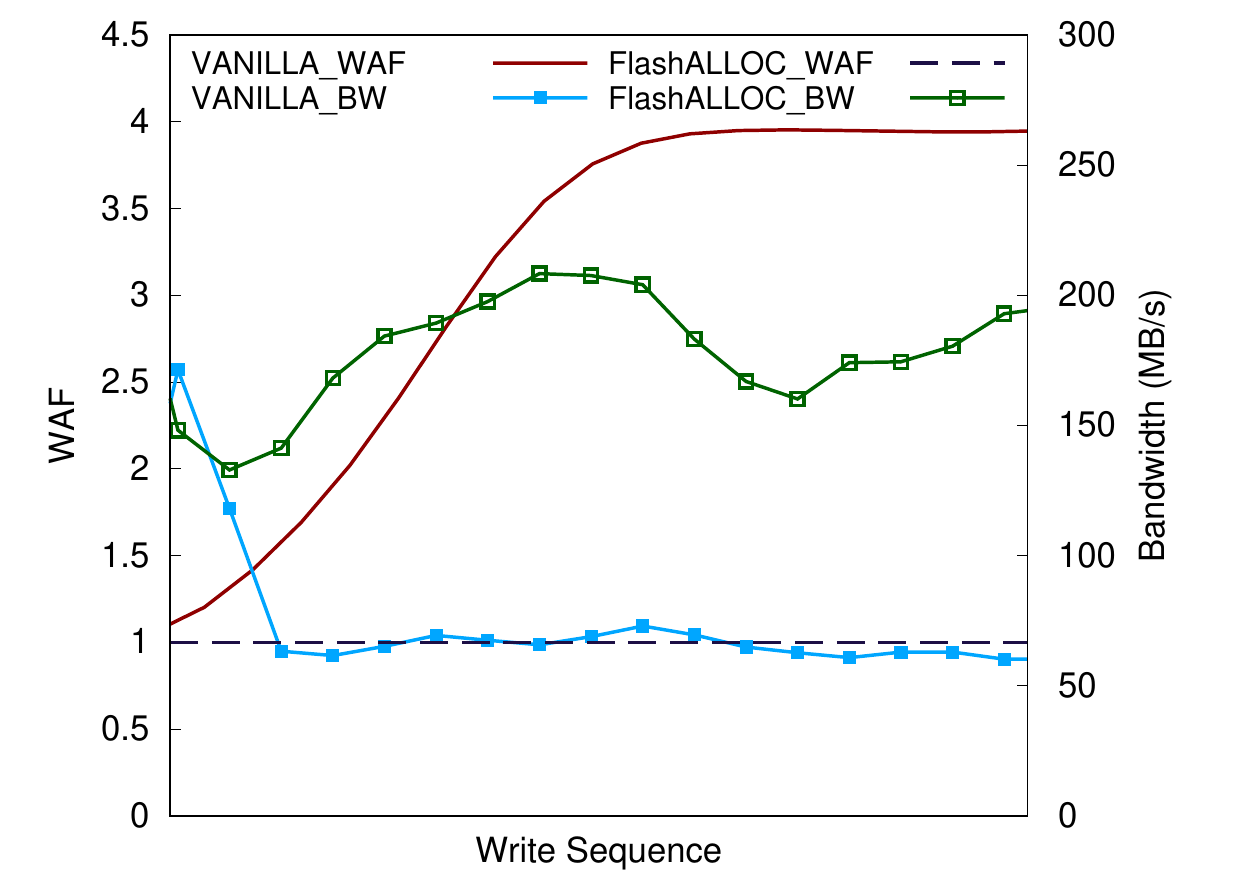}
}
\caption{WAF and Bandwidth in FIO}
\label{fig:fio-workload}
\end{figure}

\module{Synthetic FIO Workload} Before explaining the effect of \falloc on realistic workloads in Figure~\ref{fig:falloc-expr}, let us show the benefit of \falloc using a synthetic write workload. For this, using the {\tt fio} tool, we created eight 2GB files on Linux and Cosmos board and ran eight threads, each of which performs random overwrites in the unit of 2MB against its dedicated 2GB file. The same experiments were conducted in two modes, vanilla and \falloc-ed.  In the \falloc-ed mode, before invoking each 2MB overwrite, \falloc was called a prior so as to secure a dedicated flash block to store new data.  While running each experiment during one hour, we measured the device WAF and the write bandwidth and plotted the result in Figure~\ref{fig:fio-workload}. \falloc has reduced the device WAF from 3.1 to 1 and has doubled the write bandwidth (\ie, approximately from 75MB to 150MB). 
In addition, to further highlight the effect of \falloc when the multiplexing degree is increased, we carried out another experiment by increasing the number of concurrent write threads to 32 in {\tt fio} tool and thus decreasing the per-thread file size to 512MB, and present the result in Figure~\ref{fig:fio-workload}. 
Note that, under more concurrent write threads, a flash block in the Cosmos board will be multiplexed by more files with more deviating lifetimes. As clearly shown in the figure, \falloc has reduced the device WAF from 4 to 1 and thus has tripled the write bandwidth (\ie, roughly from 60MB to 180MB). The considerable gain in WAF and write bandwidth was direct reflection of reductions in the garbage collection overhead. 
\comment{ 
We can make two general observations about the results in Figure ~\ref{fig:fio-workload}. First, \falloc can achieve zero copybacks and shows consistent running WAF close to 1.0, as ideal case, when the workload is controlled so that all writes have the same life time.
Since the physical flash block size of the Cosmos+ board is 2MB, every single write requests from FIO threads are gathered via \falloc command and occurs page invalidation around the same time (\ie they have similar death time).
\falloc can obtain the free block without any copybacks 
when the garbage collection is triggered.
Second, \falloc achieves better performance bandwidth of the storage. Thanks to the \falloc as a true realization of zero copybacks,  the running WAF keeps 1.0 during the whole time. On the other hand, vanilla FTL increases the running WAF upto 2.5. Due to the huge performance difference between \falloc and vanilla FTL in running WAF, performance gap also take place in device bandwidth. Again, \falloc has no performance degradation as a result of heavy garbage collections \falloc outperforms vanilla FTL by more than two folds in terms of the write bandwidth.
}


\comment{
\falloc의 핵심 아이디어는 비슷한 life time을 갖는 페이지들을 동일한 블록에
grouping 시킴으로써 쓰기 증폭을 줄이는 것이다 (as addressed in Section 3.1?).
이러한 \falloc의 장점을 극명하게 보여주기 위해, 
우리는 FIO를 활용하여 synthetic workload를 수행하였다.
일단 \falloc명령어가 호출되면 반드시 write명령어보다 선행되기 때문에 
각 thread에서 쓰기 요청한 페이지들은 같은 erase unit (\ie block) 으로 모아진다.
각 thread 별로 지정된 영역 안에서만 overwrite가 발생하기 때문에 
그 결과, 자연스럽게 각 블록에는 life time이 비슷한 페이지들로 구성된다.
We would like to make two general observations about the results Figure ~\ref{fig:fio-workload}.
First, \flloc zero copyback을 달성하고, shows consistent running WAF close to 1.0.
ideal case로써, 모든 write가 동일한 life time을 갖도록 통제된 워크로드인 경우.
cosmos+ 보드의 물리적 블록 크기는 2MB이기 때문에 
FIO thread가 요청한 write는 반드시 \falloc을 통해 한 블록으로 grouping 되고, 
비슷한 시점에 page invalidation이 발생한다 (\ie  비슷한 death time 을 가진다).
따라서, copyback 없이 GC 과정에서 free block을 얻을 수 있다. 
Second, \falloc provides higher bandwidth of the storage.
zero copybacks을 실현하는 \falloc 덕분에 running WAF가 1로 유지되는 반면, 
vanilla FTL의 경우 Running WAF 가 2.5 까지 증가한다.
\falloc과 vanilla scheme 사이의 WAF 차이는 디바이스 bandwidth에서도 반영된다.
Again, \falloc의 경우 write amplification이 유발하는 성능저하가 없기 떄문에 
\falloc outperforms vanilla up to 2.x time in terms of the write bandwidth.
}


\module{RocksDB on EXT4}\label{sec:ext4}
To analyze the effect of \falloc on RocksDB engine, we ran 4 RocksDB instances concurrently on top of two configurations of ext4 file system and the Cosmos board, vanilla and \falloc-ed. Each RocksDB instance was run with the {\tt fillrandom} workload in the db\_bench benchmark. In order to minimize the interference from flushing the WAL log, the log file was stored in a separate storage device.
\comment{WAL log file: normal vs. falloc 성능 비교?} While running both experiments, we measured device-level WAF and average OPS of four RocksDB instances over time and present the result in Figure~\ref{fig:ext4}. In the case of vanilla mode, as SSTables at different levels are simultaneously created and populated by multiple compaction threads in RocksDB, they are multiplexed into the same flash blocks. Recall that SSTables at different levels will be compacted and thus deleted at different points of time. For this reason, in the vanilla version, device-level WAF increased steadily while RocksDB's OPS decreased inverse-proportionally, which is consistent with the result in Figure~\ref{fig:dbbench-on-micron}. In contrast, in the case of \falloc-ed mode, WAF at the Cosmos board, as expected, remained nearly one even till the end of the experiment and accordingly the average RocksDB's OPS improved by 1.5x, compared to the vanilla version. This clearly illustrates that \falloc can drastically reduce write amplification in RocksDB by de-multiplexing SSTables into different flash blocks. 

Nonetheless, write amplification was not completely removed in \falloc-ed mode. The running WAF was 1.1 at the end of the experiment. The residual physical write amplification is attributable to metadata files. RocksDB maintains several metadata files (\eg, MANIFEST and CURRENT) to keep track of database state changes whose write patterns are not log-structured instead random writes. Though the sizes of those files are relatively small, they contribute non-marginal fraction of total writes \comment{(\ie, approximately \reminder{one tenth} of all writes in our experiment)} from RocksDB. Those random writes go to the non-FA instance which is managed by the conventional Greed FTL and hence incur write amplifications.  

Note that the OPS gain by \falloc over vanilla is relatively smaller than the WAF gain. Recalling that OPS is determined by logical write amplification at the RocksDB level as well as device-level write amplifcation at the device  level, the physical WAF reduction is offset by the same logical write amplification in both modes. 



\comment{
\falloc이 optimal WAF를 가지는 이유는 다음과 같다.
RocksDB 엔진에서 SSTfile에 대한 space를 미리 확보하기 위해 fallocate() 함수를 호출한다. 
일단 fallocate()아 호출되면 \falloc이 invoke되고, 동일한 SST file 을 구성하는 페이지들은
동일한 flash block으로 모여진다.
compaction이야기 추가!

(레벨 capacity(?) threshold 도달하면; 생략), compaction수행 결과 
이때 SST파일들이 삭제되고, 이때 동일한 블록에 있는 페이지들은 
비슷한 시점에 invalidate 된다. 
따라서, cosmos+ OpenSSD에서 free block을 얻기 위해 GC가 발생할 때
victim block선택시 \falloc-ed된 블록에 있는 모든 페이지들은 invalidated 되어 있기 때문에
copyback이 발생하지 않는다.
(SST 파일을 제외하고) metadata 관리를 위해 RocksDB가 manipulate 하는 다른 파일들 (\eg LOG, MANIFEST, CURRENT, and etc)
이로부터 발생한 random write는 
SST 파일 data write 에 비해, 1)한정된 logical space를 가지며,
2) \falloc으로 인해 separating SST file data, 자연스럽게(?) meatadata영역끼리 모아지는 net effect 발생함.
따라서 vanilla FTL 에 비해 copyback 횟수가 significantly 줄어듬.

Bandwidth 성능: 초반에는 \falloc 명령어로 한 채널에 모아서 쓰기 때문에 
vanilla FTL 보다 성능이 약간 낮지만,
free block을 얻기 위한 GC가 trigger되는 시점부터 성능이 역전된다.
\falloc outperforms vanilla FTL up to XX.
}

\comment{ 
\begin{figure}[h]
\centering
\includegraphics[clip, width=0.35\textwidth]{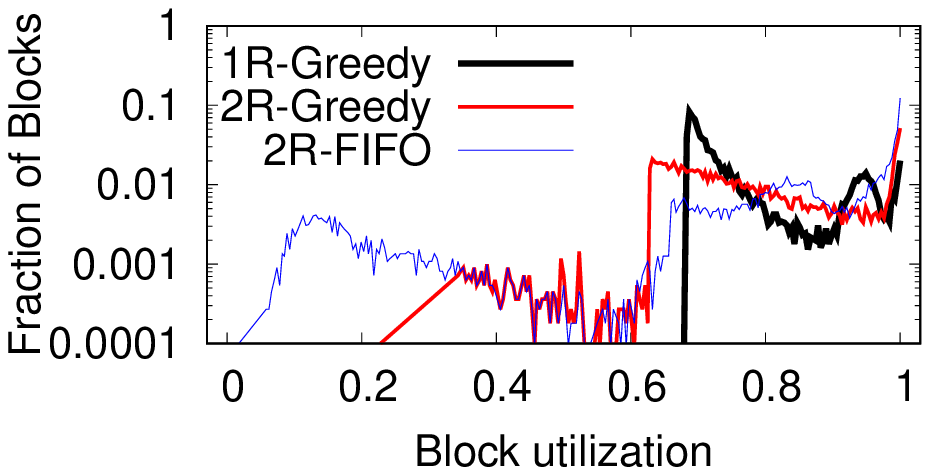}
	\caption{Bimodal Block Utilization (RocksDB) - from 2R)} 
\label{fig:blkutil}
\end{figure} 

\module{Bimodality in Block Utilizations} As discussed in Section~\ref{sec:design}, \falloc will show the bimodal distribution of block utilizations. To verify the superiority of \falloc over the vanilla FTL in bimodality, we measured the utilization of all flash blocks at the three-fourths point while running the YCSB trace for both firmwares, and plotted the result in Figure~\ref{fig:blkutil}. Note that the y-axis is log-scale. The non-\falloc-ed version shows no bimodality because it has a single block type: most block tends to linger just above the cleaning point~\cite{LFS}. \falloc show clear bimodality, which can in turn explain the WAF gap between them. \comment{Compared to the vanilla version, \falloc will choose normal blocks with quite lower utilization as victim: the average utilization of all normal victim blocks in \trf and in \trg was 0.16 and 0.41, respectively.} 
}

\comment{
\subsubsection{Adaptive Space Allocation} 
As discussed in Section~\ref{sec:operations}, the size of \falloc-ed and normal regions will adapt to the host usage. That is, \falloc 호출한 논리 객체 데이터들은 자신이 현재 사용중인 영역을 dedicated flash block을 점유하고 있고, 나머지 flash blocks은 normal writes들에 의해 사용된다. 
\comment{from 2R paper: In the TPC-C trace, as the database grows over time, the number of cold and frozen pages also increases. Therefore, the cold region has to accordingly expand to accommodate more cold pages.  2R schemes are novel in that they allocate space adaptively to flash regions, thus requiring no tuning effort.}  

To verify the GC algorithm above will adaptively allocate blocks to \falloc-ed and normal region, we measured the {\bf distribution of \falloc-ed and normal block over time}  
\comment{fraction of cold blocks over all blocks} 
while running the multi-tenant workloads used in  Figure~\ref{fig:multi-tenant} and 
present the result in Figure~\ref{fig:adaptive-space}.
\reminder{
multi-tenant실험에서 시간대 별 (WAF 계측 시점) die별 \falloc non-\falloc block비율 계측!
salloc block flag 하는 순간 전역변수++방식 동작확인
}

\reminder{From 2R: 수정: Note that Y-axis in the figure ranges up to 110\% because the OP factor\comment{percentage} is set to 10\%. While both schemes allocate more blocks to cold region over time, the fraction of cold region in \trf is less than that in \trg about by 0.2 across all ranges greater than 0.1 in X-axis. This result indicates that \trf is better than \trg in terms of the space utilization of cold region. \trf can utilize the cold region better because it prevents false cold pages from entering the cold region, and also because it takes the eager cold block merge. With larger normal region, non-cold pages in normal blocks will have more temporal chances to be invalidated before the normal blocks become victims. This is the reason why the average utilization of normal victim blocks in \trf was lower than that in \trf.}

\begin{figure}[!h]
\centering
\includegraphics[clip, width=0.31\textwidth,height=3cm]{figure/sallocate/distribution.eps}
	\caption{Adaptive Space Allocation (RocksDB and TPC-C)} 
\label{fig:adaptive-space}
\end{figure} 
}

\module{RocksDBs on F2FS}
To evaluate the effect of \falloc on log-structured file systems, we ran four RocksDB instances each with the same {\tt fillrandom} workload used in subsection ~\ref{sec:ext4} on top of two configurations of F2FS and Cosmos boards, vanilla and \falloc-ed. While 
concurrent log writes from active segments are multiplexed into the same flash block in the vanilla configuration, writes from each segment is perfectly isolated into its dedicated flash block in the \falloc-ed version. Thus, the \falloc-ed version can nearly remove write amplification due to the write multiplexing in the vanilla version; the WAF at the final phase was reduced from 3.5 to 1.1, as shown in Figure ~\ref{fig:f2fs}. The residual WAF of 0.1 in \falloc-ed version is we guess contributed by random writes for hot metadata in F2FS~\cite{F2FS}. Accordingly, \falloc-ed version outperforms the vanilla version about by three folds in terms of the db\_bench's OPS at the end of experiment. 

The result from F2FS experiment indicates that \falloc can be a fundamental solution to the {\em log-on-log} problem~\cite{DONOTSTACK, AMF} by allowing to perfectly align logical segments in higher F2FS with physical blocks in lower flash storage. Namely, \falloc enables F2FS and RocksDB to achieve an ideal WAF of 1 and thus to realize the full potential of their flash-friendly log-structured write patterns. 



\comment{ subsection ~\ref{sec:ext4}에서 동일한 조건에서 실험하였고 파일 시스템을 EXT4가 아닌, F2FS을 사용하였다. \falloc 덕분에, F2FS 파일 시스템의 경우 동일한 segment에 속하는  데이터 블록들은 physical flash block에 grouping + separated 된다.

\reminder{from 교수님 작성 부분}
By dedicating separate physical flash blocks to each segment,  F2FS can now isolate log writes from different logical segments into different flash blocks (that is, de-multiplexing) segments, instead they will be dedicated to a physical flash block of 2MB. The WAF of F2FS in Figure 6(c) will be reduced from  3.5(?) to 1.1. Who contribute to up to 0.1 (running) WAF (random writes for hot metadata [20]) 이러한 random write는 앞서 말한 것 처럼 \falloc의 net effect가 한정된 physical flahs block에 reside하기 때문에 copyback 유발을 완화시킨다. 

Thanks to the delayed discard: in F2FS [20] which discards obsolete spaces in the unit of segments only when checkpointing is triggered, Note that all pages in each segment’s block(s) will remain valid until the segment is compacted and will be invalidated altogether at once upon compaction completion. Therefore, likewise in the case of RocksDB. Thus, segments will not cause no further write amplification. This indicates that \falloc can effectively and completely resolve the LOG-ON-LOG problem [30, 34]  by PERFECTLY ALIGNING logical log segment in lower F2FS with a flash block.  Keypoint: \reminder{해당 논리 실현?} As shown in two RocksDB results on ext4 and F2FS, due to log-on-log’s physical WAF, F2FS underperforms ext4, but FlashAlloc can make F2FS truly flash-friendly (faithful to its original design goals) and thus outperform EXT4! We expect FlashAlloc-ed F2FS to be competitive for more workloads. }

\module{DWB in MySQL/InnoDB} To evaluate the effect of separating the DWB object with cyclic and sequential writes from the main database with random writes, we measured the throughput and the device-level WAF while running the TPC-C benchmark using the vanilla and \falloc-ed MySQL/InnoDB engines, and present the result in Figure~\ref{fig:dwb}. 
Recall that half of writes goes to the FA instance for DWB in the case of \falloc-ed version while the other half (that is, random writes against original database) does to the non-FA instance. Hence, the write amplification induced by the random writes in non-FA instance is inevitable even in the \falloc-ed mode.  Though, as shown in Figure~\ref{fig:dwb}, \falloc can reduce the additional write amplifications by one third (\ie, 1.2 to 0.8) and thus improve the throughput by 50\%. We believe that the benefit of \falloc on DWB will hold also on other ubiquitous journal objects including WAL files in RocksDB, SQLite and relational databases and {\tt jbd2} in Ext4 file system.

\comment{
\falloc과 normal write의 interference가 있는 경우(서로 공존하는 경우?) 
성능을 탐구하기 위해, 
한정된 영역에 대해 cyclic write가 발생하는 DWB write에 대해서만 \falloc을 적용하도록 
수정한 MySQL InnoDB engine에서 TPC-C workload를 수행하였다.
\reminder{참고: multi-tenancy에서는 동일한 application이 아닌 서로 다른 tenant에서 발생하는 것을 좀 더 부각시켜서 설명할 예정}

Unlike, RocksDB SST file 또는 F2FS segment 분리하는 것과는 다르게, 
주요 데이터 쓰기가 \falloc instance를 선 확보 후 쓰기가 발생해서
대부분 블록들이 비슷한 life time을 갖기 때문에 copyback발생이 적은 반면
DWB는 normal write도 계속 발생하고,
normal write와 \falloc-ed write 사이에서 free block occupying 경쟁이 발생함.
따라서, 데이터가 차면서 발생하는 copyback 증가를 피할 수 없다.

major performance gap between \falloc and vanilla 주된 이유: 
DWB의 very high write-density를 가짐 (즉, 지정된 영역 약 2MB)에 대한 write가 발생한다.
MySQL DWB where data are cyclicly overwritten and the write amount in each area의 경우,
\falloc 반복 호출 없이, MySQL intance가 시작할 때 기 지정된 logical space (즉, dwb file로 지정된 LBAs 영역들)에 속하는
data page들을 physical flash block grouping 할 수 있다.
}
\comment{
The performance gain by \falloc-ing DWB will hold also for SQLite WAL and ext4 journal area since DWB-like journal where data are cyclicly overwritten and the write amount in each area is significant: cycle reuse and high write-density (very hot) use cases, traditional WAL log (with some alignment?), Undo tablespace in Oracle, and Ext4 journal area. Such journal areas with fixed address는 booting시 한번 만 \falloc 후 계속 사용가능. Not frequent call to \falloc!
}

\module{Multi-Tenancy} 
As discussed in Section~\ref{sec:bg}, when run together on the same SSD, multi-tenants can interfere each other in terms of write amplification since objects from different tenants with more distant lifetimes are multiplexed in the same flash blocks. 
To demonstrate the benefit of \falloc in mitigating the WAF interference in multi-tenancy, 
we ran two databases concurrently on Ext4 file system and the Cosmos board, one RocksDB instance (used in Figure~\ref{fig:ext4}) and one MySQL instance (used in Figure~\ref{fig:dwb}), in vanilla and \falloc-ed modes, respectively, and present the results in Figure~\ref{fig:multi-tenant}. In the case of vanilla version, the WAF in multi-tenancy is much worse than that in either single tenant (\ie, Figure~\ref{fig:ext4} and Figure~\ref{fig:dwb}, which is consistent with the result in Figure~\ref{fig:multi-tenant-on-micron} obtained from commercial SSDs.  
In the case of \falloc-ed version, the WAF in multi-tenancy remains lower than that in either single tenant. As a result, both benchmarks' throughputs in \falloc-ed mode are considerably higher than in vanilla mode. 
The result in Figure~\ref{fig:multi-tenant} indicates that \falloc is not only beneficial to the calling tenant itself but also {\em altruistic} to neighbor tenants and is thus effective in isolating the performance between tenants~\cite{SeattleReport}.  

\comment{
\reminder{
jhpark(TODO): micon F2FS 추가, micron multi-tenant 추가
1. \falloc updaterandom read latency 관점 성능 이득 보여줄 수 있도록
2. 구현 완성도 높여서 YCSB 동작 (best)
3. \falloc을 위한 최소조건(?)을 궁금해 할 것 같음; 
예를 들어 실험장에서 보여준 pure한 state가 아닌, dirty state (seq + rnd write 여러번 반복 수행한) ssd에 적용했을 때 
\falloc이 제대로 동작하는가? 
}
}
\comment{

\begin{itemize}
\item (++) Effect of \falloc on WAL: WAL이 쓰기 양을 꽤 차지, 다른 SSTable과 섞여 WAF 피해(like DWB?) 
\end{itemize}

\begin{figure}[h]
\centering
    \includegraphics[clip,width=0.46\textwidth]{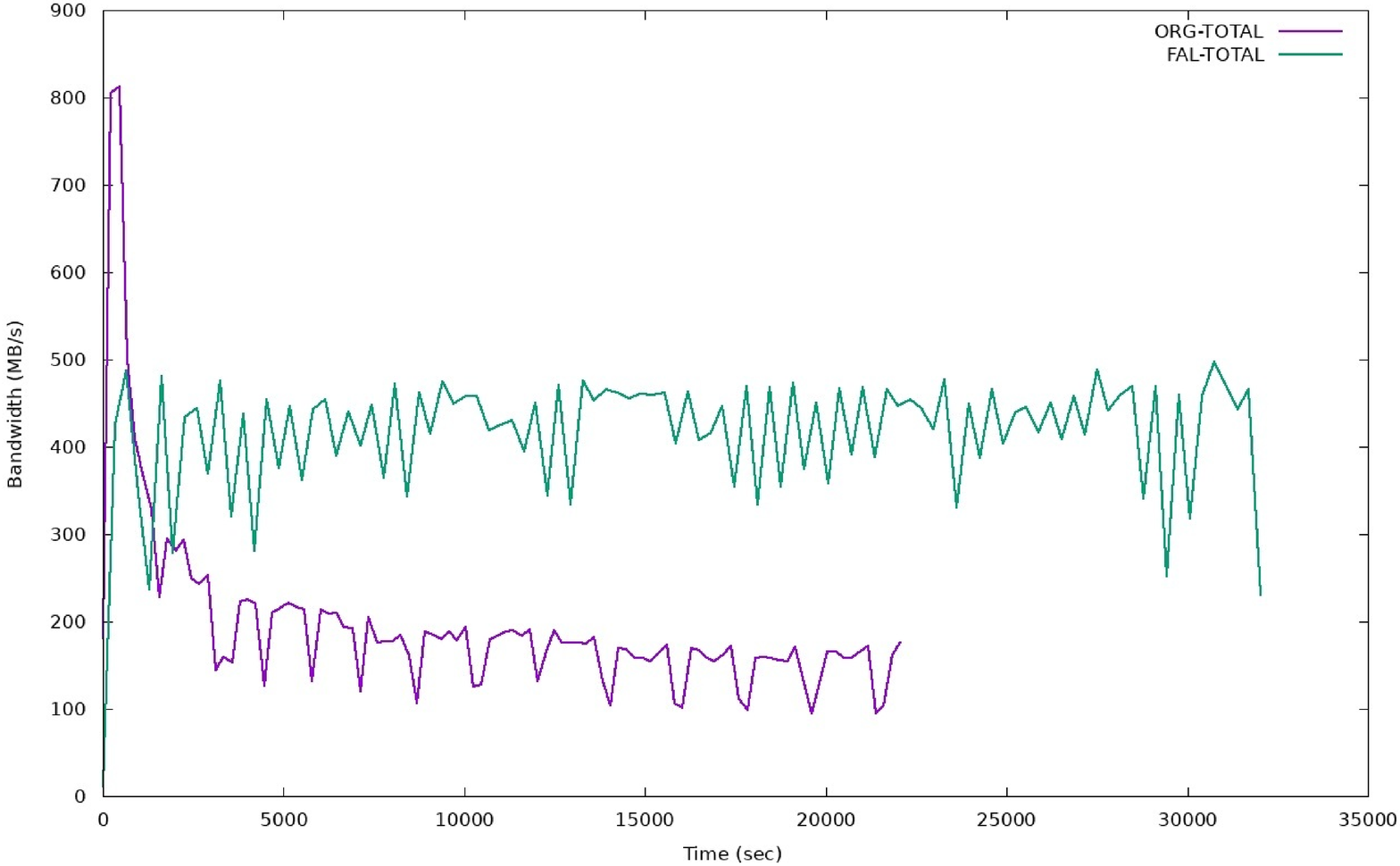}
    \label{fig:rocksdb-original-ext4}
    \caption{RocksDBs on Ext4: ORG vs. \falloc (BW-only; OPS 2x)}
\end{figure}

\begin{table}[h]
\begin{tabular}{|c||c|c|} \hline  
            & Block Erase \# & Page Trim \# \\ \hline \hline
Original    &               &               \\ \hline  
\falloc      &               &               \\ \hline  
\end{tabular}
\centering
\caption{Effect of \falloc on Trim}
\label{tab:trim}
\end{table}

\module{Trim} 
As discussed in Section~\ref{sec:design}, \falloc... ,thus removing the trim overhead.  
To XXX YYY quantitatively, while conducting the experiment in Figure XX, we measured the number of total trim operations in original and \falloc, respectively, and present the result in Table~\ref{tab:trim}.
}

\comment{
\subsection{F2FS; 이하 교수님 작성 부분}

본 실험에서는 alignment 없이 한 성능 평가; 예를 들어, RocksDB SSTable을 2MB (flash block size)에 align했을 때 성능 차이?

실험 종류: 
\begin{itemize}
    \item A synthetic workload, FileBench~\cite{FileBench}
    \item RocksDB on Ext4: 1 instances, n instances: with large n, salloc will have more effect.
    \item multiple concurrent container 환경에서 fallocate, write mix 효과, WAF 값 계산. --> 예를 들어, 동시에 rocksdb container를 4개 동시에 돌릴 경우
    \item F2FS Usecase
    \item DWB and SQLite WAL
\end{itemize}

To evaluate the effect of \falloc on a real system, we have prototyped \trf on the OpenSSD Cosmos board~\cite{Cosmos} and compared it with \org, the default FTL in the board, by running TPC-C, LinkBench, and YCSB benchmarks on each of two FTLs. All the experiments were conducted on a Linux platform with the 4.15 kernel running on an Intel Core i7-4770 3.4GHz processor with 16GB DRAM. The size of over-provisioning area on the board was set to \reminder{10\%: 종혁 확인요망! - 10\% 맞습니다.} for all experiments carried out.

+++++  By dedicating separate physical flash blocks to each segment, F2FS can now isolate log writes from different logical segments into different flash blocks(that is, de-multiplexing)  segments, instead they will be dedicated to a physical flash block of 2MB. 

The WAF of F2FS in Figure~\ref{fig:f2fs} will be reduced from 5(?) to 1.2(?). Who contribute to 0.2 WAF (random writes for hot metadata~\cite{F2FS}) 

Thanks to the {\bf delayed discard:} in F2FS~\cite{F2FS} which discards obsolete spaces in the unit of segments only when checkpointing is triggered, 
Note that all pages in each segment's block(s) will remain valid until the segment is compacted and will be invalidated altogether at once upon compaction completion. Therefore, likewise in the case of RocksDB. Thus, segments will not cause no further write amplification. 

This indicates that \falloc can effectively and completely resolve the {\bf LOG-ON-LOG} problem~\cite{DONOTSTACK,AMF} by PERFECTLY ALIGNING logical log segment in lower F2FS with a flash block. {\bf Keypoint:} As shown in two RocksDB results on ext4 and F2FS, due to {\bf log-on-log}'s physical WAF, F2FS underperforms ext4, but \falloc can make F2FS truly flash-friendly (faithful to its original design goals) and thus outperform EXT4! We expect \falloc-ed F2FS to be competitive for more workloads.

\comment{
\begin{figure}[!]
\centering
    \includegraphics[clip,width=0.46\textwidth]{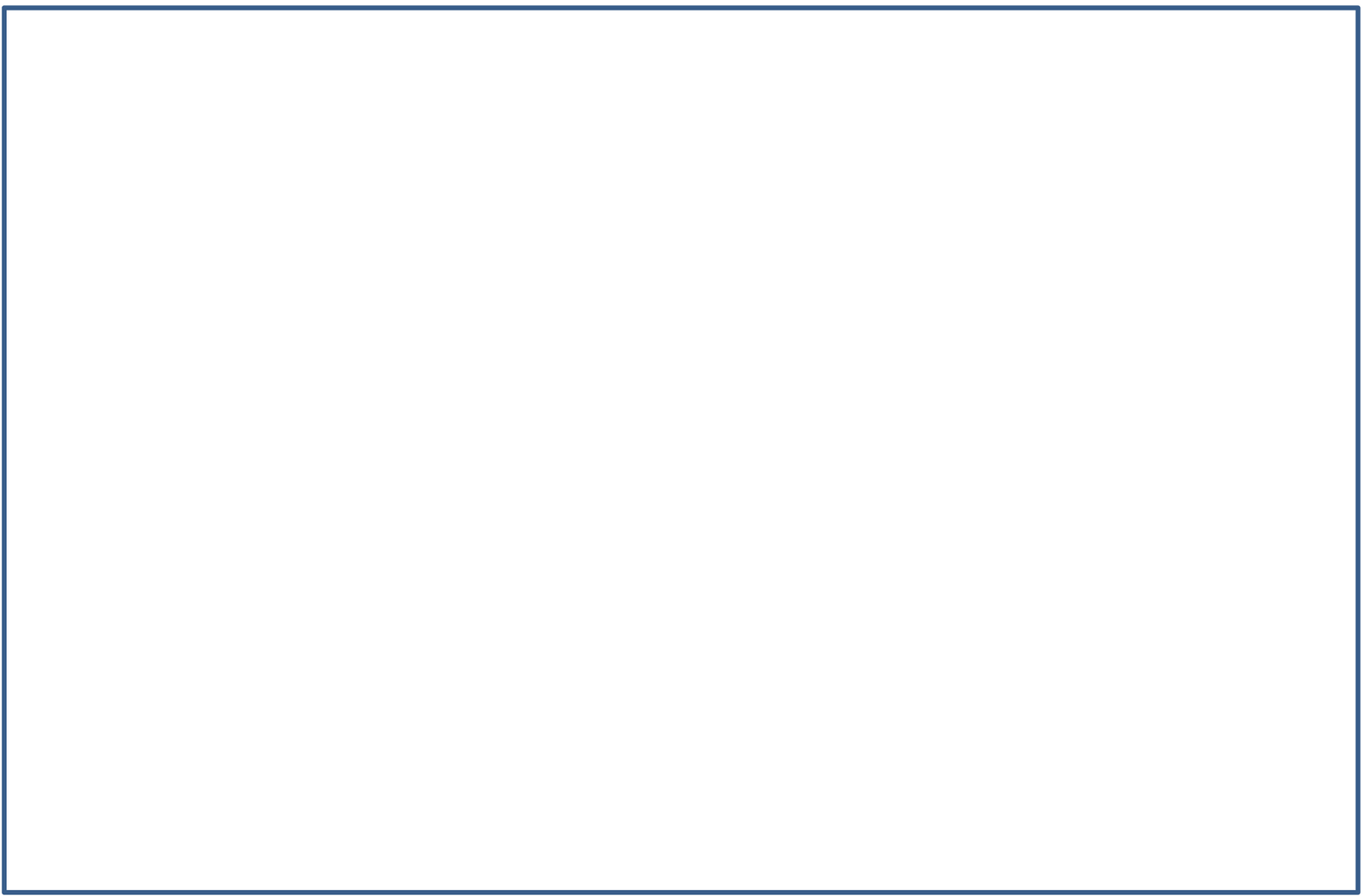}
    \caption{RocksDBs on F2FS: ORG vs. Salloc}
    \label{fig:rocksdb-original-f2fs}
\end{figure}
}

\comment{
\subsection{Synthetic FileBench}

Filebench is a flexible framework for file system benchmarking ~\cite{FileBench}. Filebench can generate a variety of workloads which is specified using Workload Model Language (WML). 파일 벤치에는 몇몇 predefined workload 들이 존재한다. 그 중 varmail workload 는 mail server 을 emulation 한 워크로드이다. 예를 들어 사용자가 이메일을 수신하면 파일이 {\it create}되고 내용이 {\it write} 되고 {\it sync} 된게 된다. 혹은 사용자가 이메일을 읽으면 파일을 {\it open}하고 {\it read} 
한후 읽었음을 표시한다. 이러한 일련의 과정들이 정의되어 있으며 파일을 delete, write append 를 포함하여 크게 4가지 동작이 정의되어있다. varmail workload 에서 write operation의 특징은 기존의 mail 내용을 overwrite 하지 않고 append 한다. 이에 logically 하게는 WML 파일에 정의된 IO chunk size 크기의 sequential write 들이 내려올 것으로 예상된다. 

\reminder{WAF 그래프 추가 예정}

Filebench 의 varmail workload를 수행하면서 정의된 IO chunck size의 IO command가 SSD까지 잘 내려와서 WAF가 낮게 유지되는지 확인하기 위해서 간단한 실험을 수행하였다. 256GB의 Micron SSD를 활용했으며 100MB size file 2000개, 즉 total 200GB, 를 생성하였다. User thread 개수는 32개 append size는 2M로 설정하였다. 약 9시간 실험을 진행하여 total about 1TB 만큼 logical write call을 내렸다.
Filebench의 워크로드 특성상 2M 단위의 write call을 내리기 떄문에 physically 2M 단위의 sequential write가 관찰될 것으로 예상되었으나 blktrace 확인 결과 대채로 672K, 1.34K, 1.28K 의 2M 보다 작은 크기의 physical IO chunck 들이 관찰되었으며 Running WAF가 약 4.5 로 매우 높게 관찰되었다. 

\begin{itemize}
\item Micron SSD (model: CT250MX500SSD1) 256GB, 
\item Initial DB config.: 2000 files of 100MB each, total  200GB; 
\item  32 IO(?) threads; append unit = 2MB, read IO unit = 1MB, DRAM memory size: 10MB;
\item The total amount of written data was 1.3TB while the benchmark was run for nine hours.  what is the total capacity of all files at the end of benchmark?
Running WAF measured in every five minutes; 
\item file 삭제가 uniform random? or skewed?; The average unit of writes measured at the block trace layer (blktrce) = 2MB? or splitted-thus-smaller?
\item Figure~\ref{fig:filebench-original-ext4} WAF 결과 분석: 
The size of physical flash blocks used in the SSD is guessed to be 2MB. But, since each append write is striped over 8-channels and the files are randomly deleted, each flash block is multiplexed with pages from different files and with distant death times. Thus, high WAF.
\item When the same benchmark is conducted against \falloc-ed SSD, WAF graph? 
\end{itemize}

\begin{figure}[!]
\centering
    \includegraphics[clip,width=0.46\textwidth]{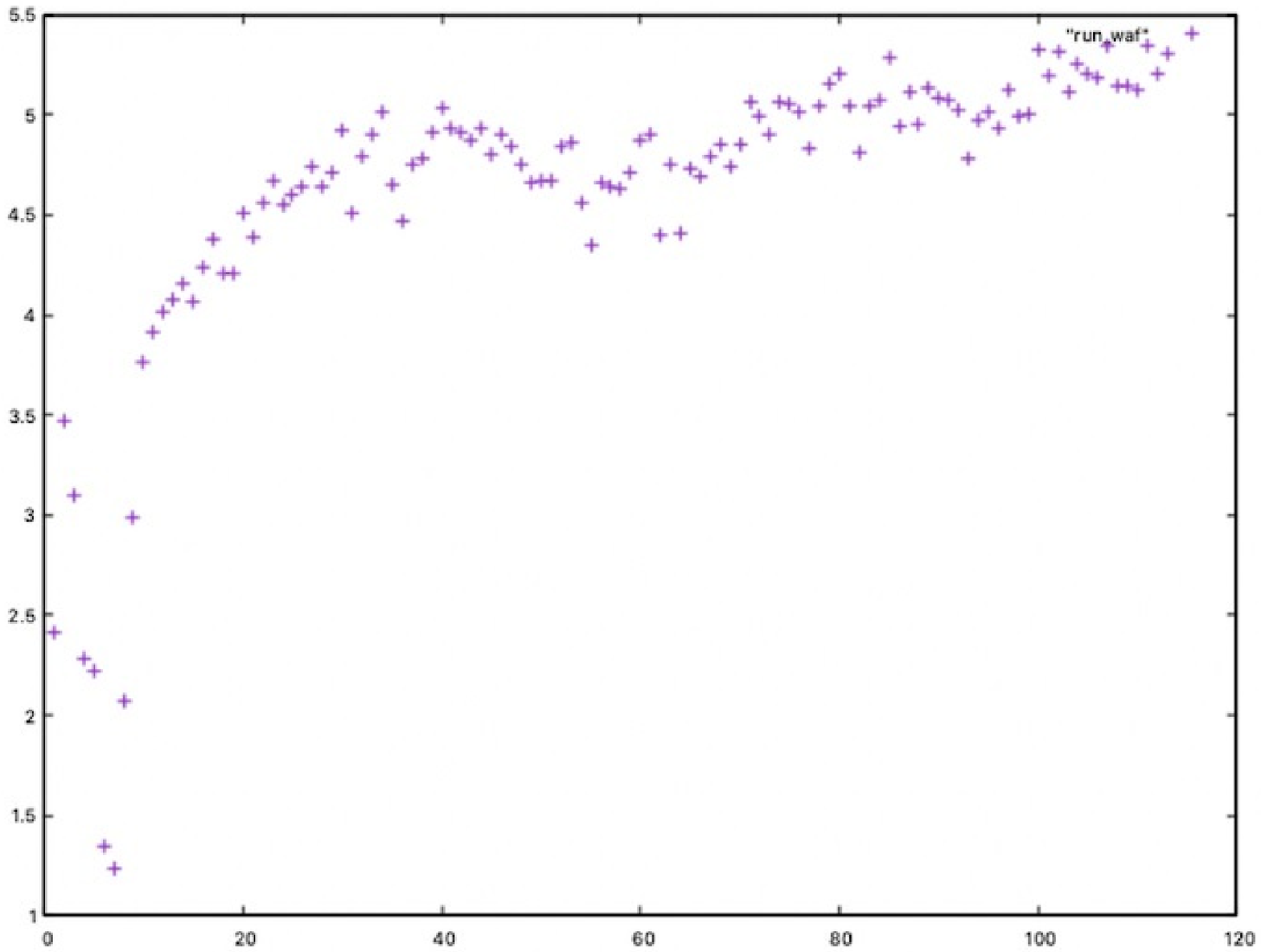}
    \label{fig:filebench-original-ext4}
    \caption{File Bench: ORG vs. \falloc}
\end{figure}
\begin{figure}[!]
\centering
    \includegraphics[clip,width=0.46\textwidth]{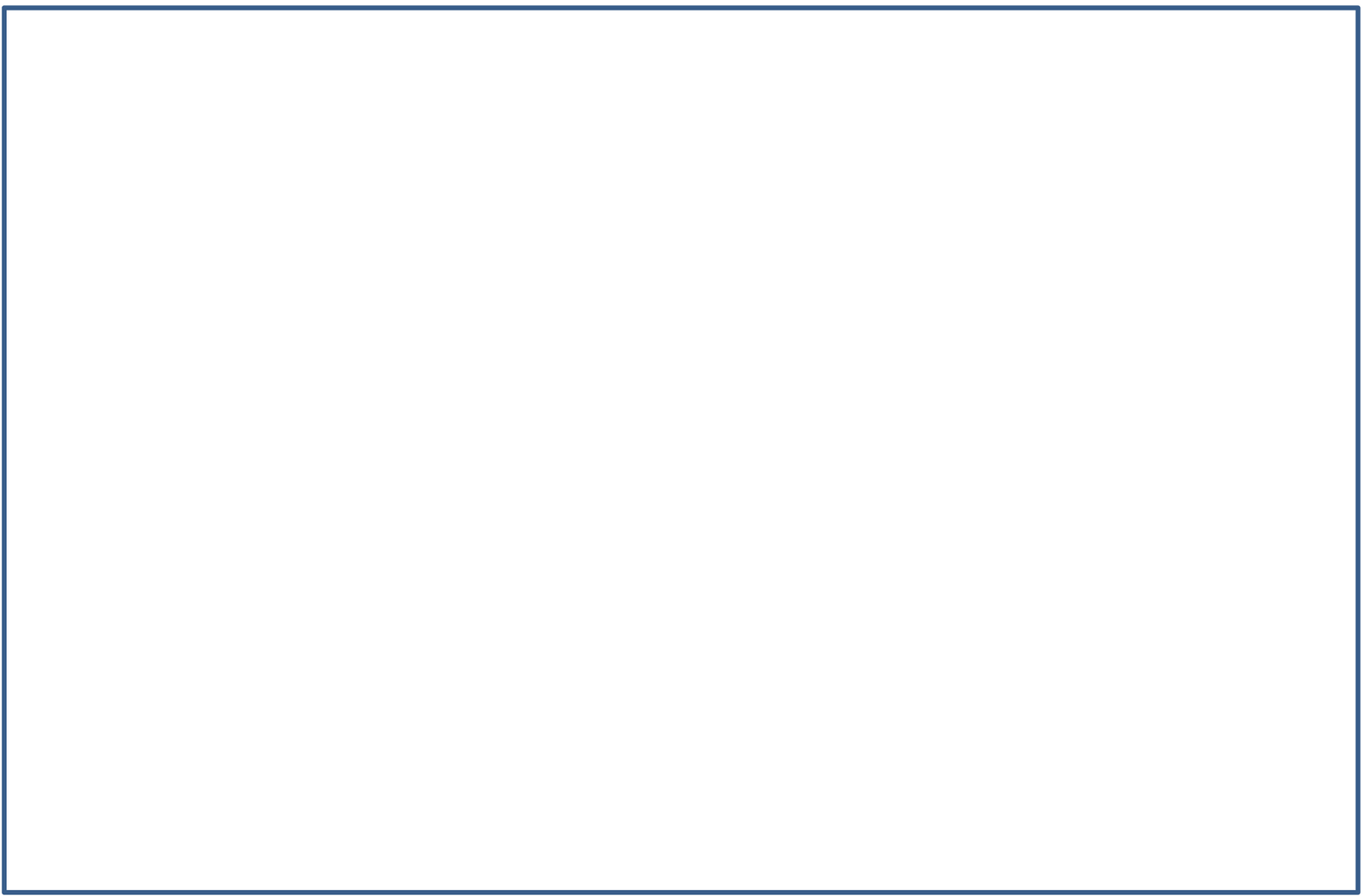}
    \label{fig:fio-original-ext4}
    \caption{FIO: ORG vs. \falloc}
\end{figure}
}

\subsection{Isolating Special Object with \falloc}

Consequently, as shown in  Figure~\ref{fig:mysql-dwb}, the WAF of InnoDB experiments in 
was reduced from 2(?) to 1.1(?).

{\bf The performance gain by \falloc-ing DWB will hold also for SQLite WAL and ext4 journal area since}  DWB-like journal where data are cyclicly overwritten and the write amount in each area is significant: cycle reuse and high write-density (very hot) use cases, traditional WAL log (with some alignment?), Undo tablespace in Oracle, and Ext4 journal area. Such journal areas with fixed address는 booting시 한번 만 \falloc 후 계속 사용가능. Not frequent call to \falloc! 

\comment{
\begin{figure}
\centering
    \includegraphics[clip,width=0.46\textwidth]{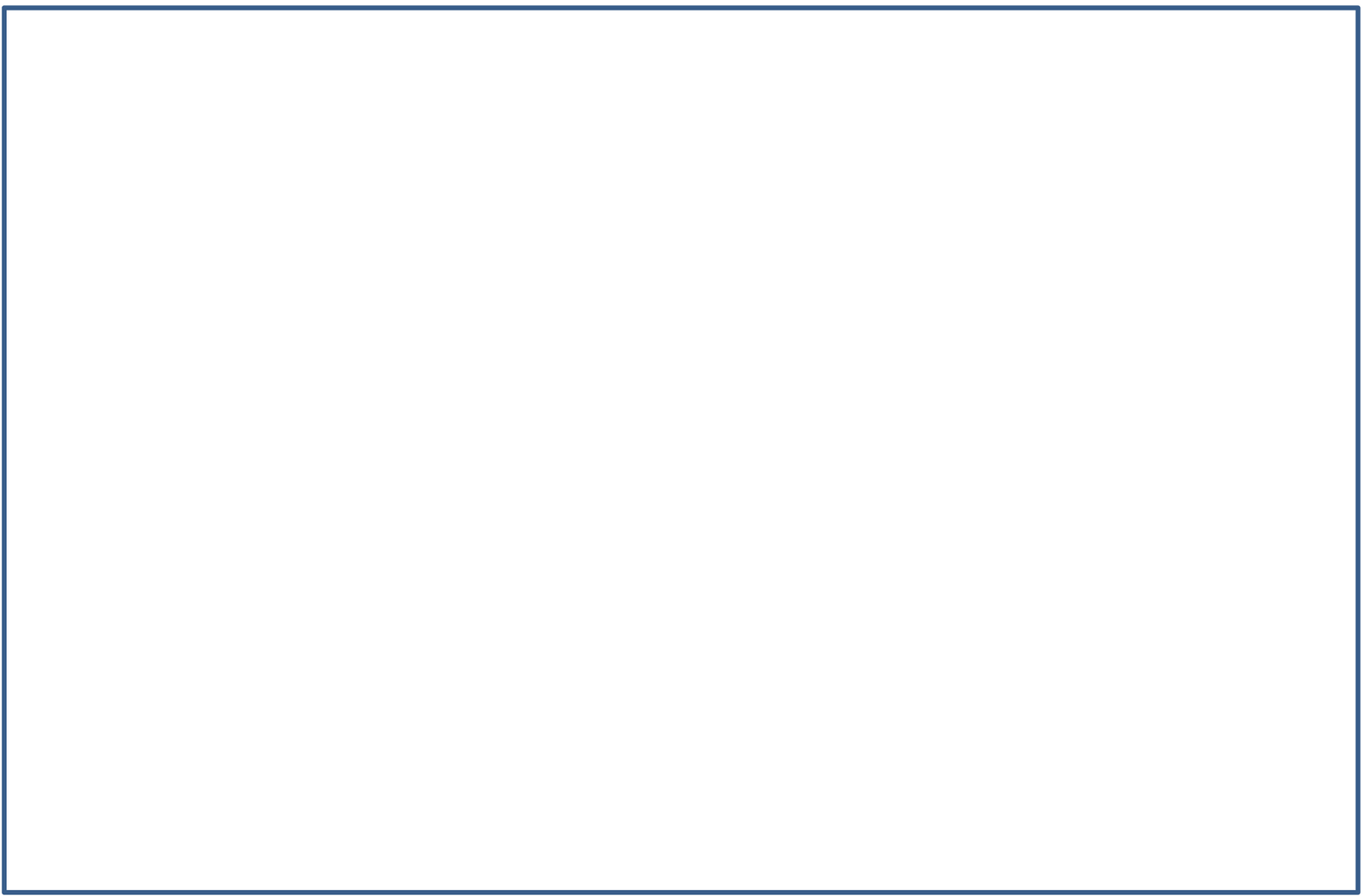}
    \label{fig:mysql-dwb}
    \caption{Isolating InnoDB DWB} 
\end{figure}

\begin{figure}[h!]
\centering
    \includegraphics[clip,width=0.46\textwidth]{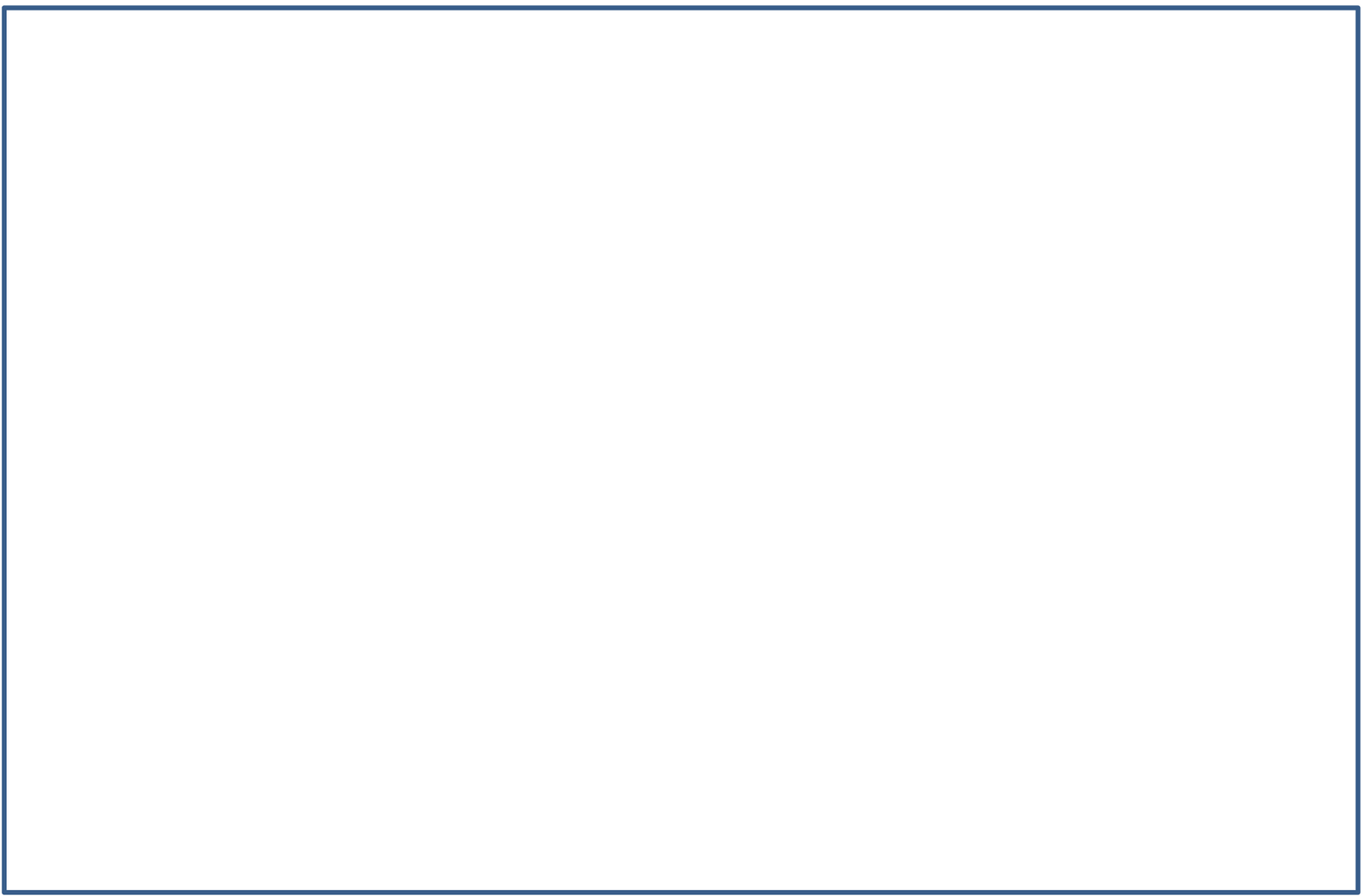}
    \label{fig:sqlite-wal}
    \caption{Isolating SQLite WAL} 
\end{figure}
}

\module{Tail Latency}

\module{Read performance}

\subsection{Performance Isolation in Multi-Tenancy} 

장장치의 용량이 커지고, 가상화 기반의 클라우드 데이터센터환경에서 복 수 개의 응용이나 데이터베이스가 단일 저장장치를 공유하는 멀티테넌트 환경이 보편환되고 있다. 예를 들어, RocksDB와 MySQL 데이터베이스가 단일 SSD에 동시에 저장관리 될 수 있다. 이러한 멑티테넌트 환경에서 기존 온디맨드할당 방식의 경우, 앞서 설명한 이유로 (2.1절), 서로 다른 테넌트 데이터가 동일 플래시메모리 블록에 저장될 수 있으며, 각 테넌트의 데이터의 생명주기가 다르기 때문에 SSD 쓰기증폭 측면에서 상호 간섭을 일으킬 수 있다. 
\module{Multi-tenancy} As discussed in Section~\ref{sec:bg}, when two tenants are run together on the same SSD, the performance of a tenant with low WAF (\ie, YCSB on RocksDB) can be severely degraded by the noisy neighbor with high WAF (\ie, TPC-C on MySQL), which is undesirable in terms of {\bf performance isolation}. \falloc can be an effective solution to making multi-tenants less interfere with each other. Let us illustrate the benefits of \falloc using the example of concurrent YCSB and TPC-C benchmarks. If RocksDB tenant isolates its SSTables into dedicated flash blocks using \falloc, we expect that the performance of YCSB will be quite less interfered from the noisy neighbor, TPC-C. Next, since data pages of two tenants with quite different death times are not multiplexed on the same flash blocks, the overall WAF at the flash storage device will be considerably lowered, which can in turn boost the performance of other non-\falloc-ed neighbors.  
From this, we would like to say that, when properly used, \falloc is not only beneficial to the corresponding tenant itself {\bf but also altruistic} to neighbor tenants. 
Thus, we recommend to use \falloc whenever applicable. In summary, the benefits of \falloc will stand out in {\bf multi-tenant environments.} than when each application is run alone and in particular it is an effective solution to performance isolation among tenants.  

\comment{\falloc used in one application would be beneficial not only to itself but also to other applications. In this regard, \falloc is {\bf not selfish but altruistic} mechanism!} 


Org vs. (vs. \falloc-ed RocksDB + Org MySQL??) vs. \falloc-ed RocksDB/MySQL - 중간 실험 결과 제시가 의미?? 

MySQL/DWB의 경우, 1MB로 쓰고 clustering도 힘들고, 따라서, DWB write latency도 특정 block에 수십개 page를 쓰니 latency도 안좋게 됨. 따라서,  차라리 미리 64MB를 잡아 두고 cyclic하게 dwb write를 반복해서 64MB write 다쓰게 되면, 다시  \falloc 하게 함. 또는 SSD내에서 알아서 realloc받음. DBMS엔진에서 호출하니까 충분히 수정 가능함. 

\module{Less Interference thus performance isolation} 
RocksDB, TPC-C 각각 따로 수행할 때의 WAF대비, RocksDB가 isolate됨으로써, TPC-C가 여유있는 OP활용해서 TPS 높아짐. 
TPC-C의 RocksDB에 대한 interfernce가 사라지고, 반대로 \falloc-ed RocksDB가 TPC-C에 대해 altrustic한 효과를 가짐. 

\comment{
\begin{figure}[h!]
\centering
    \includegraphics[clip,width=0.46\textwidth]{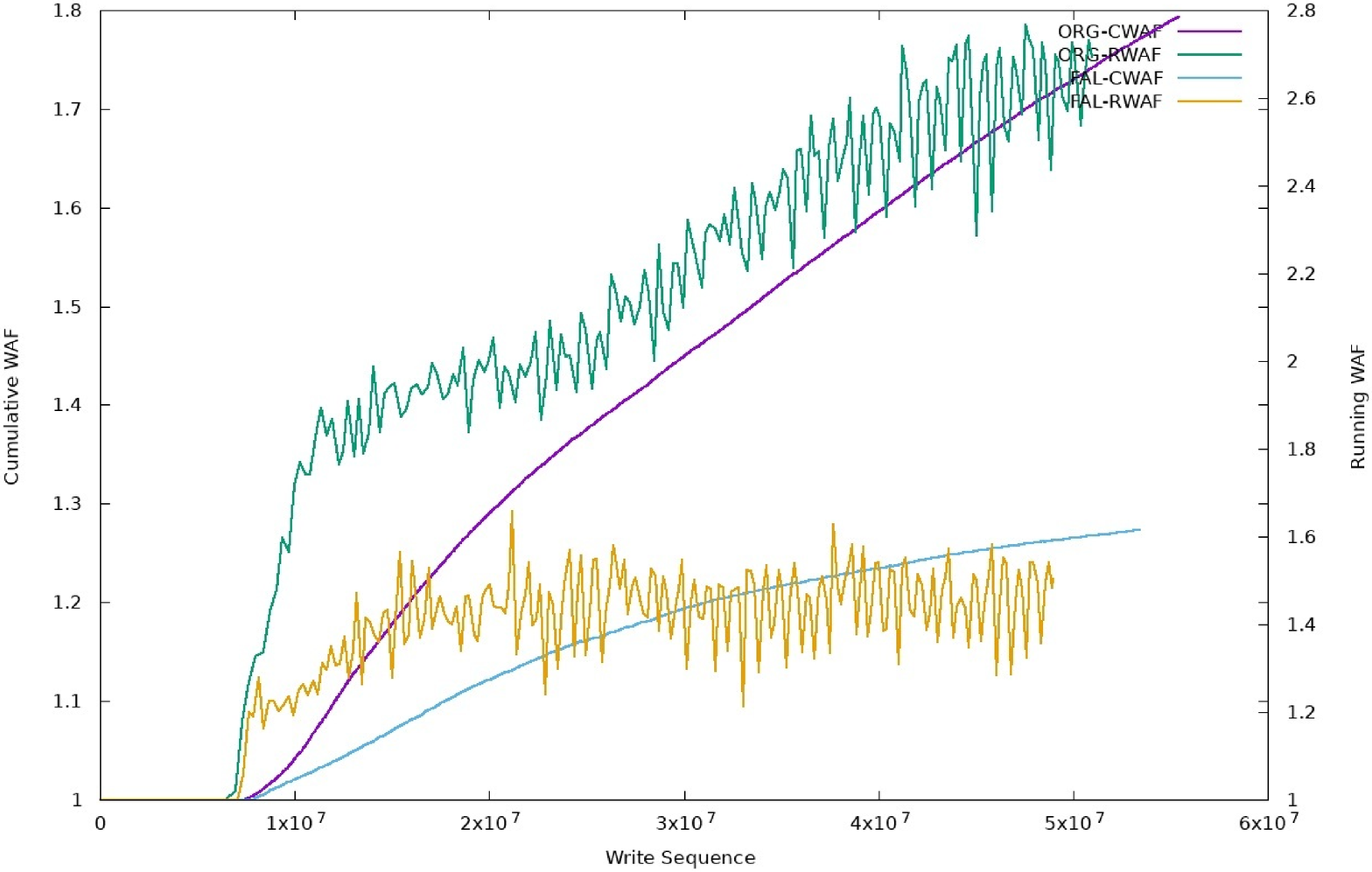}
    \label{fig:rocksdb-mysql-multitenant}
    \caption{Multi-tenant Databases: Isolating RocksDB from MySQL/InnoDB (WAF only)} 
\end{figure}
}

\begin{figure}[!h]
\centering
\includegraphics[clip, width=0.31\textwidth,height=3cm]{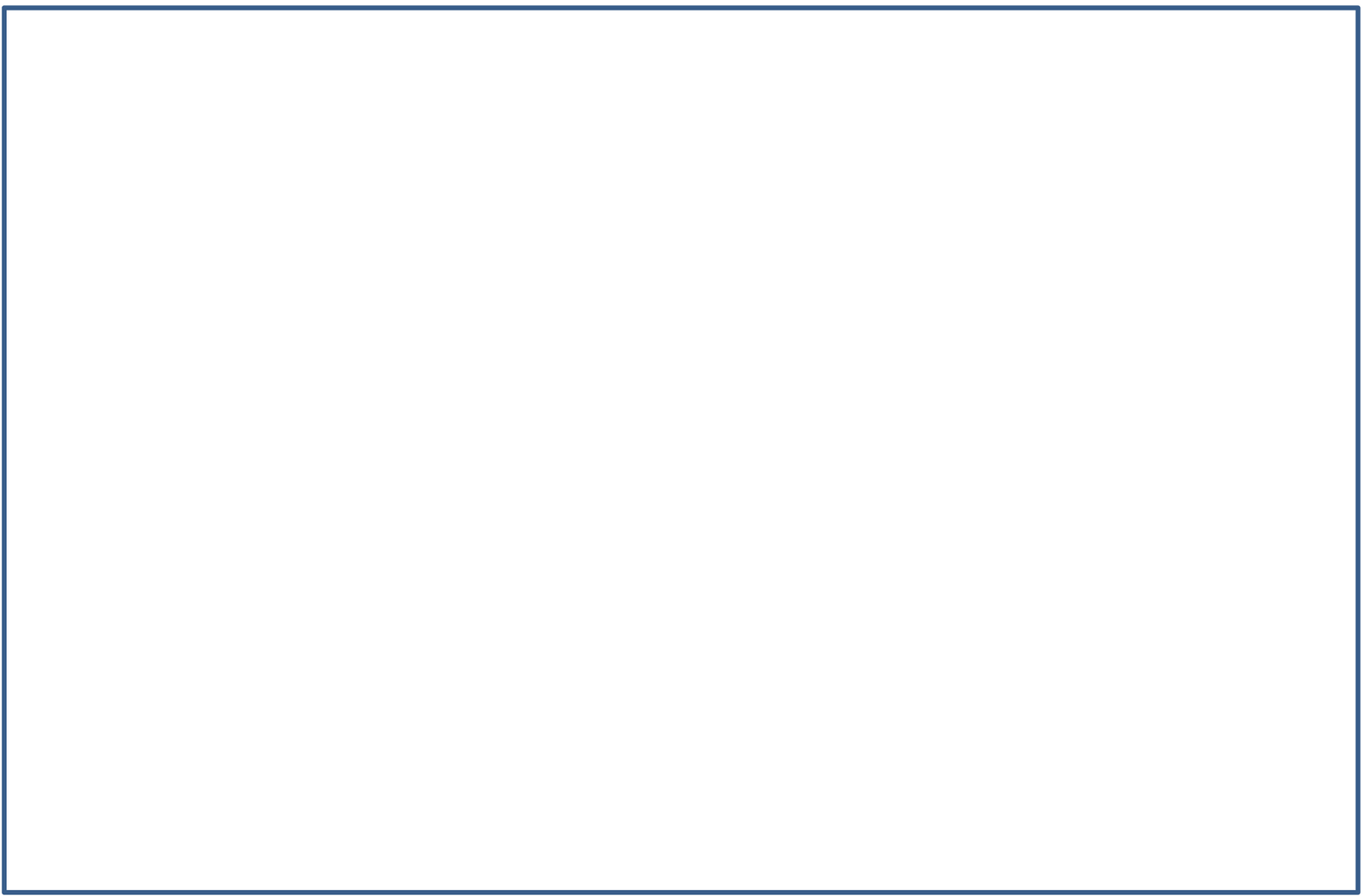}
	\caption{Adaptive Space Allocation (RocksDB and TPC-C)} 
\label{fig:adaptive-space}
\end{figure} 

\module{Adaptive Space Allocation} 
As discussed in Section~\ref{sec:operations}, the size of \falloc-ed and normal regions will adapt to the host usage. That is, \falloc 호출한 논리 객체 데이터들은 자신이 현재 사용중인 영역을 dedicated flash block을 점유하고 있고, 나머지 flash blocks은 normal writes들에 의해 사용된다. 
\comment{from 2R paper: In the TPC-C trace, as the database grows over time, the number of cold and frozen pages also increases. Therefore, the cold region has to accordingly expand to accommodate more cold pages.  2R schemes are novel in that they allocate space adaptively to flash regions, thus requiring no tuning effort.}  To verify the GC algorithm above will adaptively allocate blocks to \falloc-ed and normal region, we measured the {\bf distribution of \falloc-ed and normal block over time}  
\comment{fraction of cold blocks over all blocks} while running the multitenant workloads used in  Figure~\ref{fig:rocksdb-mysql-multitenant} and 
present the result in Figure~\ref{fig:adaptive-space}.
\reminder{From 2R: 수정: Note that Y-axis in the figure ranges up to 110\% because the OP factor\comment{percentage} is set to 10\%. While both schemes allocate more blocks to cold region over time, the fraction of cold region in \trf is less than that in \trg about by 0.2 across all ranges greater than 0.1 in X-axis. This result indicates that \trf is better than \trg in terms of the space utilization of cold region. \trf can utilize the cold region better because it prevents false cold pages from entering the cold region, and also because it takes the eager cold block merge. With larger normal region, non-cold pages in normal blocks will have more temporal chances to be invalidated before the normal blocks become victims. This is the reason why the average utilization of normal victim blocks in \trf was lower than that in \trf.}

}

\section{Related Work}~\label{sec:related}

In that \falloc aims at reducing physical WAF by passing the host semantic to flash devices, three interfaces are closely related to it:  Trim~\cite{trim}, Multi-stream SSD~\cite{MultiStreamSSD} and Zoned Name Space~\cite{ZNS}.


\comment{A discard command gives a hint to the underlying flash storage device that a specified address range has no valid data. For instance, upon a file deletion, the file system calls the TRIM command to explicitly inform SSDs of the corresponding locations of the deleted data so that they can be treated as the dead data.} 
\comment{which is used to inform flash storage of dead pages. This command is sometimes called ``trim'' or ``unmap''~\cite{trim}.The command takes an address and informs the device that page(s) specified by the address are no longer valid; the device thus will not relocate those pages upon garbage collection, thus preventing unnecessary write amplification~\cite{OST}.} 

\module{Trim} Even when a file is deleted, the old storage interface (\eg, SATA) provides no mechanism to pass the host semantic about the file deletion and thus flash devices regard pages from the deleted file as still valid and unnecessarily relocate them during GC, causing write amplification. To address this, the {\tt trim} command was proposed to inform flash devices that page(s) specified by a logical address range are no longer valid (\ie, dead) at the host~\cite{trim}; the trim-hinted device will not relocate those pages upon GC~\cite{OST}. \falloc and {\tt trim} are common in that both explicitly provide flash storages with host-side semantic so as to lower write amplification: \falloc is used to pass the semantic about the logical address space constituting an object and the \falloc-hinted device will de-multiplex different objects into different blocks. In addition, \comment{though each scheme is effective individually,} they are {\em synergetic} to each other: when a file is \falloc-ed, the {\tt trim} command can complete simply by erasing all the \falloc-ed blocks (that is, nearly zero-overhead {\tt trim}), instead of invalidating all pages individually~\cite{ToTrimOrNot}. \comment{the trim overhead upon its deletion~\cite{TrimOverhead} is, as discussed in Section~\ref{sec:archi}, negligible}
Meanwhile, they differ in their invoke time: \comment{they are called at different times of object life cycle:} \falloc is called at object creation while trim is at object deletion. 
Lastly, let us remark that \falloc and trim are not in need for harddisks:
since overwrites are allowed, logical space management implies physical one as well. 

\comment{In fact, the idea of Trim (novel and practical) has been rapidly adopted by the OS kernels~\cite{window,linux}, \comment{https://events.static.linuxfound.org/sites/events/files/slides/discard\_0.pdf)} which is attributable to three factors. First, its abstraction is simple and intuitive, therefore can be easily leveraged by the developers. Second, as the trim command is in most scenario called transparently by OS kernels (\eg, file deletion), most application can benefit from the command without making any change to itself.  Third, the command is very effective in reducing write amplification, boosting the application performance, and prolonging the lifespan of flash storage. Since \falloc is we believe comparable to the Trim command in terms of those factors and thus as novel and practical as trim, we expect \falloc can also be quickly adopted by OS community as well as SSD vendors.}
\comment{ 
HDD가 overwritable해서, GC overhead 없어 trim  필요없었지만, flash는 no overwrite라써, GC overhead 크기 때문에, logical operation 수행시 physical operation 도 같이 수행해야함.; \falloc도 이와 비슷}
\comment{ TRIM & \Falloc: Sync vs. Async Impl. 
In addition, as vendors can take either synchronous or asynchronous implementation for trim/DISCARD (https://events.static.linuxfound.org/sites/events/files/slides/discard\_0.pdf), the physical space can be allocated in either synchronous or asynchronous way in \falloc.;;  \falloc + SHARE: + As F2FS logical waf can be reduced by SHARE+Hotstorage 19, we can expect to reduce physical as well as logical waf by combining {\bf share + \falloc}.
}

\comment{
- write시 stream id 부여 vs. object creation시 \falloc
- to lower the garbage collection overhead with the help of explicit hint on data lifetime from applications
- Inform the belongs-to relationship b/w pages and logical object {\bf indirectly via stream id} (explicit?)  
- Multi-Stream SSD is also user-hint interface 
- It allows applications to place data pages .. 
- host write를 플래시내 물리적으로 분리저장케하는 대표적인 기술로 MS-SSD. MS-SSD는 복수 개의 쓰기 스트림 제공하고, 
쓰기연산시 스트림번호를 힌트로 제공하면 (예를 들어, write(LBA, data, stream\_id)), 스트림번호에 따라 데이터를 
다른 물리공간에 저장한다. 
- MS-SSD (open->fadvise(fd, streamid)) 차이점: 1) stream 구분말고는, 동일 stream에 write 되는 page들에 대해 fine placement control은 못함; In addition, due to limited  number of physical streams, two or more files have to be mapped to the same stream.; Heavy solution( SSD resource is divided into M units, which is inefficient unless every stream evenly use the resource.  
- Meanwhile, the effectiveness of MS-SSD is highly dependent on the accuracy of the classification of the lifetime of each data pages~\cite{AutoStream}. Therefore, it is critical for any MS-SSD applications to precisely classify logical data pages according to their update intervals, which is a daunting task except in a few applications such as LSM-based KV Store~\cite{MultiStreamSSD}.
- For this reason, it is a daunting task to leverage MS-SSD for OLTP workloads, and the performance gain is quite limited with an enormous tuning effort~\cite{soyee:edbt18} or even storage engine modification~\cite{hwpark:damon19}.  In contrast, without any user hint or any complicated tuning effort, 2R can isolate cold pages into cold region and at the same time performs nearly optimally.
}


\module{Multi-Stream SSD} A novel interface for flash storages, Multi-Stream SSD (MS-SSD), was proposed and standardized~\cite{MultiStreamSSD, t10-2015}, which allows applications to place pages with different lifetimes to different streams (\ie, flash blocks). More specifically, when invoking a write system call, applications can assign a proper stream identifier (\ie, stream-id) for the data being written, and, on receiving the write command with a stream-id, MS-SSD will place the data into the corresponding physical stream. This interface performs effectively when correctly hinted by applications~\cite{MultiStreamSSD}. 

While both commonly aim at streaming writes to reduce write amplification, MS-SSD and \falloc are in stark contrast in their abstractions for write streaming. The MS-SSD interface has introduced the additional concept of stream-id and mandates applications to statically bind stream-id to each write call. \comment{write call에 physical str-id (일종 semantic hint)를  application-level binding함. } The static binding of stream-id, combined with the limited number of physical streams available in commercial MS-SSDs (\eg, 8), will raise several practical issues. First of all, it is a {\em non-transparent} abstraction in that every write call has to come with static stream-id. Next, it would be a non-trivial task for developers to estimate the number of physical streams for their applications and to correctly group numerous objects with different lifetimes into the limited streams. Third, the static stream-id assignment is {\em non-adaptive}. As the lifetimes of objects can change over time, programmers need scrutinize those statistics and periodically re-assign stream-ids to objects.\comment{, which task is we believe neither possible nor practical.} Lastly, the effect of write streaming would be useless due to the {\em stream-id conflict} in the multi-tenant environment~\cite{AliCloud}. Different applications which were independently developed might have assigned the same stream-id to their objects. Therefore, different tenants' objects with quite distant lifetimes might share the same physical stream. 

In contrast, \falloc supports the {\em per-object write streaming} abstraction, which provides several benefits over MS-SSD. First, it enables transparent write streaming: since it simply requires the logical address range of each object upon object creation, application can achieve write streaming with no or minimal change. Second, since it provides 
fine-grained per-object streaming, the application developers are free from burdens of managing stream-ids to objects and further need not care about the stream-id conflict in multi-tenant environment. 
In summary, \falloc is we believe more advanced streaming mechanism than MS-SSD.


\comment{ no stream id, fine-grained (thus, in theory unlimited number of streams) streams, thus MS-SSD단점 모두극복: even in multi-tenant environment
  that is, simply need to identify logical objects and 필요시 \falloc 호출
  (file application의 경우, FS에서 자동 호출). 
  1) no program modification없이 kernel에서 falloc시 transparently streaming
  2) fine-grained physical flash block / object -> no mapping overhead 
3) naturally adaptive! ++ no stream-number conflicts 등 MT서도 문제없음.
:: thus does not require additional hint from the applications.   

While \falloc can {\em transparently} provide write streaming simply by invoking the {\tt fallocate()} call upon object creation, MS-SSD requires to assign a stream-id for every write,
assign id: user responsibility! optimization problem. limited # --> the overhead of estimating object's lifetime and grouping them into the limited number of streams: challenging. 
\falloc pre-allocation vs. MS-SSD are still on-demand allocation(?). Basic differences. 1. hint 시점: on write vs. logical space alloc, 2) thus, physical space allocation: pre-alloc vs. on-demand(?0, 3) physical stream  vs. fine-grained logical object. In addition to this shallow difference, 
MS-SSD has a few practical and critical drawbacks: 1) complicated programming abstraction, 2) optimization overhead, 3) non-adpative, and 4) impractical in multi-tenant scenario. In contrast, \falloc is free from all such problems.

The MS-SSD abstraction requires considerable development and tuning effort to assign proper streams to various data types~\cite{MultiStreamSSD} mainly for two reasons: 1) it needs the intermediate abstraction of stream-id and 2) limited number of physical streams.  With MS-SSD, since SSD-internal limited hardware resource including write buffer should be partitioned, the number of physical streams exposed to users should be limited. For instance, the number of physical streams provided in a commercial SSd is limited to 8. Stream-id has to be determined and assigned to every distinct files. The fact that it requires to assign stream id to every write for write streaming brings two shortcomings~\cite{fstream}.\comment{(stream 구분 overhead@INTRO) - 1) program modification and 2) proper stream id assignment overhead.} First, MS-SSD mandates every write to be assigned stream-id and thus require {\bf program modification (for XX)}. Second and more crucially, users are solely responsible for classifying logical objects in terms of their lifetimes/deathtimes, grouping them, and in particular assigning proper/correct stream to each group.(Logical Object/Physical Stream Binding/Coupling) Given limited number of streams and numerous logical objects with diverse death time, this overhead of assigning stream ids to objects are {\bf non-marginal rather challenging optimization task}~\cite{autostream}. 
\reminder{note that the number of streams are rather limited(8 by default).} 
\comment{For instance, Kang et. al. showed how to lower WAF by assigning different streams to files belonging to different levels  in LSM-based key-value stores~\cite{MultiStreamSSD}. tuning and development effort with MS-SSD abstraction is considerable! 즉, 필요 stream 갯수, level 별 stream id assignment 등에 대한 쉽지않은 tuning effort 필요로 함.} 
In addition, MS-SSD is {\bf NOT ADAPTIVE} when the lifetimes of logical objects are dynamically changing. For instance, if the death times of two objects which were once grouped to the same stream are diverging over time, MS-SSD will also suffer from multiplexing. In this case, one object have to migrate to other or new stream, causing program modification for stream split or merge. 
To be worse, this limited number of streams will make {\bf MS-SSD less practical in multi-tenant} where multiple databases share a  large SSD and where active write streams from multiple applications are dynamic and unpredictable. 
When number of logical objects with different death times are greater than this number, different objects have to share physical stream (N:1 mapping!) -> thus, multiplexing.  When two or more files or logical streams have to be grouped into the same physical stream because of limited physical streams, multiplexing within the stream is unavoidable so that WAF persist. --> developers는 L2P stream mapping 부담/최적화 등 고민. In particular,  we suspect that it would be extremely challenging, hard or in practice impossible for MS-SSD with limited streams to properly (that is, correctly and effectively) stream writes from numerous logical object from different tenants in multi-tenant environment where the number of active logical objects are dynamic and unpredictable in terms of 수행 시점. {\bf multi-tenancy 문제점} different tenants which are completely irrelevant each other and developed independently are very likely to share same stream ids; this totally invalidate the assumption that objects with same lifetime are streamed in the same stream id. In fact, the same lifetime is at best a relative guess-work by developer within each specific workload. Therefore, among multi-tenants, developed by different developers for different workloads, it is almost impossible to assume that objects from different tenants but with the same stream id  will have the same lifetime. In this respect, MS-SSD is not a proper abstraction for separating different objects in multi-tenant environment. 

{\bf 사용자 data - logical stream id - physical stream (larger grain) @MS-SSD VS. logical - physical clustering @ finer segment grain; In addition to the programming difficulty, limited stream # and hw partitioning / stream can in practice limit the effect of write streaming.}

IN CONTRAST, \falloc provides {\bf better abstraction}  in that write streaming can be achieved at minimal or even no program modification.  and at no intermediate concept such as stream-id. This is mainly attributable to two factors: {\bf 1)  
no stream id concept, and 2) fine-grained streams}, which allow to avoid MS-SSD 단점들. That is, logical objects is decoupled from stream id. 내부동작입장에서도 비슷하거나 lighter하다(?). 
Unlike MS-SSD, intermediate stream id 할당, (finer grain이라써) stream 갯수 제한 등 overhear 없이, 선할당에 따른, thus enable dedicated streaming of logical object to physical flash block and thus preventing multiplexing of logical objects. Therefore, application developers need to identify proper logical segments suited for \falloc, and simply invokes the {\tt fallocate} call, which FS kernel will then automatically call \falloc for users. With \falloc, as explained in Section~\ref{sec:eval}, without any tuning or modification for its codebase, RocksDB can reduce WAF to 1. 
In addition, with \falloc where each logical object has its dedicated FA instance and physical flash space, \comment{that is, can support the (in theory/practice) unlimited number of physical streams.}  developers/administrators need not worry about the deathtime difference/mismatch among different tenants. They only need to decide when/how to use fallocate properly both in usage as well as size (Use case item 6.). In addition, \falloc without the notion of stream id, in stark contrast, 개별 range로 요청하므로, does not have any 단점 of multi-stream SSDs. In addition, with \falloc where each logical object has its dedicated physical flash space, no program change is required, as long as each logical object's pages have same deathtime. In this regard, {\bf \falloc is adaptive to dynamically changing lifetimes} of logical objects. 

\falloc vs. MS-SSD can be thought as orthogonal and thus they can be combined. \falloc이 좀 더 fine-grained control로써..  In this regard, we argue that {\bf \falloc is much better abstraction for write streaming than MS-SSD:} \falloc is based on the existing file system eco, thus can improve performance without requiring further hint form the users (e.g. stream id) (like TRIM). We believe that \falloc be simple and proper but more powerful abstraction than MS-SSD.  



\comment{
Statically Assign stream id according to the each SSTable's level. Provided that the number of level tend to dynamically change, this {\bf static stream-id assignment is not such desirable}.}
\comment{\module{Adaptiveness} 또한, MS-SSD can not adapt to the stream changes (new stream need, stream merge/split etc) over times.  In contrast, with \falloc, fine grained per logical object라써 변화 무관하게 streaming 효과유지됨.}

--1/N architecture vs no overhead: Heavy-weight architectural approach (ranging from write-buffer to FTL and garbage-collection vs. light-weight scheme to place data in a flash-optimal way simply with minimal changes in FTL mapping 
-- stream-oblivious GC 문제점- 실험 통한 MS-SSD 한계점제시?

Impl. Different vs. MS-SSD: stream-id 전달을 위해 kernel stacks 수정 vs. \falloc 통해 logical object's address range 정보만 전달}
\comment{In contrast to this per-file stream management, Choi et. al. demonstrated that it is possible to reduce WAF by storing some hot objects with different lifetime in one file into several different streams in ForestDB KV store (that is, per-object stream assignment)~\cite{EDBT18}. Recently, Kim et. al. proposed an automatic stream management technique, called PCStream, which can fully automates the whole process of stream allocation within the OS kernel with no manual work by identifying dominant I/O activities using the concept of program contexts~\cite{PCStream}. However, this scheme works only when there are multiple writer processes each of which has its own context distinguishable from others. For this reason, the scheme is inapplicable to database engines where one database writer process usually is responsible for flushing data pages. As far as we know, there has been no prior work which can properly stream data pages in OLTP workloads with random write patterns using MS-SSDs.}
\comment{By taking advantage of the copy-on-write mechanism of flash storage, X-FTL can efficiently guarantee that all the pages updated by a transaction are successfully propagated or no pages are written at all upon a failure. In this regard, three types of existing work are closely related to X-FTL: shadow paging technique [15], journaling file systems [7, 19, 21, 25], and a few FTL techniques for atomic write of file system journal data [17, 18, 20]. Now, let us briefly explain each work and compare it with X-FTL.}
\comment{\item 멀티스트림 SSD의 경우, SSD 내에서 쓰기버퍼 등의 HW 리소스를 파티셔닝해야 하므로 하드웨어 리소스 제약으로 인해 SSD에서 제공할 수 있는 스트림 개수는 제한적이다. 따라서, 파일 개수가 스트림 갯수보다 많은 경우 서로 다른 파일들이 하나의 스트림을 공유해야만 하기 때문에 서로 다른 파일의 페이지들이 동일 플래시메모리 블록을 공유하게 된다. 반면, \falloc의 경우 서로 다른 fallocate 된 논리주소공간은 다른 플래시블록에 저장하기 때문에 스트림 개수 등의 제약으로부터 자유롭다.
\item 멀리스트림 SSD의 경우, (현재는) 파일 단위의 스트림번호 부여만 지원하기 때문에, 한 파일 내의 생명주기가 다른 데이터들을 각기 다른 스트림으로 지정하기 위한 별도 작업이 필요함. 또한, 분리해야 할 파일 개수 대비 스트림 갯수 반면, \falloc은 다른 fallocate 호출시마다 다른 물리적 영역에 분리 저장하게 된다.}

\module{Zoned Name Space} To overcome the write amplification problem in the conventional SSDs with block interface while obviating the need for in-device GC, DRAM resource for page mapping FTL, and the over-provisioned physical space, the system community has recently proposed a new interface for flash storage, ZNS (Zoned Name Space)~\cite{ZNS}, which exposes {\em zones} (a set of logical blocks) to the host as the unit of data management. 
\comment{
With the advent of the ZNS, Open-Channel SSD~\cite{OCSSD} approaches have remained a just transient step in the evolution of SSD.
ZNS has replaced them and is now a standardized interface in the LightNVM~\cite{LightNVM}.
}

Though novel and worth investigating, however, the interface imposes  strict write-ordering rules: all writes to zones should be to be sequential and also in their LBA order. In other words, ZNS disallows out-of-LBA-order random writes to each zone.  Such strict rules will bring two drawbacks. First, all the software stacks from applications, database engines to file system should be modified to meet the sequential write ordering. A storage interface which mandates the whole software stacks to adapt to it
is unlikely to succeed. Second, while exempting from the block interface tax, ZNS instead introduces yet-more-expensive tax of log-structured writes (\eg, compaction in RocksDB and segment cleansing in F2FS). Such operations are known to induce application-level logical WAF of more than 10~\cite{VLDB21FlashSSDs,DONOTSTACK}. In addition, log-structured write inevitably needs over-provisioned logical space. 
In contrast, \falloc does not ask the existing software stacks to adapt to it; they can achieve transparent write streaming. 
Also, flash storage supporting \falloc can support random writes as well simultaneously.


\comment{
\module{Physical Page Addressing in OC-SSD}

\module{Application-managed Flash~\cite{AMF}:} similar to OC-SSD or ZNS, thus host eco 모두 바뀌어야 함? 목표/개념적 유사성? AMF assume that applications should be refactored to take sequential writes -- for this reason, Block mapping FTL; But, with flashalloc, we do not assume such one. Instead, all the existing SW eco need not to change at all.   
\module{ZNS:} 
 \begin{itemize}
 \item NVMe Zoned Namespace(ZNS)~\cite{ZNS} = OC-SSD + ((MS-SSD + AMF))??  
 \item [1] page 20-21 -- file --> zone 단위 preallocation 효과? 차이점? ZNS가 logical address range에서 zone이다? physical address range와 무관? ~\cite{SDC2019}
 \item (기환) ZNS 에코서만 사용가능. 기존 kernel 레이어부터 수정 필요하며, 경우에 따라 애플리케이션 레이어까지 수정 필요하고, 호스트 레벨 FTL 레이어 추가됨. zone의 alloc은 application 컨텍스트 무관하게 할당되고 open zone 갯수 제약사항 존재.
 \end{itemize}

full control가능하고, 사용자 잘 사용하면 성능개선 가능; 단점: 기존 software stack 전면적 수정해야 하고, SSD도 수정해야함.
In contrast, \falloc can provide the same benefit simply by introducing minimal API to page placement (, hint SSD about clustering logical pages into the same physical flash block) while preserving the existing eco, including both host software and SSD. 
}

\comment{ \subsection{File system} 
\begin{itemize}
\item Defragger~\cite{defragger}: mobile phone에서 logical and physical frag에 따른 overhead 분석; 그리고, logical defrag (using remap) + physical defrag를 통한 성능 개선; vs. \falloc은, 이 관점에서 (despite logical frag) physical frag avoidance / preemption 기법으로, 주로 waf 개선목적이고, logical frag에 따른 read overhead는 개선못함.
\comment{F2FS관련: do not stack your log on my log ~\cite{DONOTSTACK}. FlashKV, ORCFS~\cite{FlashKV,ORCFS} - Though the goal and eventual effect are similar to our \falloc, they require to change the DBMS engine, eco, or rewriting  file system. In contrast, \falloc, application, DB engine, and file system need not to be changed at all. Only minimal changes in kernel layer and DBMS are required.}   
\end{itemize}

\comment{ \module{File system fragmentation:} \reminder{포함?}  fallocate 을 하기 때문에 그 정도가 cdf 그래프 대시 훨씬 완화된 양상이고 평균값도 높긴하지만 SAF 증가에 따라 fragmentation 증가가 맞습니다. 특히 level0 file 은 256MB 정도라서 더욱 발생. fallocate 하지 않는 f2fs는 그 정도가 훨씬 심함. Figure~\ref{fig:filemap_Ext4_f2fs}.

\begin{figure}[]
\centering
    \includegraphics[clip,width=0.42\textwidth]{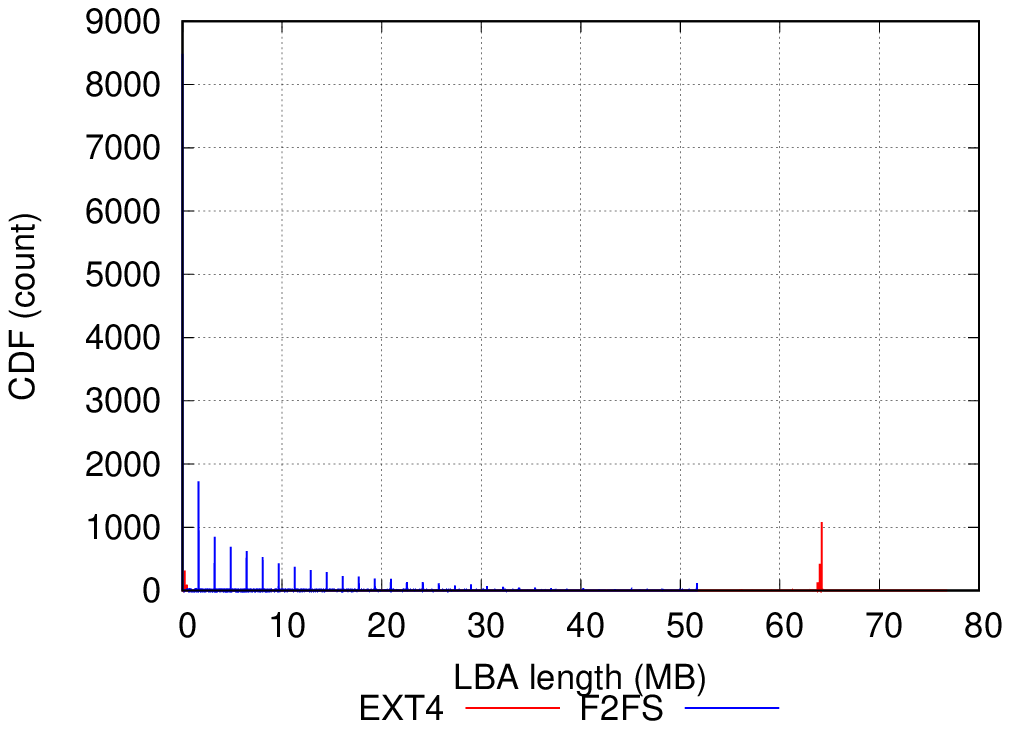}
    \label{fig:filemap_Ext4_f2fs}
    \caption{filemap EXT4 F2FS }
\end{figure}

With the help of explicit hint provided via \falloc , flash storages can maximize the performance by physically separating logically frozen pages from other pages and thus minimizing the write amplification. Four existing works are related to the \falloc interface: hot-cold separating FTLs~\cite{DAC,Stoica}, defragmentation in file systems and database~\cite{OST,defragmentation, defragger}, {\tt trim}~\cite{trim}, and multi-stream SSD~\cite{MultiStreamSSD}. None of them are proposed with the problem of logical frozen pages in mind. We briefly review each and compare it with GATHER. \comment{ NVMe spec: write hint mechanism(?)}}

\comment{
(( I’m not sure if adding another interface is only the solution to mitigate the write amplification problem caused by frozen pages. For the past 20 years, lots of GC techniques have been proposed. One of the well-known approaches is a hot-cold separation technique which separates hot data from cold ones.)) -- {\bf The GATHER interface might perform better than the hot-cold separation heuristic because it is able to directly deliver application-level information to the SSD level.(부각)}; Considering the property of frozen pages (which are never overwritten again), frozen pages can be easily detected by the hot-cold separator and separated from hot or warm pages even with naive hot-cold separation heuristics. It would be good if the authors {\bf directly compare} GATHER with existing hot-cold separation heuristics.}

{\bf Defragmentation in file system} Since file system fragmentation hurts performance on harddisks, defragmentation tools were introduced to reorganize on-disk data to place files contiguously and make free space for one or a few contiguous regions, moving data around and then rewriting inodes and such to reflect the changes~\cite{OST,degragger}
In contrast, with flash storages, file system fragmentation is not as much of a performance problem, as there is no mechanical disk head movement, and thus defragmentation can shorten their lifespans at no performance benefit~\cite{fragmentation,degragger}.

\comment{While both defragmentation tool and GATHER are common in that they aim at improving performance by reorganizing data, the existing defragmentation tool will be of no use in solving the problem of scattered frozen pages since it can not guarantee the complete isolation of frozen pages into a separate flash area. }

\comment{ On the other hand, copy-on-write·Î ÀÎÇØ LBA-PBAÀÇ mappingÀº ever-changingÇØ¼­, LBA-PBA°¡ alignµÇÁö ¾Ê´Â flash¿¡¼­´Â object¿¡ ´ëÇØ ¹Ì¸® contiguous allocation of LBAs at no fragmentation or ÁÖ±âÀû defragmentation is not effective at all in flash storage because XXX.  \reminder{SSD¿¡¼­ LBA fragmentation ÀÚÃ¼´Â Å« ¹®Á¦ ¾ÊµÊ. ´ë½Å frozen page°¡ physically scatteredµÇ¾î WAF¸¦ ³ôÀÌ´Â °ÍÀÌ ´õ ¹®Á¦ÀÓ.} µû¶ó¼­, ÀÌ·± ¹®Á¦Á¡ °³¼±À» À§ÇÑ solutionÀ¸·Î, GATHER clusters and thus isolate objects of frozen pages into physical flash blocks so that the objects to be gathered are necessarily neither to be contiguous in terms of LBA nor to belong to the same object. The source of physical scattering of 'log structured write+FTL'
Defragmentation - cluster data pages belonging to an object in terms of LBA so that their LBAs are to be contiguous as much as possible. vs. Gather - to cluster and thus isolate objects of frozen pages into physical flash blocks; the objects to be gathered are necessarily neither to be contiguous in therms of LBA nor to belong to the same object. 
The goal of defragmentation tools and file system defragmentation is to cluster data pages belonging to an object(e.g., file) in terms of LBA so that place object contiguously and make free space for one or a few contiguous regions. Note that ÀÌ°Ô µ¿ÀÛÇÏ´Â ÀÌÀ¯´Â,  in hdd, LBA/PBA°¡ alignµÇ±â ¶§¹®.
what is the source of LBA fragmentation? LBA - PBA 1:·Î mappingµÇ±â ¶§¹®¿¡ fragmented LBAs means more movements of mechanical disk head and thus hurts read/write performance in HDD.  To rectify the performance degradation, ... }
}

\comment{
\reminder{OSD와 비교? DFS 논문 참고/비교 필요?}
\begin{itemize}
\item Object-based SCM: An Efficient Interface for Storage Class Memories~\cite{MSST2011}
\item DFS: A file system for virtualized flash storages@ ACM TOS 2010 Sep.
\item De-indirection for Flash-based SSDs with Nameless Writes@FAST 2012 -- similar to the paper, FS and fallocate 걷어내고, SSD가 \falloc 및 space management 수행하면?
\item Extending the lifetime of flash-based storage through reducing write amplification from file systems@FAST2013: 
\end{itemize}
}

\comment{ 
\subsubsection{DFS: A file system for virtualized flash storages}
OCSSD와 유사한 접근법. But, simpler host level ftl 지향. 
\module{Idea}
\begin{itemize}
    \item VFSL - Virtualized Flash Storage Layer
    \item DFS - A filesytem leveraing VFSL
    \item ioDrive - A flash memory SSD that provides flash operations explicitly to the host
\end{itemize}

기존에 SSD 내부에서 관리하는 FTL을 호스트 레벨에서 구현하였음. 이것이 VFSL. L2P, P2L, GC, Wear-leveling 등을 모두 이곳에서 관리함. 
단, SSD의 physical parallelism 은 관장하지 않음. L2P 역할을 하되, mapping은 단순하게 append-only write 로 (LFS 스타일) 처리함. 
실제 Physical IO는 ioDrive내부에서 알아서 병렬성을 구현하도록 하였음. 사실상 VFSL이 파일시스템의 space management를 모두 제공하는 격임.

DFS는 단순히 VFSL 인터페이스를 사용하기만 하는 파일 시스템임. 파일의 Logical Sequential Allocation을 보장함. 

ioDrive는 위에서 언급한대로 physical 병렬성, physical checksum/error detection 만을 지원함. 

\module{차이점 및 문제점}

역시나 Logical Sequentiality 만 보장함. space allocation 레벨에서 physical sequentiality 를 고려하지 않음. 
기존 File system들이 지닌 문제가 그대로 남아있음. => outstanding thread들의 write가 모두 physical level에서 섞임

심지어 VFSL 디바이스 드라이버도 사용해야 하고, fusion IO의 ioDrive에서만 동작함. 제대로 활용하기 위해선 kernel 레벨의 
대대적인 수정과 지원이 필요함. => fallocate는 device driver 레벨 혹은 module 레벨의 코드 추가만 하고,
실제 파일시스템의 코드는 10줄 미만의 fallocate 호출 코드만 추가하면 됨 (예상)

OCSSD와 일반 ssd의 중간지점이라서 애매모호한 위치. 성능에서도 그렇게 큰 향상이 없음. 

\subsubsection{OSD and Object-based SCM: An Efficient Interface for Storage Class Memories}

OSD:  A more expressive interface such as object-based storage (OSD) [10, 11, 22] can improve the current state of the art. First, OSD delegates finer details of block management to the SSD, thereby preventing any new storage-specific assumptions. Second, OSD expresses the intentions of the higher layers clearly, thereby improving the internal SSD operations.

Several previous works note this deficiency and propose more expressive interfaces [5–7, 9]; ... , . One approach is to use an object-based interface [10, 11, 22], which exports the abstraction of an object as a collection of bytes. Structures such as trees, tables, files, and directories can be represented as objects, reflecting the higher-level semantics better than a block-based interface; the device controller performs block allocation and layout for the object 


기존 표준으로 존재하는 OSD 모델을 SCM (본 논문에서는 SSD)에 적용하자가 주된 내용임. 

\module{Idea}

역시나 파일시스템 매핑과 FTL의 중복을 제거하는게 큰 목표이고, 데이터의 hint를 수월하게 전달하는것도
부가적인 효과로 볼 수 있음. OSD이므로 CRUD 를 전부 지원할 뿐만 아니라 데이터에 대한 힌트도 모두 제공할 수 있음. 

결국 기존 fs의 역할을 OSD SSD가 전부 하고, fs는 단순히 VFS를 거친 데이터를 object 형태로 변환하여 OSD로 전달하는 역할을 함. 

Object 데이터에 데이터 타입이나 힌트를 추가할 수 있으므로, 효율적인 Hot/cold 분리를 할 수 있음을 주장. 
Inode, Data 블록, atime 데이터를 따로 분류하여 서로 다른 segment에 저장함. atime이 가장 많이 invalidation 되고, 
inode가 그다음 invalidation, data 블록이 가장 늦은 invalidation이 되므로 GC할 때 이 순서대로 접근해서 victim 선정하면
least valid pages를 가진 블록을 선정하기 쉬워짐. 

append-only B+ tree (wandering tree) 를 사용하여 object 데이터 페이지를 append 형태로 기록함. 
작은 object들이 마구잡이로 쓰이는 경우 심화되는 wandering tree를 막기 위해 buffering을 함. 
큰 object의 데이터 페이지들이 쓰일 때마다 wandering tree가 업데이트 되는것을 막기 위해, (physical offset, length)로
미리 할당하여 tree 업데이트를 한번만 하도록 했음. 

\module{차이점 및 문제점}
일단 파일 시스템도 OSD 여야 하고 디바이스도 OSD 여야 함. => 그래도 ceph 등에서 잘 쓰이고 있음. 
hint 를 전달 할 수 있다는 걸 강조하지만, DSM 명령으로 같은 힌트들을 전달할 수 있음. 
OSD 인터페이스 제공하려면 SSD 내부 리소스가 너무 많이 들어가게 됨. 
Data segment 에는 여전히 서로 다른 파일의 데이터가 섞임. GC에 안좋음. 
atime 데이터를 유지하기 때문에, read만 해도 write가 유발될 가능성이 있음. (물론 버퍼링해서 atime write를 최소화하긴 함)
infrequent, large 파일에 대해서 fallocate와 유사한 아이디어인 extent-based allocation을 지원함. => 주된 목적이 매 page 에 대한
physical 매핑 할당으로 인한 wandering tree 문제를 방지하는 것임. 실험 결과에서도 GC를 전혀 언급하지 않음. 
\begin{itemize}
    \item 파일시스템이 애초에 object write명령을 할 때 대형 write로 만들어서 내려야 함. => 너무 큰 단위의 write는 동시 수행되는 다른 io를 막음
    \item 같은 object에 대해서 따로 extent를 할당할 수 없음 => fallocate는 같은 파일 내에서도 서로다른 fallocate 호출 가능. 
    \item 서서히 늘어나는 (append 되는) 오브젝트에 대해서 extent 할당 불가 => rocksdb에서 사용 불가. 
\end{itemize}

결국 OSD 환경에서만 사용가능하고, extent-based 할당 기법도 제한적인 사용만 가능함. 

\subsubsection{Extending the Lifetime of Flash-based Storage through reducing Write Amplification from File Systems}
OSD와 비슷한 아이디어. Object file system을 효과적으로 지원하는 OFTL 만들자. 기존 block IO 인터페이스에 object 모델을 추가함. 
주된 내용은 OSD와 비슷, 데이터 관리는 일반 B+tree로, 메타데이터 관리만 Appen-only 로 관리. B+tree의 인덱스 업데이트에 따른 인덱스 write를
줄이기 위한 추가적인 최적화. 메타데이터 update 양을 줄이기 위한 추가적인 최적화. 

\module{Idea}

특별한 아이디어는 없는듯. Object level atomicity가 보장된다는 장점은 있음. 
당연히 FS의 space management 를 걷어내고 전부 FTL 내부에서 space management를 함. 
Sequentiality, hot/cold separation에 대한 아이디어는 없음.

Byte-unit access 인터페이스 추가하자 => 메타데이터 변경이나 페이지의 일부 변경은 이걸로 전달해서, block IO 없이
효과적으로 전달하자. => 이런 데이터들은 diff-page라는 따로 분리된 영역에 log처럼 기록하고 나중에 merge. => IPL 과 유사한 아이디어

object update에 따른 metadata 업데이트도 전부 diff-page에 기록. 이걸로 metadata write 양을 줄임.

\module{차이점 및 문제점}

역시나 object file system이 필요함. FTL도 OS 커널상에 존재하는 거라서 DFS와 동일한 오버헤드가 존재함. 

Object 단위로 space management를 함. => 역시나 OSD와 같은 문제점이 발생. 

{\bf In summary,} the idea of fallocate can be seamlessly embeded into all the frameworks above. 

Compared to existing file system approaches, fallocate has several benefits. First, it can be easily implemented by modifying the FTL, which is much light-weight than other approaches. Second, once implemented, many existing applications leverage the fallocate (적용, 활용 절대적 간단). For instance, in In the case of an application already with fallocate, no need to chage its code. Third, fallocate can be used for (or can be adpated to) various and diverse workloads, almost at no or minimal changes. 
}
\comment{
\subsection{Hot/Cold Separation FTLs} 

Since the inception of flash storage, numerous hot/cold separation FTL (flash translation layer) schemes~\cite{DAC,Stoica,2R} have been proposed, which classify pages according to the update hotness and then place them into different regions. While those FTLs and \falloc commonly aim at reducing write amplification \comment{by placing different pages into different streams} transparently (\ie, without any hint from applications), they are proposed to address different problems. \comment{which is hard to effectively address with other scheme.} 
Hot/cold separation FTLs focus on ... while the main usage of \falloc is to prevent a logical object with almost similar lifetime from
being multiplexed with other pages to the same flash block. Therefore, hot/cold separating FTLs are not such effective in reducing write amplification in RocksDB (mainly because XXXX)~\cite{2R} and \falloc is useless for workloads with random and unpredictable hot/cold pages (\eg, OTLP workloads). In this sense, it is desirable for flash storages to support both schemes which are complementary to each other.
}
\section{Conclusion}\label{sec:conc}

Existing flash devices are {\em object-oblivious} in handling writes and thus allow to colocate data from different objects \comment{with different lifetimes} in the same flash block, causing high write amplification. 
To remedy such write multiplexing problem, we proposed a novel interface, \falloc, which is used to enlighten flash devices to de-multiplex writes from different logical objects into different flash blocks (\ie, {\em object-aware} in handling writes), thus minimizing write amplification. 

To verify the effect of \falloc, we have prototyped \falloc on a real SSD board by extending its conventional FTL firmware and also modified a set of representative software stacks 
so as to use the \falloc interface. Experimental results have confirmed that \falloc can enable RocksDB and F2FS to eliminate write amplification, realizing their full potential of flash-friendliness. Also, we have demonstrated that \falloc is effective in mitigating the performance interference between multi-tenant applications. 

\comment{\subsection{Future Work} \falloc은 비단 GC 줄이기 위해 clustering 이외에도, dynamic page allocation~\cite{DynamicPageAllocationInSSD,PageAllocationsInSSD}을 위한 정보로도 활용가능. 즉, \falloc 정보를  활용해서 어떤 page가 어떤 segment에 속하는지 알면 이를 활용해서 channle/way/die/plan Allocation을 semantic하게 수행 자능함.  
혹시 OS fallocate() (OS는 이 할당 단위를 뭐라 (extent/segment) 부르나? fd 정보도 주면 segment/extent 더 큰 file 단위로 clustering 포함해 관리 가능!!}

\comment{
\section*{Acknowledgment}
We thank the anonymous PVLDB reviewers for their valuable comments.
This work was supported by Samsung Research Funding \& Incubation Center of Samsung Electronics under Project Number SRFC-IT1802-07.}

\bibliographystyle{ACM-Reference-Format}
\balance
\bibliography{sigproc}

\end{document}